\documentclass[reprint,aps, pre, footinbib, superscriptaddress,amsmath,amssymb,floatfix,nobibnotes]{revtex4-2}
\usepackage{graphicx}% Include figure files
\usepackage{dcolumn}% Align table columns on decimal point
\usepackage{bm}% bold math
\usepackage{hyperref}
\usepackage{color}
\usepackage{float}

\newcommand{\matnot}[1]{{\bm{\mathrm{#1}}}}

\newcommand{\vecnot}[1]{\bm{#1}}
\newcommand{\hatnot}[1]{\bm{\hat{#1}}}

\begin{document}
\preprint{APS/123-QED}

% Used that for the PRL instead
%\title{Critical transition to oscillations in self-aligning active disordered solids}% Force line breaks with \\
\title{Active elastic theory of self-aligning solids}

\author{Sander C. Kammeraat}
\affiliation{Instituut-Lorentz for Theoretical Physics, Universiteit Leiden, 2333 CC Leiden, Netherlands}

\author{Silke Henkes}%
 \email{shenkes@lorentz.leidenuniv.nl}
\affiliation{Instituut-Lorentz for Theoretical Physics, Universiteit Leiden, 2333 CC Leiden, Netherlands}

\date{\today}% It is always \today, today,

\begin{abstract}
Active disordered solids including dense human crowds and epithelial cells under confinement exhibit striking system-scale oscillatory spatiotemporal patterns. These are linked to a local feedback mechanism, self-alignment, that aligns the direction of a particle's motility vector to the total force. Simulations and experiments of solids made of such agents show these spontaneous oscillation patterns. Theoretically, they have been linked to both a nonlinear bifurcation and to selection of long-wavelength normal modes. Here we derive a closed nonlinear equation for the displacement field of active self-aligning 2d solids subject to angular noise. Along the normal modes of the solid, the dynamics of the mode amplitudes correspond to nonlinearly damped and stochastically driven harmonic oscillators. 
To linear order, we show that the system transitions from Active Brownian type correlated motion to oscillatory motion that increasingly condenses onto the lowest modes of the solid. We compare the analytical predictions for the mode spectra with simulations, finding excellent agreement approaching the transition from the disordered side. Strong nonlinearities manifest deep in the oscillating phase at strong alignment and small noise, consistent with the previous observations. At the continuum level, we derive a closed-form nonlinear wave equation for self-aligning solids. The transition to undamped oscillations is a second order dynamical phase transition driven by the competition between noise and alignment.  At the linear level, we predict travelling acto-elastic waves, together with the emergence of system-scale oscillations for confined systems, consistent with observations in tissues and crowds. Our framework extends the understanding of self-aligning solids, which are pervasive among artificial and biological systems across multiple scales.
\end{abstract}

\maketitle

%\tableofcontents

\section{Introduction}
Active matter displays a wide variety of emergent collective motion patterns~\cite{marchetti2013hydrodynamics,gompper20252025}, with flocking as the most familiar example \cite{cavagna2014bird}. More recently, the emergent behavior of soft \emph{active solids} has gained significant attention, as a distinct class of active material. In these active matter systems, the activity at the level of the constituents, coupled to soft elastic interactions between them, allows for rich dynamical behavior \cite{dauchot2026active}. A better understanding of soft active solids is a prerequisite for gaining insight into developmental biology, where mechanics, signalling and activity all coordinate to shape a single cell into the developing embryo \cite{petridou2019tissue}. Similarly, engineering functionality at the micro- or centimeter scale requires the active (self)-assembly of components into units capable of generating both forces and targeted motion \cite{wei2026life,xi2024emergent}. 

Active solids have been studied intensively for simple types of activity, especially Active Brownian and Ornstein-Uhlenbeck types of driving, giving rise to active versions of the glass and jamming transitions \cite{janssen2019active}. While these phases are modified through the emergence of mesoscopic space and time correlations \cite{henkesDenseActiveMatter2020,caprini2020spontaneous,keta2024emerging}, they phenomenologically resemble their passive and sheared counterparts in the classic jamming diagram \cite{berthier2014nonequilibrium,nandi2018random} except for very long active persistence times \cite{mandal2020extreme,morse2021direct,keta2023intermittent}. To a large extent, this is due to the absence of feedback on the active driving, allowing formal mapping to a memory kernel, with the result that systems recover thermal-like properties at long wavelengths. In contrast, the more striking emergent active patterns like flocking, moving topological defects and wave propagation, when studied using hydrodynamics \cite{doostmohammadi2018active} all rely on coupling of the activity with the fluid or solid properties of the material. Yet, at the microscopic scale, the universality of hydrodynamic couplings breaks down, with a zoo of possible active interactions that are not easily classifiable as simply `polar' or `nematic' \cite{bruckner2022geometry,vagne2025generic}.
A different perspective on active solids and liquids has in recent years emerged through the study of odd and non-reciprocal active materials \cite{fruchart2023odd,dauchot2026active}. For odd materials, the Onsager relations and symmetry constraints on the stress tensor \cite{saarloos2024soft} do not apply \cite{scheibner2020odd}, allowing for a non-symmetric stiffness matrix with exotic couplings between shear and rotations, an idea with wider implications in biological materials \cite{vagne2025generic}. In addition to computational implementations \cite{huang2023odd}, experimental versions of these materials have recently been engineered \cite{tiwari2026reentrant,du2026metamaterials}. The related but distinct idea of \emph{non-reciprocal} interactions \cite{fruchart2021non}, though likely paraphyletic as a concept, explores asymmetric couplings between different degrees of freedom, e.g. chiral interactions \cite{tan2022odd}, and non-symmetric coupling matrices in spin systems \cite{loos2023long}. Deep connections to the theory of the nonlinear dynamics of pattern forming systems have also recently been recognised \cite{veenstra2025wave}.

However, one frequently observed phenomenon in both artificial and biological systems, across multiple scales,  has remained without clear explanation so far:
%On this background of recognition of active solids as a distinct class of material enters a phenomenon that has been observed many times but has remained without clear explanation: 
active mechanical wave propagation and oscillations of the constituent's movements in a confined solid. The confinement is a crucial ingredient: at lower density or in the absence of boundaries  the system flocks instead \cite{baconnier2025self}. A dangerous example of the emergence of these active mechanical waves was first described in crowd crush conditions at the  Hajj \cite{helbing2007dynamics}, where they take the form of pressure-waves. Furthermore, the emergence of these oscillations in crowds has recently been confirmed and carefully characterized in an ingenious observational study of the San Fermin festival \cite{gu2025emergence}. Arguably the same phenomenon also arises in epithelial cell sheets \cite{peyretSustainedOscillationsEpithelial2019,petrolli2019confinement} and biofilms \cite{xu2023autonomous}. It can be engineered too, for example in the form of active mechanical networks \cite{baconnier2022selective,baconnier2023discontinuous,baconnier2024noise,baconnier2025collective,baconnier2025reentrant}, where it has been dubbed `collective actuation'.  

These oscillations in confinement emerge from a very generic coupling, called self-alignment, by which an active particle locally aligns its motility vector to the total force it experiences. Initially regarded as an alternative mechanism that induced flocking \cite{szabo2006collective}, collective oscillations were only discovered later \cite{henkes2011active}. For a rigid driven body, such as hexbots and engineered active particles, it emerges from either anisotropic force or friction distributions. For cells, self-alignment emerges from processes such as contact inhibition of locomotion or plithotaxis, involving feedback between stress and the polarization of the cytoskeleton, and is thought to be responsible for the effect of global oscillations in epithelial cell sheets.  
Self-alignment has been added to vertex models to understand both flocking epithelia \cite{malinverno2017endocytic,giavazzi2018flocking}, and also oscillations in confinement \cite{barton2017active,petrolli2019confinement}. The glass transition of self-aligning systems is only beginning to be understood \cite{paoluzzi2024flocking}.
Please see \cite{baconnier2025self} for a detailed review of systems that exhibit self-alignment. 

However, how exactly these oscillations emerge from the microscopic equations of motion has remained a challenging question to address analytically due to the nonlinearity of the alignment torque and the fact that the naturally present noise on the particle's orientation is multiplicative.
For ordered, noiseless systems, nonlinearities dominate the dynamics, and in a normal modes formalism, a drift-pitchfork instability occurs in a select sequence of modes as a function of active coupling strength \cite{baconnier2022selective, baconnier2023discontinuous,baconnier2025reentrant}. Meanwhile, a mean-field version of the model \cite{baconnier2024noise, gu2025emergence} is able to recover the flocking transition (as the lowest normal mode), and with a symmetry breaking mechanism akin to a Goldstone mode. More recently, steps have been taken towards a more formal field theoretical treatment of the transition, albeit without the elastic interaction terms \cite{musacchio2026flocking}.

The first analytical treatment of self-alignment, by one of us \cite{henkes2011active}, also employed a normal modes formalism. By linearising the active dynamics, it mapped the system to a set of damped harmonic oscillators, with the damping vanishing for the lowest normal modes. This mechanism then explained the condensation on the lowest normal modes, with an oscillation frequency determined by the mode stiffness and the alignment strength.

\begin{figure}[t]
    \includegraphics[width=\linewidth]{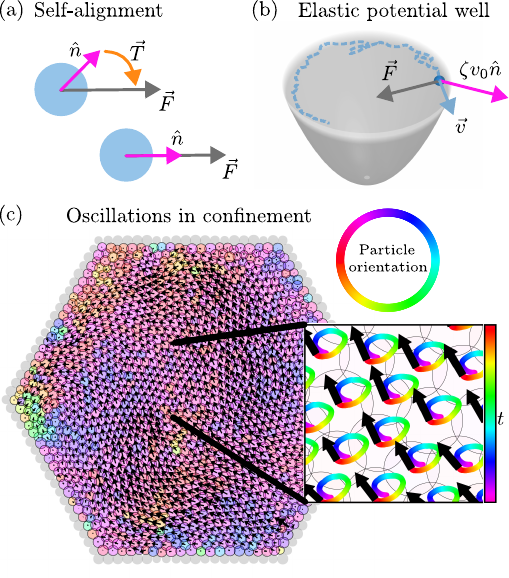 }
    \caption{(a) Self-alignment refers to the torque $\vecnot{T}$ on a particle that aligns its polarity vector $\hatnot{n}$ to the total force $\vecnot{F}$ it experiences, which in the overdamped limit corresponds to alignment to the velocity vector. (b) Close to the transition to oscillations, particles orbit the effective elastic potential well defined by its neighbors, with the elastic force $\vecnot{F}$ almost balancing the self-propulsion force $\zeta v_0 \hatnot{n}$, giving the resulting velocity vector a small non-zero tangential component. This dynamical state is the starting point of the linearization procedure. (c) These particle oscillations appear synchronized over the span of the system, which in simulations ($J=0.1, \tau=100$, SI Movies \cite{SI}, S7) is made out of soft-repulsive disks with pinned boundary particles (grey disks). In disordered solids, these collective oscillations occur along the lowest modes of the solid. Black arrows indicate velocity vectors and the disk color is the orientation of the polarity vector $\hatnot{n}$. The inset represents the particle trajectories over time (time indicated by color) in the oscillating phase.}
    \label{fig:schematics}
\end{figure}

Here, we pick up these early results and derive a fully general, nonlinear equation for the overdamped self-aligning dynamics of a two dimensional solid, including the multiplicative noise terms. This allows us to reconcile and unify the disparate theoretical approaches to the phenomenon into one generic active solids framework. At the linear level along the normal modes of a discrete solid, we recover damped driven harmonic oscillator dynamics, and derive an exact fluctuation spectrum for each mode and frequency. In the limit of low active coupling, it matches the analytical results for active Brownian solids. The transition to oscillations corresponds to a condensation of the spectrum on the lowest modes, leading eventually to a nonlinear regime. Our predictions are in excellent agreement with numerical simulations, for both confined and flocking system with periodic boundary conditions, and for two variants of self-alignment coupling. Our results also allow us to identify nonlinearities as arising from mode coupling.
We derive the equivalent continuum theory using linear elasticity, and show that self-aligning systems have spectra that include both diffusive and propagative parts. In this framework, oscillations emerge in a critical phase transition where the diffusive terms vanish for the longest wavelengths in the system, heralding the onset of acto-elastic wave propagation. Through a self-consistent calculation of the mean square velocity of the system, we are able to derive the boundaries between diffusive, underdamped and then a critical transition towards freely oscillating regimes. The emerging phase diagram as a function of coupling strength and noise amplitude explicitly depends on system size and the specific alignment torque implementation, and is qualitatively different from earlier results due to treating both noise and elasticity within our framework. 

Our results make quantitative predictions, including velocity and polarisation spectra, for dense crowds and epithelial cell sheets. They are applicable to all systems near the transition to collective oscillations and open the way to a more formal study of its critical properties. We also shed light on the mode-coupling mechanisms responsible for the strongly nonlinear oscillations seen in small, low-noise systems. Simultaneously, the nonlinear continuum equation we derive appears distinct from known pattern forming ones, though there are likely deep formal connections \cite{dauchot2026active}. 

This article is organized as follows: We begin with a derivation of the theory for discrete solids, using a normal modes formalism, followed by testing it against simulations and exploring the phase transition and nonlinearities. We then derive the continuum version of the theory and focus on active elastic wave propagation and the critical phase transition, deriving the phase diagram and ending with an outlook and discussion. 

%%%%%%%%%%%%%%%%%%%%%% readability %%%%%%%%%%%%%%%%%%%%%%%%%

\section{Theory of discrete self-aligning disordered active solids}
\subsection{Equations of motion}
Physical realizations of active solids in nature or engineering exhibit different repulsive and/or attractive interactions between the particles, agents or individuals that make up the solid. Here we consider a very general form of a polar active solid, consisting of $N$ Active Brownian particles (ABPs) moving over a surface, with overdamped Langevin dynamics for the position vector $\vecnot{r}_i$, $\zeta \dot{\vecnot{r}}_i = \sum_{j=1}^N \vecnot{F}_{ij} + \zeta v_0 \hatnot{n}_i$ \cite{marchetti2016minimal}, where the overdot denotes differentiation with respect to time, $\zeta$ is the substrate friction coefficient, $v_0$ is the self-propulsion speed, while $\hatnot{n}_i(t)$ is the polarity vector of particle $i$, which defines the direction of the self-propulsion. The sum is over all $N$ particles that make up the solid and the interaction force of particle $j$ on $i$ is written as $\vecnot{F}_{ij}$.   
To keep our discussion generally applicable, we assume that the mechanical (or social, or active) pair interactions between agents can be cast in the form of a dynamical matrix $\matnot{D}$ of the solid, which for conservative forces consists of the second derivatives of the potential, and takes the form of a stiffness matrix on the graph Laplacian of the interaction network. Denoting the reference positions $\{\vecnot{r}_{i,0}\}$ in our 2-dimensional solid in the $xy$-plane with normal $\hatnot{z}$, the equation of motion for the displacement vector $\delta\vecnot{r}_i(t) \equiv \vecnot{r}_i(t) - \vecnot{r}_{i,0}$ of particle $i$ around the reference state to linear order becomes

\begin{equation}
    \zeta\delta \dot{\vecnot{r}}_i(t)= -\sum_{j=1}^N \matnot{D_{ij}} \ \delta\vecnot{ r}_j(t) + \zeta v_0 \hatnot{n}_i(t),
\label{eq:discrete_displacement_eom}
\end{equation}
where $\matnot{D_{ij}}$ is the 2 by $2$ submatrix of $\matnot{D}$ that couples the movement of particle $j$ to its force on $i$.

The agent's polarisation is tracked by the unit vector $\hatnot{n}_i$, which makes an angle $\theta_i$ with the $x$-axis. Its overdamped dynamics corresponds to an equation of motion $\dot{\theta}_i = \frac{1}{\zeta_r} \vecnot{T}_i \cdot \hatnot{z} +\sqrt{\frac{2}{\tau}}\eta_i(t)$, where $\vecnot{T}_i$ is the total torque on the agent and $\eta_i(t)$ is an angular white noise with zero mean and unit variance. We work in units where the rotational mobility $\zeta_r^{-1}=1$. In the absence of torques, we recover the dynamics of an active solid made of Active Brownian Particles (ABPs) with persistence time $\tau$ (the inverse of the rotational diffusion constant: $1/\tau = D_r$), as considered in Ref.~\cite{henkesDenseActiveMatter2020}.

Self-alignment enters as the alignment of the polarity vector of each particle to its velocity vector $\vecnot{v}_i \equiv\dot{\vecnot{r}}_i = \delta \dot{\vecnot{r}}_i$, which in the overdamped limit is equivalent to alignment to the total force on the particle, see Fig.~\ref{fig:schematics}a. There are two common formulations for the torque $\vecnot{T}_i$ that is responsible for this alignment \cite{baconnier2025self}. The first is alignment based only on the orientation of the  velocity vector, \textit{i.e.} $\vecnot{T}_i = J \hatnot{n}_i \times \hatnot{v}_i$, with $J$ the strength of alignment. Taking respectively $\phi$ and $\theta_i$ as the polar angle of $\hatnot{v}_i$ and $\hatnot{n}_i$  in the $xy$-plane, the same torque is equivalently written as $\vecnot{T}_i = J\sin(\phi_i - \theta_i)\hatnot{z}$ \cite{szabo2006collective,henkes2011active}. The second is alignment where the magnitude of the velocity vector influences the strength of the alignment, \textit{i.e.} $\vecnot{T}_i = J \hatnot{n}_i \times \vecnot{v}_i= J|\vecnot{v}_i|\sin(\phi_i - \theta_i)\hatnot{z}$. 
We consider here the former case, hereafter referred to as unit-alignment, explicitly, but all stated results carry over to the latter case, referred to as full-alignment, by substituting $J\rightarrow J|\vecnot{v}_i|$. 
Altogether, the equation of motion for the polarity unit vector of each particle reads
\begin{align}  
    \dot{\hatnot{n}}_i (t)= &J\left[\hatnot{n}_i(t)\times\hatnot{v}_i (t)\right]\times \hatnot{n}_i(t) \nonumber \\
    &- \frac{1}{\tau}\hatnot{n}_i(t)+\sqrt{\frac{2}{\tau}}\eta_i(t) \hatnot{z}\times\hatnot{n}_i(t),
    \label{eq:discrete_polarity_eom}
\end{align}
where we note that the terms without $J$ follow from It\^{o} calculus due to the presence of the angular white noise with zero mean $\langle \eta_i(t)\rangle=0$ and variance $\langle \eta_i(t) \eta_j(t') \rangle = \delta_{i,j}\delta(t-t')$. The choice for the It\^{o} formulation allows to conveniently calculate ensemble averages, denoted by $\langle \ldots \rangle$, as we do below, since the variables and noise are uncorrelated at the same time $t$.

\subsection{Nonlinear displacement dynamics}
Having introduced the general governing equations, we now derive an exact decoupled equation for the displacement vectors of each particle. First, following an approach developed in \cite{jung1987dynamical,
caprini2020spontaneous, capriniChiralActiveMatter2023,caprini2025odd,musacchio2026flocking}, and that mirrors the first theory of self-alignment \cite{henkes2011active}, we take the time derivative of the displacement equation Eq.~\eqref{eq:discrete_displacement_eom}, substituting in Eq.~\eqref{eq:discrete_polarity_eom}, and obtain

\begin{align}
    \zeta \delta\ddot{\vecnot{r}}_i =&-\!\sum_j \matnot{D_{ij}} \  \delta \dot{\vecnot{r}}_j %\label{eq:nonlinear_displacement_with_n} \\ 
    + \zeta v_0 \left[ \frac{J}{v_i} \left(\delta \dot{\vecnot{r}}_i - (\hatnot{n}_i \cdot \delta \dot{\vecnot{r}}_i)\hatnot{n}_i\right) \!-\! \frac{1}{\tau} \hatnot{n}_i \right]  \nonumber \\
& + \zeta v_0 \sqrt{\frac{2}{\tau}} \eta_i(t) \hatnot{z}\times \hatnot{n}_i, \label{eq:nonlinear_displacement_with_n}
\end{align}
where $v_i = |\delta \dot{\vecnot{r}}_i|$ is the particle velocity magnitude and we have taken advantage of the identity $(\hatnot{n}_i \times\hatnot{v}_i)\times \hatnot{n}_i  = (\hatnot{n}_i \cdot \hatnot{n}_i) \hatnot{v}_i - (\hatnot{n}_i \cdot \hatnot{v}_i ) \hatnot{n}_i=\hatnot{v}_i -(\hatnot{n}_i \cdot \hatnot{v}_i)\hatnot{n}_i $ since $\hatnot{n}_i$ is a unit vector.
This establishes that the active coupling is nonlinear, with the nonlinearity coming from the the projection of the velocity vector onto the polarity vector, $(\hatnot{n}_i \cdot \delta \dot{\vecnot{r}}_i)\hatnot{n}_i$. The last term shows that the driving noise is always along $\hatnot{z}\times \hatnot{n}_i$, the direction orthogonal to the director. Both nonlinearities are due to the normalisation constraint on $\hatnot{n}_i$, and will change if other angular equations are used \cite{lacroix2024emergence}.

We then use Eq.~\eqref{eq:discrete_displacement_eom} again to eliminate $\hatnot{n}_i$,
and eventually obtain (Appendix, \ref{apx:nonlinear_displacement_dynamics}) the following decoupled equation, using repeated index summation convention for index $j$ and $k$:

\begin{align}
\delta \ddot{\vecnot{r}}_i =& -\left[ \frac{\matnot{D_{ij}} }{\zeta} \!+\!\frac{\matnot{\delta_{ij}}}{\tau} +J \matnot{\delta_{ij}} \left(\frac{v_i}{v_0}\!-\!\frac{v_0}{v_i} \!+\!\frac{\matnot{D_{ik}} \ \delta \vecnot{r}_k}{\zeta v_i v_0} \!\cdot\!\delta \dot{\vecnot{r}}_i \right) \right] \delta \dot{\vecnot{r}}_j \nonumber\\
&-\left[ \frac{1}{\tau} + J \left(\frac{v_i}{v_0} + \frac{\matnot{D_{ik}}\  \delta \vecnot{r}_k}{\zeta v_i v_0}  \cdot \delta \dot{\vecnot{r}}_i\right)\right] \frac{\matnot{D_{ij}}}{\zeta}   \delta \vecnot{r}_j  \nonumber \\
&+ \sqrt{\frac{2}{\tau}} \eta_i \hatnot{z}\times \left(\delta \dot{\vecnot{r}}_i + \frac{\matnot{D_{ij}}}{\zeta} \delta \vecnot{r}_j\right).
\label{eq:discrete_decoupled}
\end{align}

The resulting equation for full alignment is readily obtained by replacing  $J\rightarrow Jv_i$.

\subsection{Linearization}
\paragraph*{Linearization around elastic and active force balance.}
Eq.~\eqref{eq:discrete_decoupled} is an exact decoupling of the displacements from the polarity dynamics equation, representing coupled, nonlinear damped-driven oscillators. It is difficult to analyze due to the nonlinear terms stemming from self-alignment. The simplest approximation is to  neglect all nonlinear elastic terms, which is a good approximation for the flocking state, as we will demonstrate below. Alongside also neglecting all elasticity-related terms of $\matnot{D_{ij}}$ with the displacement vectors, this is the approach taken by Ref.~\cite{musacchio2026flocking}. As we will show below, this is equivalent to a mean-field description of the flocking phase. 
Inside the solid phase, however, there are significant elastic distortions due to the competition between active driving and interactions. In the case of a single particle in a harmonic well, this manifests itself in two possible dynamical regimes, depending on the strength of the self-alignment \cite{dauchot2019dynamics}. The first is the \emph{climbing} state in which the particle stays at a finite distance from the center of the well and diffuses in the azimuthal direction. The second is the \emph{orbiting} state in which the particle orbits the well, again at a finite distance from the center. In both cases, the active force and the elastic restoring force cancel up to a remnant that is orthogonal to the elastic force, as shown in Fig.~\ref{fig:schematics}b. Inspired by this almost-force-balance for a single particle in a harmonic well, we develop a linearization for the multi-particle solid, which we detail in (Appendix, \ref{apx:linearization}). The result is that for the deviations $\delta \vecnot{r_i}$ around this almost-force-balance state (compare Fig.~\ref{fig:schematics}b to SI Movies \cite{SI}, S1-S3 to observe these almost-force balance-states in simulations), we obtain 
\begin{align}
\delta \ddot{\vecnot{r}}_i =& -\left[ \frac{\matnot{D_{ij}} }{\zeta} \!+\!\frac{\delta_{ij}}{\tau} +J \matnot{\delta_{ij}} \left(\frac{v_i}{2v_0}\!-\!\frac{v_0}{2v_i} \right) \right]\delta \dot{\vecnot{r}}_j \label{eq:discrete_decoupled_linearized}\\
&-\left[ \frac{1}{\tau} + J \frac{v_i}{2v_0}\right]\frac{\matnot{D_{ij}}}{\zeta} \  \delta \vecnot{r}_j  %\nonumber \\
+ v_0\sqrt{\frac{2}{\tau}} \eta_i \hatnot{z}\times \hatnot{n}_i, \nonumber
\end{align}
in which we have reintroduced $\hatnot{n}_i$ in the noise term for the next step: the projection onto the normal modes of the solid.

\paragraph*{Projection onto the normal modes.}
Inspired by the observed steady state dynamics in simulations for both ABP-like and oscillating regimes, we consider a constant, spatially homogeneous particle speed $v_i=v$, which by assumption is then equal to the root mean square (RMS) velocity of the particles. This allows us to write Eq.~\eqref{eq:discrete_decoupled_linearized} as a linear matrix equation, which we can diagonalize by projecting the particle displacements onto the eigenvectors of the dynamical matrix $\matnot{D}$.

For a solid in confinement, $\matnot{D}$ is symmetric and positive-definitive with $2N$ eigenvalues $\{\lambda_{\nu} > 0 \}_{\nu=1,2,..,2N}$ (ordered by increasing value), and the corresponding eigenvectors $\{\vecnot{\chi}_{\nu}\}_{\nu=1,2,..,2N}$ form a basis. Consequently, we can decompose the displacement vectors into a linear combination of the normal modes, \textit{i.e.} $\delta\vecnot{R} = \sum_{\nu} a_{\nu}(t) \vecnot{\chi}_{\nu}$, where $\delta \vecnot{R}$ is the $2N$-dimensional displacement vector constructed using the tensor products of the individual displacement vectors: $\delta \vecnot{R}=\delta\vecnot{r}_1 \otimes \delta \vecnot{r}_2 \otimes \ldots \otimes \delta \vecnot{r}_N$ and in which $a_{\nu}(t)$ are the mode amplitudes. Substituting this decomposition into Eq.~\eqref{eq:discrete_decoupled_linearized}, and projecting onto a single $\vecnot{\chi}_{\nu}$ we obtain the dynamics of each mode amplitude $a_{\nu}$,
\begin{align}
\ddot{a}_{\nu}(t) = &- \left( \frac{\lambda_{\nu}}{\zeta}+ \frac{1}{\tau} - \frac{J v_0}{2v}  + \frac{J v}{2v_0}  \right) \dot{a}_{\nu}(t) \nonumber \\
    &-\left( \frac{1}{\tau} + \frac{J v}{2v_0} \right)\frac{\lambda_{\nu}}{\zeta} a_{\nu}(t) + v_0 \sqrt{\frac{2}{\tau}} w_{\nu}(t),
\label{eq:discrete_mode_dynamics}
\end{align}
Here we have approximated the noise projection term $\vecnot{\mathcal{N}}(t)\cdot \vecnot{\chi}_{\nu}\equiv  \sum_i \eta_i(t)(\hatnot{z} \times \hatnot{n}_i) \cdot \vecnot{\chi}_{\nu}^i$, with $\vecnot{\chi}_{\nu}^i$ the part of $\vecnot{\chi}_{\nu}$ that lies in the subspace spanned by the displacements of particle $i$, by white noise $\vecnot{\mathcal{N}}(t)\cdot \vecnot{\chi}_{\nu} \approx w_{\nu}(t)$ with $\langle w_{\nu}(t) \rangle=0$ and $\langle w_{\nu}(t) w_{\rho}(t') \rangle = \frac{1}{2} \delta_{\nu, \rho}\delta(t-t')$. 

The reasoning behind this simplification of the noise is that in the ABP limit with $J=0$, the polarity vectors are uncorrelated between particles. The projection of the noise term onto the eigenvector $\vecnot{\chi}_\nu$ is then equivalent to summing $N$ independent identically distributed random variables $\eta_i$ in some deterministic way. The factor of $\frac{1}{2}$ is due to there being $2N$ normal modes, but only $N$ angular degrees of freedom. Note that this is different from simplifying the noise at the very beginning, as we retain the persistence-related $\tau$-terms.

\subsection{Linearized mode dynamics}
\paragraph*{Damped driven harmonic oscillator dynamics.}
The key insight of Eq.~\eqref{eq:discrete_mode_dynamics} is that the self-aligning solid can be understood in terms of its normal modes as a collection of damped harmonic oscillators that are driven by noise. They map to Brownian oscillators of the form $\ddot{a}_{\nu} = - \Gamma_{\nu} \dot{a}_{\nu} + \Omega^2_{\nu} a_{\nu} + v_0 \sqrt{\frac{2}{\tau}} w_{\nu}(t)$, with a non-trivial, mode-dependent, damping term $\Gamma_{\nu}$  and natural frequency $\Omega_{\nu}$, both of active origin: 
\begin{align} 
&\Gamma_{\nu} = \frac{\lambda_{\nu}}{\zeta}+ \frac{1}{\tau} - \frac{J v_0}{2v}  + \frac{J v}{2v_0} \label{eq:gamma_nu}, \\
&\Omega_{\nu}^2 =\left(\frac{1}{\tau} + \frac{J v}{2v_0} \right)\frac{ \lambda_{\nu}}{\zeta}. \label{eq:omega_nu} 
\end{align}
The qualitative dynamics of each normal mode is then determined by the sign of $\phi_{\nu} = \Omega_{\nu}^2 -\Gamma_{\nu}^2/4 $, just as for Brownian oscillators (\textit{cf.} Refs. \cite{chaikin2000principles,norrelykke2011harmonic} or Appendix, \ref{apx:velocity_autocorrelation}). For $\phi_{\nu}<0$, the normal mode $\nu$ behaves as an overdamped oscillator; meaning that there are no oscillations. If $\phi_{\nu}=0$, the mode $\nu$ is critically damped with no oscillation, and finally, when $\phi_{\nu}>0$, the normal mode is an underdamped oscillator, which oscillates at frequency $\omega_{\nu}\equiv\sqrt{\phi_{\nu}}= \sqrt{\Omega_{\nu}^2 -\Gamma_{\nu}^2/4}$.

Crucially, the mode-dependent damping $\Gamma_{\nu}$ and natural frequency $\Omega_{\nu}$ depend on the eigenvalue $\lambda_{\nu}$ of the mode and also on the alignment strength $J$, noise persistence time $\tau$ and finally the speed $v$. Therefore, the specific values of these parameters determine the nature of each mode. We note that the lowest mode, i.e. the one with the smallest eigenvalue $\lambda_{\nu}$, always has the least amount of damping $\Gamma_{\nu}$. Qualitatively, a growing $J$ decreases $\Gamma_{\nu}$, and at a critical $J$, the damping $\Gamma_{\nu}$ becomes zero, implying undamped oscillations. Since the damping is smallest for the lowest mode, both the transition to oscillations $\phi_{\nu}>0$ of the least damped mode and the transition to undamped oscillations $\Gamma_{\nu}=0$ occur first for the lowest mode when increasing $J$.

\begin{figure}
    \includegraphics[width=1\linewidth]{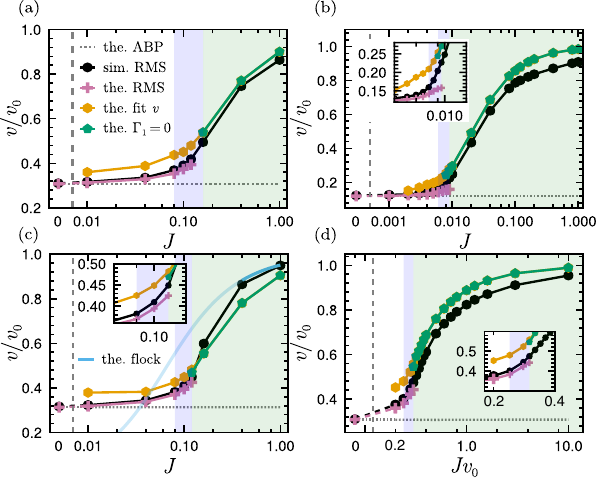}
\caption{Scaled root mean square (RMS) velocity as a function of alignment strength $J$. Shown are the empirical RMS velocity from simulations (black dot markers), the RMS velocity according to the exact ABP theory for $J=0$ (horizontal dotted line), which due to the horizontal log scale is placed on an arbitrary $J$-value on the left. Also shown is the fitted $v$-value obtained by fitting the velocity projection of (only) the lowest mode (orange hexagon markers). Furthermore we show the resulting RMS value from the linearized theory obtained by summing the mode amplitudes, \textit{i.e.} the RHS of the consistency relation (pink plus markers) and finally the values of $v$ that make the lowest mode undamped, \textit{i.e.} $\Gamma_{1}=0$ (green pentagon markers). Blue shade on the left correspond to the underdamped regime, green shade on the right to the undamped and nonlinear regime.  Panels correspond to different systems (a) Confined system with unit alignment, $D_r=0.1$ and (b) $D_r=0.01$. (c) Periodic (flocking) system with unit alignment, $D_r=0.1$. The blue line shows the continuum prediction for $v$ in the undamped regime obtained by expanding around the deformation free state (flocking state) instead of the elastic and active force balance.  (d) Confined system with full alignment, $D_r = 0.1$. }
\label{fig:a_Dr0.1_w_pred}
\end{figure}

\paragraph*{Role of the mean-square velocity.}
The root mean square velocity (RMS) $v$ is determined by a self-consistency relation for the system's kinetic energy, $v^2 = \langle v_i^2(t) \rangle_{i,t} = \frac{1}{N} \sum_{\nu} \langle |\dot{a}_{\nu}(t)|^2 \rangle $, which is valid in linear response and at steady-state.
Here $\langle |\dot{a}_{\nu}^2(t)| \rangle$ corresponds to the equal-time auto-correlation function of a Brownian damped oscillator, with a standard analytical solution (see Refs.~\cite{chaikin2000principles,norrelykke2011harmonic} and Appendix, \ref{apx:velocity_projections}). However, to compute the sum over the modes we first need to know the normal modes of the particular solid packing. This is not analytically tractable for disordered jammed packings \cite{silbert2005vibrations}, and we instead leave $v$ as a fit parameter to compare to simulations. 
In contrast, in the continuum theory discussed later, we are able to solve this self-consistency equation and obtain the full solution to the mode dynamics. As we show there, and in Fig. \ref{fig:a_Dr0.1_w_pred}  for four different simulated systems (to be discussed in detail later),  $v$ acts as an order parameter that increases with $J$ and goes through a critical transition towards $v\rightarrow v_0$ when oscillations appear.

\paragraph*{Steady-state projections.}
To compare to simulations, we calculate the steady state dynamics of Eq.~\ref{eq:discrete_mode_dynamics} by Fourier transforming it to the frequency domain, and then extracting the velocity projection amplitudes in temporal Fourier space  (Appendix, \ref{apx:velocity_projections}). As long as all $\Gamma_{\nu}>0$, this yields the spectrum of a damped, driven harmonic oscillator,
\begin{align}
    \langle \tilde{\dot{a}}_{\nu}(\omega)  \tilde{\dot{a}}_{\nu}(\omega') \rangle= \frac{v_0^2 \frac{1}{\tau} 2 \pi \delta(\omega+\omega') \omega^2}{ \left(\Omega_{\nu}^2 - \omega^2 \right)^2+ \omega^2 \Gamma_{\nu}^2}.
    \label{eq:discrete_mode_velocity_spectrum}
\end{align}
Inverse Fourier transforming Eq.~\ref{eq:discrete_mode_velocity_spectrum} back to the time domain (Appendix, \ref{apx:velocity_projections}), one obtains the remarkably simple expression for the equal-time steady state velocity projections, 
\begin{equation}
     \langle |\dot{a}_{\nu}(t) |^2 \rangle= \frac{v_0^2}{2 \tau\Gamma_{\nu}}.
\label{eq:discrete_mode_velocity_equal_time}
\end{equation}
We note that for $J=0$, $\Gamma_{\nu}= \frac{\lambda_{\nu}}{\zeta}+ \frac{1}{\tau}$, and Eq.~\ref{eq:discrete_mode_velocity_equal_time} becomes the known exact result for Active Brownian particles \cite{henkesDenseActiveMatter2020,caprini2020spontaneous}. As is apparent from Eq.~\ref{eq:discrete_mode_velocity_equal_time}, there is an explicit $\lambda_{\nu}-$dependence to the square velocity projection on each mode. In contrast, for equilibrium thermal systems at temperature $T$, we simply have equipartition over the mode velocity amplitudes, i.e. $ \langle |\dot{a}_{\nu}(t) |^2 \rangle=\frac{1}{2} k_bT$ \cite{henkes2012extracting,saarloos2024soft}, independent of $\lambda_{\nu}$. Equipartition of energy over the mode velocity amplitudes is thus clearly violated, a hallmark of active matter systems.

Using the Fourier transform of Eq.~\ref{eq:discrete_polarity_eom}, one can also calculate the Fourier spectrum of (for example) the $x$-component of the polarity vector (Appendix, \ref{apx:polarity_spectrum}) as
\begin{equation}
    \langle \tilde{{n}}_x(\omega) \tilde{n}_x(\omega') \rangle= \frac{1}{2 N}\sum_{\nu}\frac{1}{v_0^2}\left[1 + \frac{ \lambda_{\nu}^2}{\zeta^2\omega^2}\right]\langle {\tilde{\dot{a}}}_{\nu}(\omega) \tilde{\dot{a}}_{\nu}(\omega') \rangle.
\label{eq:discrete_polarity_spectrum}
\end{equation}
As we show below, this provides an exceptionally sensitive measure to detect oscillations in the simulated system. We note that the predictions of Eqs.~\ref{eq:discrete_mode_dynamics}-~\ref{eq:discrete_polarity_spectrum} exactly carry over to the case of full-alignment by replacing $J\rightarrow Jv$ in these equations.

The critical conclusion of this picture is that the lowest modes, which involve global motions of the solid, will oscillate with the largest amplitude, explaining why oscillations observed in active self-aligning disordered solids tend to be system-scale. In the remainder of this article,  we first test the validity of the linearization procedure and the insight it provides by comparing to numerical simulations. After that, we develop a continuum theory and solve its linearized dynamics in order to predict the transition to oscillation and understand its nature. 

%%%%%%%%%%%%%%%%%%%%%% readability %%%%%%%%%%%%%%%%%%%%%%%%%
\section{Comparison to simulations}
We simulate a disordered self-aligning solid by confining repulsive self-propelled and self-aligning polydisperse disks in a hexagonal enclosure at high density (packing fraction $\phi = 1.3$, polydispersity = 0.3), where confinement is achieved through a layer of pinned boundary particles (see Fig.~\ref{fig:schematics}c for a snapshot). For flocking systems, we simulate an equivalent system without pinned particles and use periodic boundary conditions. 

The simulated translational equation of motion is $\zeta \dot{\vecnot{r}}_i(t)= \sum_{j=1}^N \vecnot{F}_{ij} + \zeta v_0 \hatnot{n}_i(t)$, while the rotational equation of motion for the angle $\theta$ of $\hatnot{n}_i = \cos(\theta_i)\hat{x}+ \sin (\theta_i)\hat{y}$ is given by $\dot{\theta_i} = J (\hatnot{n}_i \times \hatnot{v}_i)\times \hatnot{z} + \sqrt{\frac{2}{\tau}} \eta_i(t)$, replacing the coupling term with $J (\hatnot{n}_i \times \vecnot{v}_i)\times \hatnot{z}$ for full alignment dynamics. These equations are fully equivalent to Eqs.~\eqref{eq:discrete_displacement_eom}-\eqref{eq:discrete_polarity_eom} after linearization of the potential with the dynamical matrix. The simulation is implemented in the open-source simulation package JAMs~\cite{Kammeraat_JAMs}.  We work in units where $\zeta=1$ and where the mean of the uniform radius distribution is $R=1$. Our particles interact with a repulsive one-sided harmonic spring potential, with stiffness $k=1$ between interior particles and stiffness $k=2$ between boundary and interior particles. This simple potential has well known linear response properties, and has extensively been used to study the granular jamming transition \cite{o2003jamming,wyart2005effects,van2010jamming}.

\begin{figure*}[t]
    \centering
    \includegraphics[width=0.8\linewidth]{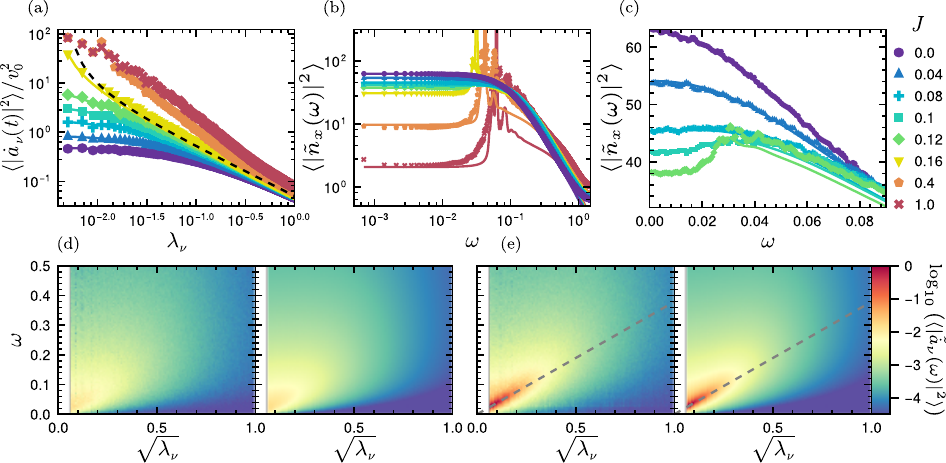}
    \caption{Steady state results for unit-alignment simulations (markers) with $\tau = 10$ compared to theory (lines). (a) Velocity projections. The dashed line corresponds to the critical line with $\Gamma_1=0$. (b) Polarity spectrum (log-log scale). (c) Polarity spectrum zoomed (log-lin scale) to see the emergence of the oscillation peak. (d) Mode spectra of simulations and theory, $J=0.04$. (e) Mode spectra of simulations and theory, $J=0.16$. }\label{fig:predictions_ua}
\end{figure*}

The high density combined with the polydispersity of the disks generates a stable, yet spatially disordered packing of particles at low activity $v_0=0.01$, well below the melting or unjamming transition, so that the system remains solid at all times \cite{fily2014freezing}. This is equivalent to a Hermitian dynamical matrix $\matnot{D}$ with all positive eigenvalues $\lambda_{\nu}$, corresponding to a minimum in the potential energy landscape that is stable in all directions \cite{silbert2005vibrations}. We use an Euler-Maruyama integration scheme with time step $\delta t = 10^{-2}$ and simulate 5 to 10 different configurations per parameter combination with $N=312$ up to $N=6375$ to check consistency between theory and simulations (results shown for $N=1473$) particles in the confinement systems (excluding boundary particles) and $N=2000$ particles in the system with periodic boundary conditions. These system sizes are small, but mimic those found in experimental systems such as cell collectives, human crowds and artificial solids \cite{gu2025emergence,baconnier2025self}. Simulations were run up to $t=10^4$, with sampling frequency $\Delta t = 2$ or $\Delta t = 1$, to generate sufficient dynamically independent samples. After a relaxation step without activity to generate the disordered initial particle positions, we turn on activity and run the simulation for various values of $J$ (either with unit- or full-alignment) well past the time it takes to reach steady state, as indicated by the convergence of the root mean square velocity. Afterwards, we turn off the activity, let the system relax again and use this final state as reference state $\{\vecnot{r}_{i,0}\}$ to extract the dynamical matrix $\matnot{D}$ (Appendix, \ref{apx:simulations}). Finally, we obtain the normal modes of the simulated systems by numerically constructing and diagonalising $\matnot{D}$ to obtain eigenvalues $\lambda_{\nu}>0$ and eigenvectors $\vecnot{\chi}_{\nu}$ (see Appendix, \ref{apx:simulations} for details). 
By projecting the velocity vectors of the simulated particles on the so-extracted normal modes, we obtain the velocity projections shown in Fig.~\ref{fig:predictions_ua}a. 
To compare to theory, we fit the single fit parameter $v$ in the theoretical prediction of Eq.~\ref{eq:discrete_mode_velocity_equal_time} to match the velocity projection of just a single mode in Fig.~\ref{fig:predictions_ua}a, thereby obtaining the theory lines for all modes $\nu$ in  Fig.~\ref{fig:predictions_ua}a-e. This fit is carried out for $\nu=1$, \textit{i.e.}  the lowest mode, except for the case of periodic boundary conditions, in which case the first non-zero mode is used.
%As noted earlier, in the continuum formulation, it is possible to find $v$ using a self-consistency equation. 
We group the results into the following regimes based on the value of $J$, and we begin our analysis with a confined system with unit alignment at moderate levels of angular noise, $D_r=0.1$.

%%%%%%%%%%%%%%%%% readability %%%%%%%%%%%%%
\subsection{Weak alignment: ABP regime}
Starting at $J=0$, we find perfect agreement between the linear prediction and the simulation, for both the velocity projections (Fig.~\ref{fig:predictions_ua}a) and the polarity spectrum (Fig.~\ref{fig:predictions_ua}b-c).
This is expected since the $J=0$ limit corresponds to soft-repulsive Active Brownian Particles, where the exact calculation presented in Ref.~\cite{henkesDenseActiveMatter2020} results exactly in Eq.~\eqref{eq:discrete_mode_velocity_equal_time} with $J=0$, as discussed above. 

From the absence of a peak at non-zero $\omega$ in the polarity spectrum (Fig.~\ref{fig:predictions_ua}b-c), we conclude that the system does not exhibit oscillations. This is consistent with theory, as Eq.~\ref{eq:discrete_mode_dynamics} for $J=0$ reduces to a set of overdamped noisy harmonic oscillators.

For small $J$ (SI Movies \cite{SI}, S4), we do not detect oscillations in simulations (Fig.~\ref{fig:predictions_ua}b-c), implying that the alignment strength $J$ and the resulting $v$ are not large enough to tune the lowest mode from an overdamped oscillator to an underdamped oscillator, that is,  all $\phi_{\nu}<0$. After fitting $v$, the theory predictions are indistinguishable from the numerical results (Fig.~\ref{fig:predictions_ua}a-c). In Fig.~\ref{fig:a_Dr0.1_w_pred}a, we compare the RMS velocity from simulations (black dots) to the fitted $v$ (orange hexagons), and to the theory RMS velocity obtained by summing the modes, i.e. the RHS of the consistency equation (pink plusses). 
The fitted $v$-value is slightly higher than the RMS velocity in simulations, yet the theory RMS velocity is slightly lower. 
This discrepancy originates from ignoring the different nonlinear mode couplings in both the dynamics and in the mechanical interactions, since they contribute positively to the spectra, leading to an over- or  underestimate for the driving needed for a single mode and the linear spectral sum, respectively.  
%This discrepancy likely originates from the  juxtaposition of the mean-field assumption of homogeneous particle speed assumptions and the non-homogeneous displacements prescribed by the eigenvectors $\{\vecnot{\chi}_{\nu}\}$. \sander{$\leftarrow$ Silke, what do you think of this hypothesis?}

We also compare the full prediction for the mode spectrum Eq.~\ref{eq:discrete_mode_velocity_spectrum} to its numerical counterpart, as shown in Fig.~\ref{fig:predictions_ua}d for $J=0.04$. Beyond the clearly excellent agreement, we can also discern that the modes, as represented in the  $(w,q\sim\sqrt{\lambda})$-plane, are propagating diffusively, with the $\omega^2\Gamma_{\nu}^2$ term in Eq.~\ref{eq:discrete_mode_velocity_spectrum} dominating.

\subsection{Intermediate alignment: Onset of oscillations}

Increasing $J$ further past $J=0.08$ (SI Movies \cite{SI}, S5), we begin to see the first hint of a peak in the polarity spectrum at non-zero $\omega$ (Fig.~\ref{fig:predictions_ua}b-c). 
This means that the damping term $\Gamma_{\nu}$ has now become small enough for some of the low modes to have $\phi_{\nu}>0$, changing them into underdamped oscillators. This behaviour starts with only the lowest mode $\lambda_1$  but includes more and more modes as $J$ increases, making it a distinct region of the phase diagram, located here between $J=0.1-0.16$. In this region, the presence of the noise drives oscillations.
Indeed, when we plot the spectrum of the modes and the prediction according to Eq.~\ref{eq:discrete_mode_velocity_spectrum} for $J=0.12$, we see a clear condensation of the dynamics on the low modes with the lowest mode being the most excited mode (Fig.~\ref{fig:predictions_ua}a,e) compared to the overdamped case of $J=0.04$ (Fig.~\ref{fig:predictions_ua}a,d). 

Although the system is oscillating in this \emph{under}damped regime, the oscillations are still significantly damped. For the system to exhibit \emph{un}damped and thus sustained oscillations, the damping $\Gamma_{\nu}$ should go to zero.
As the damping term depends on $\lambda_{\nu}$ this first happens for the lowest mode, leading to the simple criterion $\Gamma_1 =0$ for sustained oscillations. We will show below that this corresponds to a critical phase transition with scaling laws and a diverging correlation length. Numerically, the criterion is crossed near $J=0.16$ (SI Movies \cite{SI}, S5), where we see a distinct condensation of the dynamics on the lowest mode, together with the appearance of a dramatic peak in the polarisation spectrum (Fig.~\ref{fig:predictions_ua}a-b, yellow lines). In this region, both the measured and the fitted values of $v$ begin to noticeably increase from the ABP prediction (Fig.~\ref{fig:a_Dr0.1_w_pred}a), while the match between numerics and linear theory remains excellent.
Furthermore, we clearly see that the approximation of the oscillation frequency of each mode by its natural frequency, \textit{i.e.} $\omega_{\nu} = \sqrt{\Omega_{\nu}^2- \gamma_{\nu}^2/4}\approx \Omega_{\nu} = \sqrt{\lambda_{\nu}/\zeta}\sqrt{1/\tau + Jv/(2v_0)} $ (the dashed lines in Fig.~\ref{fig:predictions_ua}d-e) starts to work well for the low modes, indicating that they have very small damping.  As we show in the continuum theory below, this corresponds to a linear dispersion relation, indicating that the system supports traveling elastic waves.

\subsection{Strong alignment: Nonlinear regime}
At large $J>0.16$ (SI Movies \cite{SI}, S6), the linear theory breaks down formally, because it predicts negative damping which is incompatible with the steady state assumption used to derive this linearized theory. Hence the only possible fit to the linear theory is $\Gamma_1=0$, corresponding to the critical line in Fig.~\ref{fig:predictions_ua}a. While this is a good fit for $J=0.16$, for larger values of $J$ the numerically observed velocity projections lie wholly above this line, while showing a strong oscillatory signature in the polarisation spectra in Fig.~\ref{fig:predictions_ua}b, and we term this region of phase space the one of undamped oscillations. 

\begin{figure}[h]
    \centering
    \includegraphics[width=0.9\linewidth]{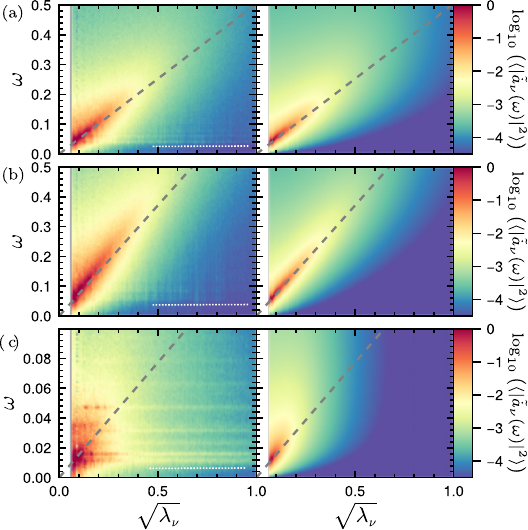}
    \caption{Nonlinear oscillation regime. Mode spectra for $\tau=10$, in simulations (left column) and linearized theory (right column). (a) $J=0.4$. (b) $J=1.0$. (c) At much lower noise, $\tau=100$, $J=0.04$ the nonlinearities appear much more strongly. The white dashed lines are guides to the eye to showcase the nonlinear mode coupling effects.}
    \label{fig:ua_mode_spectra_strong_alignment}
\end{figure}

% For $J\geq 0.16$, the only possible fit to the linear theory is $\Gamma_1=0$, corresponding to the critical line in Fig.~\ref{fig:predictions_ua}a. While this is a good fit for $J=0.16$, for larger values of $J$ the numerically observed spectra lie wholly above this line, while showing a strong oscillatory signature in the polarisation spectra in Fig.~\ref{fig:predictions_ua}b, and we term this region of phase space the one of undamped oscillations. 
At the same time, increasing signs of nonlinearities show up as horizontal stripes in the mode spectrum (white dashed lines in Fig.~\ref{fig:ua_mode_spectra_strong_alignment}a-c), which resemble, from the perspective of the linear theory, driving of high modes by the low, most excited modes at the frequencies of the low modes. 
We can explain this by revisiting the exact nonlinear equation (Eq.~\ref{eq:discrete_decoupled}), which shows that all the nonlinearities involve coupling of modes via their mode amplitudes (both via the dependence on $v_i$ and the $\left[ \matnot{D_{ik}} \ \delta \vecnot{r}_k \right] \cdot \delta \dot{\vecnot{r}}_i$ term). Therefore, the coupling strength of a specific mode to the others depends on its own amplitude and is thus nonreciprocal between the modes. This is why we see mostly horizontal stripes corresponding to the frequencies of the most excited (\textit{i.e.} the low) modes, which then add to the linear spectra of the other modes when integrating over frequencies.

It appears that the neglected nonlinearities rescue the existence of a steady state at larger $J$, as evidenced by the steady state in simulations. This can partly be understood as corrections to the linear theory that keep the damping for the lowest mode zero and for the higher modes positive. Indeed, the linear theory with $v$ chosen such that $\Gamma_{\mathrm{1}}=0$, reproduces the polarity spectra (Fig.~\ref{fig:predictions_ua}b) and the mode spectra (Fig.~\ref{fig:ua_mode_spectra_strong_alignment}a-b) of the strong alignment regime quite well, and the value of $v$ obtained using this criterion is a good match to its empirical value all the way up to $J=1$ (Fig.~\ref{fig:a_Dr0.1_w_pred}a). 

Although the linear prediction for the dispersion relation does not fit quantitatively anymore, it still predicts the salient features of the dynamics accurately, in particular the location and the slope of the effective dispersion relation (Figs.~\ref{fig:ua_mode_spectra_strong_alignment}b-c, red region), which increases with $J$, showing that the wavespeed increases.
\begin{figure}
    \centering
    \includegraphics[width=1\linewidth]{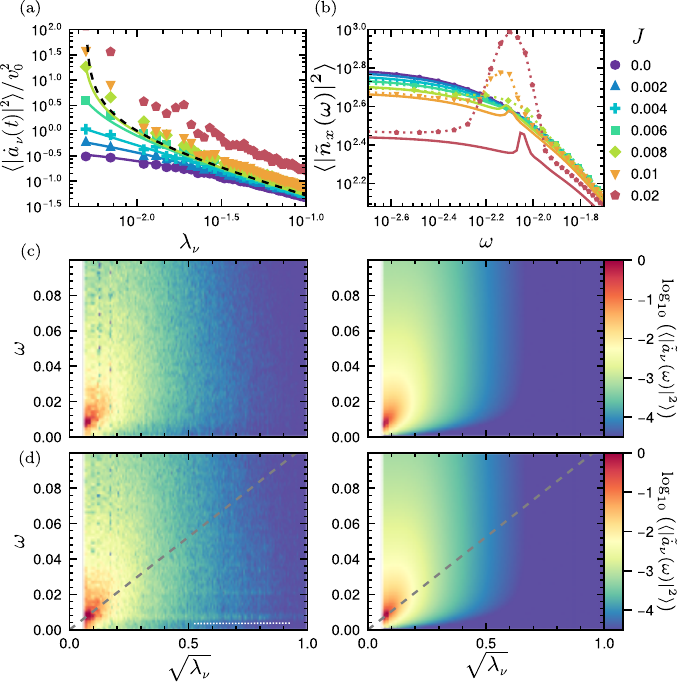}
    \caption{Steady state results for unit-alignment simulations with $\tau = 100$ compared to theory. (a) Velocity projections. The dashed line corresponds to the critical line with $\Gamma_1=0$. (b) Polarity spectrum (log-log scale) of simulations (dashed lines + scatters) and linearized theory (full lines). (c) Mode spectra of simulations (left) and theory (right), $J=0.006$. (d) Mode spectra of simulations (left) and theory (right), $J=0.008$. The white dashed line is a guide to the eye to highlight the nonlinear mode coupling effects. }
    \label{fig:Dr_0p01_ua}
\end{figure}

\subsection{The role of rotational noise}
Now that we have an overview of the different regimes as function of alignment strength $J$, we turn our attention to the role of the rotational noise $D_r$ or equivalently the persistence time $\tau = D_r^{-1}$. Specifically, we consider $\tau=100$ ($D_r=0.01$), a tenfold decrease in noise amplitude compared to the previous discussion, and run the same analysis as before (Fig.~\ref{fig:Dr_0p01_ua}).
Again for small $J$ compared to $D_r$, namely $J\leq 0.006$, we observe good agreement between linear theory and simulations in the velocity projections (Fig.~\ref{fig:Dr_0p01_ua}a) and polarity spectrum (Fig.~\ref{fig:Dr_0p01_ua}b) and the system is in the overdamped ($\phi_{\nu}<0$) regime. The comparison of the  mode spectrum for $J\leq0.006$ (Fig.~\ref{fig:Dr_0p01_ua}c) with the linear theory corroborates this view.
When it comes to larger $J \geq 0.008$, we observe that although the theory predicts correctly that $\phi_{\nu}>0$, \textit{i.e.} the existence of oscillations (as evidenced by the peak in the polarity spectrum (Fig.~\ref{fig:Dr_0p01_ua}b), the predicted polarity spectrum does not fit as well to the simulations as in the previous discussion. A reason for this is likely that nonlinearities play a larger role due to the smaller rotational noise. Indeed, when we consider the mode spectrum for $J=0.008$ in Fig.~\ref{fig:Dr_0p01_ua}d, we observe nonlinear effects where the low modes are coupling to higher modes at the frequency of the low mode, resulting in the horizontal lines in the spectrum. This also becomes visible in Fig.~\ref{fig:Dr_0p01_ua}a, where distinct resonance peaks appear for intermediate eigenvalues. This emergence of strong nonlinearities is consistent with the spectra dominated by nonlinearities that were observed at low or no noise in the small regular networks of Ref.~\cite{baconnier2022selective}. At very large $J$, the nonlinearities become extremely apparent in the mode spectrum (Fig. \ref{fig:ua_mode_spectra_strong_alignment}c), but the condensation on the low modes is still clearly visible, so that over the whole range of $J$, the numerical RMS velocity (Fig.~\ref{fig:a_Dr0.1_w_pred}) remains close to the theory RMS $v$ for the over- and underdamped regime $\Gamma_1>0$ and close to the $v$ that makes $\Gamma_1=0$ for the undamped regime.

We therefore conclude that for small noise the linearized theory loses its precision, but that it still has predictive power about the onset of oscillations and the general shape of the mode spectrum. In particular, we note that the paper that originally proposed condensation on the lowest modes, \cite{henkes2011active}, was precisely in this highly nonlinear regime at $D_r = 0.005$ and $J=1$. Indeed, at this low noise, the presence of the system scale oscillations in particle displacements is easily identifiable by eye (Fig.~\ref{fig:schematics}c and SI Movies \cite{SI}, S7).

\subsection{Detecting the onset of oscillations using experimental observables}
\begin{figure}
    \centering
    \includegraphics[width=1\linewidth]{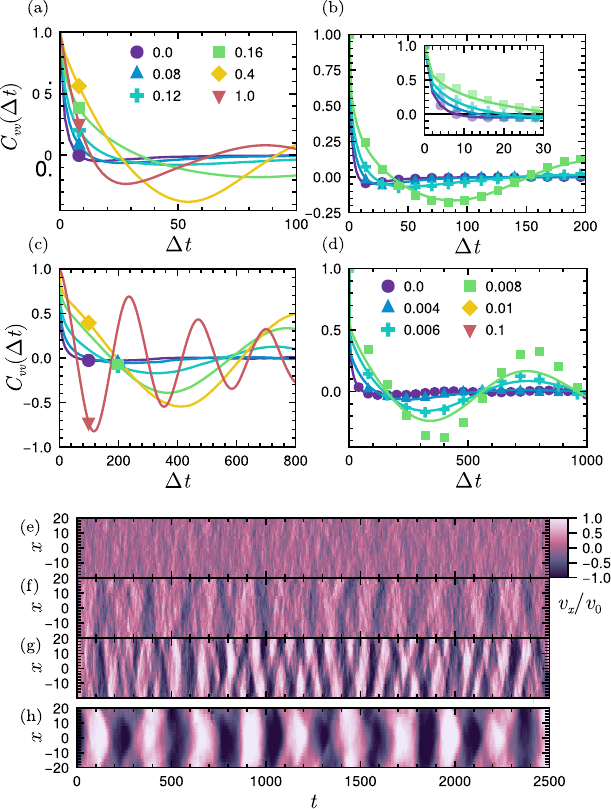}
    \caption{Velocity autocorrelation functions from  simulations (left column), and linearized theory + simulations (right column) for different values of $J$: (a) Simulations with unit alignment, $D_r=0.1$. (b) Linearized theory (lines) + simulations (markers) with unit alignment, $D_r=0.1$. For clarity, not all markers are shown. Inset: zoom for small $\Delta t$. (c) Simulations with unit alignment at much smaller noise, $D_r=0.01$. (b) Linearized theory (lines) + simulations (markers) with unit alignment, $D_r=0.01$. For clarity, not all markers are shown. Kymographs of the $x$-component of the velocity vectors in a thin strip around $y=0$ along the $x$-axis. (e-h) shows the emergence of oscillations as a function of $J$ for $D_r=0.1$. (e) Overdamped regime $J=0.04$. (f) Oscillations around the critical transition to undamped (sustained) oscillations, $J=0.16$. (g) Oscillations in the undamped, nonlinear regime, $J=0.4$. (h) At lower noise $D_r=0.01$ these sustained oscillations are more clearly visible, $J=0.1$.}
    \label{fig:velocity_autocorrelations}
\end{figure}
Although the velocity projections and the polarity spectrum are very sensitive to the emergence of oscillations, they are not easily accessible experimentally. Instead, one typically considers correlation functions of the velocity field. One way to detect oscillations this way is to use the velocity autocorrelations function, defined by  $C_{vv}(\Delta t)\equiv \langle \vecnot{v}^*_i(t+\Delta t) \cdot \vecnot{v}^*_i(t) \rangle_{i,t }$, where the velocity vectors are normalized by the RMS velocity $\vecnot{v}^*_i(t) \equiv \vecnot{v}_i(t)/v_{\text{rms}}(t)$. In the overdamped and underdamped regime (\textit{i.e.} when $\Gamma_{\nu}>0$ for all modes $\nu$), we can calculate $C_{vv}(\Delta t)$ analytically, by summing the unequal-time mode autocorrelations (Appendix, \ref{apx:velocity_autocorrelation}). For the case of $D_r=0.1$, as $J$ increases to $J=0.16$, $C_{vv}(\Delta t)$ starts to exhibit oscillatory behaviour (Fig.~\ref{fig:velocity_autocorrelations}a-b) and the linearized theory follows the simulations nicely. For larger $J$, in the nonlinear regime, the oscillatory behavior in fact becomes less clear, which is consistent with the extra contribution of multiple modes and the effects of mode coupling (see Fig.~\ref{fig:ua_mode_spectra_strong_alignment}b-c). At low noise $D_r=0.01$ we observe, just as in the case of the polarity spectrum, that the linearized theory picks up on the onset of oscillations ($J=0.006$ and $J=0.008$ in Fig.~\ref{fig:velocity_autocorrelations}d), but does deviate more from the simulations. At larger $J$ we again enter the nonlinear regime.

Another direct probe for oscillations is looking at the dynamics of the velocity vector components over time in the form of a kymograph (Fig. ~\ref{fig:velocity_autocorrelations}e-h), showing the emergence of (noisy) oscillations for $D_r=0.1$ as $J$ increases (Fig. ~\ref{fig:velocity_autocorrelations}e-d). Again, oscillations are more easily seen in the velocity field at lower noise $D_r=0.01$ (Fig. ~\ref{fig:velocity_autocorrelations}h).

\begin{figure*}
    \centering
    \includegraphics[width=0.8\linewidth]{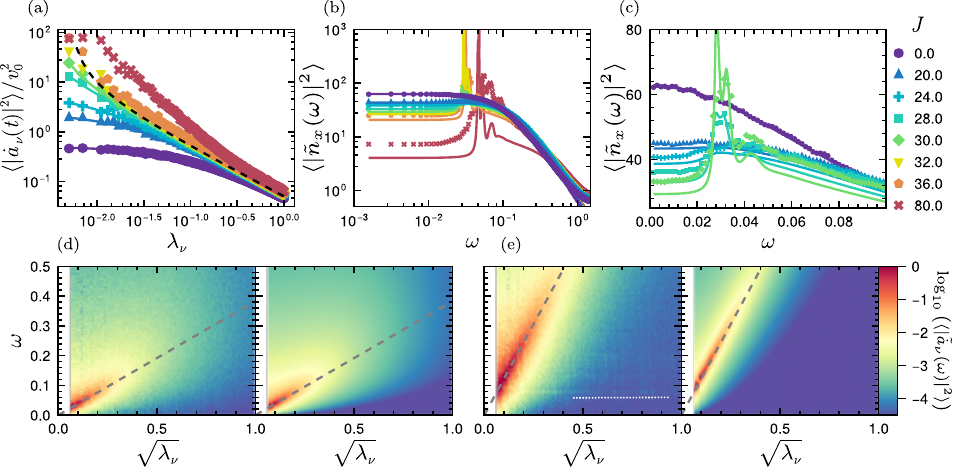}
    \caption{Steady state results for full-alignment simulations with $\tau = 10$ compared to theory. (a) Velocity projections. (b) Polarity spectrum (log-log scale) with simulations indicated by scatters and the linearized theory by full lines. (c) Polarity spectrum zoomed (log-lin scale) showing the emergence of the oscillation peak. (d) Mode spectra of simulations and theory, $J=30.0$. (e) Mode spectra of simulations and theory, $J=300$. The white dashed line indicates the presence of nonlinear mode couplings.}
    \label{fig:predictions_fa}
\end{figure*}

\subsection{Full alignment}
Full alignment,  where the particle aligns to the velocity vector, including magnitude, corresponds to the the substitution $J\rightarrow J v$ in our linear theory, which we now test in simulations with full alignment (Fig.~\ref{fig:predictions_fa}). Again, we find excellent agreement for small $J$ up to the first sign of oscillations around $J=24$, in the overdamped regime (Fig.~\ref{fig:predictions_fa}a-d). 
We then enter the underdamped regime, until we observe a distinct critical line where $\Gamma_1=0$ around $J=30$, and the linear theory remains a very good match to the numerics. Then when $J$ is further increased, nonlinearities appear. Like with unit-alignment, for significantly larger $J$, \textit{e.g.} $J=300$, we see that many more low modes are excited (Fig.~\ref{fig:predictions_fa}e) which is consistent with the observation that instead of system-scale oscillations, one observes much smaller domains of correlated oscillations. Like in the case of unit alignment, in the overdamped and underdamped regime, the fitted $v$ is close, but again systematically above, the RMS velocity of simulations (Fig. \ref{fig:a_Dr0.1_w_pred}d). Then, in the undamped regime, the $v$ that sets $\Gamma_1=0$ is again a good match to the observed RMS velocity in simulations.

Having verified that the linear theory also works to describe the onset of oscillations in case of full-alignment, we will use the linearized continuum theory to discuss the qualitative difference between unit- and full-alignment in more detail below.
\subsection{Periodic boundary conditions}
\begin{figure*}[t]
    \centering
    \includegraphics[width=0.8\linewidth]{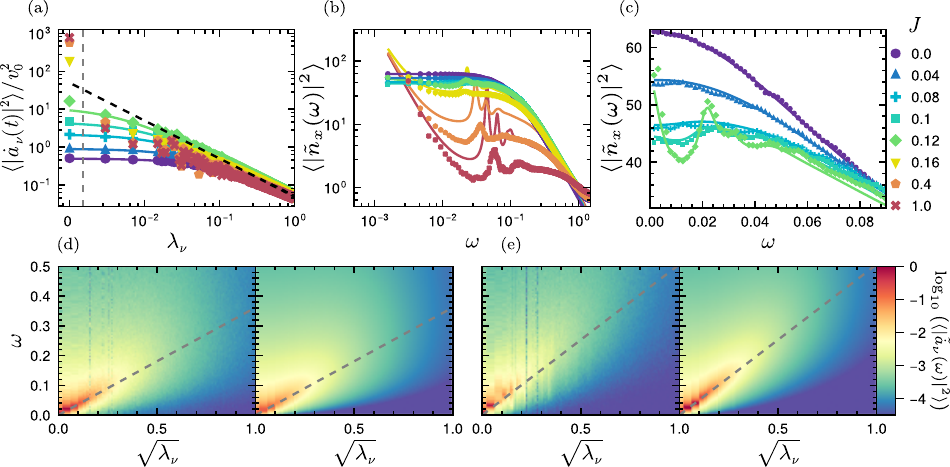}
    \caption{Steady state results for unit-alignment simulations with periodic boundary conditions with $\tau = 10$ compared to theory. (a) Velocity projections. The zero eigenvalues are placed at some non-zero location for compatibility with a log-log scale, this is indicated by the grey dashed line.(b) Polarity spectrum (log-log scale). (c) Polarity spectrum zoomed (log-lin scale) to see the emergence of the oscillation peak. (d) Mode spectra of simulations and theory, $J=0.12$. (e) Mode spectra of simulations and theory, $J=0.4$.}
    \label{fig:predictions_pbc}
\end{figure*}
One can repeat the same procedure for a solid with periodic boundary conditions instead of a pinned boundary. The two lowest modes are then the two translational modes, which have zero eigenvalue, \textit{i.e.} $\lambda_1 = \lambda_2 = 0$. Periodic boundary conditions do not allow for a rotation mode, which would couple to the higher elastic modes at the nonlinear level unless corrections are applied \cite{hernandez2024model,melio2024soft}. This allows us to treat the system as a translating solid, despite the two flat directions in the potential energy landscape, and all of our predictions carry through. Again we obtain excellent agreement with the linear theory (Fig.~\ref{fig:predictions_pbc}) for small $J$ (SI Movies \cite{SI}, S8) up to the transition to, in this case, flocking at $J=0.12$ (SI Movies \cite{SI}, S9), with the transition from overdamped to underdamped dynamics being around $J=0.008$. For small $J$, the velocity projections (Fig.~\ref{fig:predictions_pbc}a) of the periodic solid look very similar to its confined counterpart (Fig.~\ref{fig:predictions_ua}a). However,  for larger $J$, starting with $J=0.12$, the Fourier spectrum of the polarity vector  (Fig.~\ref{fig:predictions_pbc}b-c), shows not only a large $\omega=0$ component (stemming from the condensation of the dynamics on the translational modes), but also a peak at non-zero $\omega$. This is striking because it implies that even the flocking state exhibits oscillations and thus the elastic terms in the linearized theory are crucial for a correct prediction of this aspect of the dynamics in the flocking state.
At $J=0.12$, when we start to see the first hints of a peak at $\omega=0$ (flocking), the velocity projections of the linear theory still fit nicely, as is also apparent in the velocity mode spectrum (Fig.~\ref{fig:predictions_pbc}d).

For larger $J>0.12$ (SI Movies \cite{SI}, S10), the velocity projections (Fig.~\ref{fig:predictions_pbc}a) drastically change: from a condensation distributed on low modes, to an almost exclusive condensation on the two translational modes. This is also visible in the mode spectrum (Fig.~\ref{fig:predictions_pbc}e), of which the region of excited modes is much more concentrated than the linearized theory with $\Gamma_{1}=\Gamma_{2}=0$, predicts. This effect is potentially responsible for the non-monotonic glassy dynamics as a function of alignment strength around the flocking transition that was observed in \cite{paoluzzi2024flocking}.  
Interestingly, this condensation seems to be a different nonlinear effect than the nonlinearities observed in the confined solid discussed before, because the horizontal mode coupling lines in the mode spectrum observed in Fig.~\ref{fig:ua_mode_spectra_strong_alignment}b are absent in Fig.~\ref{fig:predictions_pbc}e. In fact, at the larger $J$ the square amplitudes of the higher modes in Fig.~\ref{fig:predictions_pbc}a lie below those in the linear regime, unlike in all of the confined system where they lie above, implying that the nonlinearities here instead drive the dynamics towards the translation modes. The RMS velocity of simulations is again close to the extracted $v$ based on the velocity mode projection of the first non-zero mode and the theory RMS velocity (Fig. \ref{fig:a_Dr0.1_w_pred}c) in the over- and underdamped regime. In the undamped regime however, in contrast to the confined systems, the $v$ obtained by requiring $\Gamma_1 = \Gamma_2=0$ actually underestimates the RMS velocity of simulations. 

As we discuss below in the continuum theory, the force free, \textit{i.e.} distortion-free, flocking state, would be a better state to linearize the dynamics around, shifting the prediction for $v$.
All in all, the linear theory can describe the transition up to flocking and captures oscillations in this phase, but does not capture the sudden dramatic concentration on the two lowest modes.

Having charted the regimes of validity of the normal modes formulation for the discrete system, we now turn to a continuum theory, in which we can fully solve the linearized mode dynamics (including $v$) to construct parameter predictions for the transition to oscillations/flocking and study the nature of the transition.

\section{Continuum theory for self-aligning active solids}
So far, the analytical treatment has been based on having knowledge of the dynamical matrix of the solid. However, determining the dynamical matrix for experimental systems like cell sheets or crowds may be impossible from a practical point of view. We can sidestep this issue by considering a continuum formulation of the solid, which only requires knowledge of the elastic moduli of the material. 

From a theoretical point of view, this step to a continuum formulation also turns out fruitful, as it allows us to fully solve the linearized continuum dynamics and predict the transition. We begin by deriving the exact nonlinear displacement equation of the continuum theory and then apply a similar set of approximations as the discrete case to arrive at a linear theory.

\subsection{Equations of motion}
Assuming linear elasticity with bulk modulus $B$ and shear modulus $\mu$, and introducing a coarse-graining length scale $b$, we write the equivalent active elastic continuum equations for the displacement field $\vecnot{u}(\vecnot{r},t)$,
\begin{equation}
    \zeta\dot{\vecnot{u}}=  B  \nabla\left( \nabla \cdot \vecnot{u}\right) +  \mu \nabla^2 \vecnot{u} + \zeta v_0 \hatnot{n},
    \label{eq:co_displacement_eom}
\end{equation}
and the polarity field $\hatnot{n}(\vecnot{r},t)$,
\begin{equation}
    \dot{\hatnot{n}} = \frac{J}{v} \left( \hatnot{n} \times \dot{\vecnot{u}}\right) \times \hatnot{n}-\frac{1}{\tau}\hatnot{n} + \sqrt{\frac{2}{\tau}}\eta(\vecnot{r},t)\hatnot{z}\times \hatnot{n},
    \label{eq:co_polarity_eom}
\end{equation}
in which we defined $v\equiv | \dot{\vecnot{u}}|$, and where $\zeta$ is a scalar friction term. 
%which in principle can be tensorial instead. (move to discussion)
The rotational noise field $\eta(\vecnot{r},t)$ has zero mean and variance $\langle \eta(\vecnot{r},t) \eta(\vecnot{r}',t')\rangle=b^2 \delta(\vecnot{r}-\vecnot{r}')\delta(t-t')$. As before, inserting $J\rightarrow Jv$ recovers the full alignment dynamics.
This continuum model, but without self-alignment, has been studied before in \cite{henkesDenseActiveMatter2020}.

\subsection{Nonlinear displacement dynamics}
The same decoupling steps as for the discrete system can be taken for this continuum formulation of the self-aligning solid.  We obtain the exact, decoupled, equation for the displacement field $\vecnot{u}(\vecnot{r},t)$,
\begin{align}
    \ddot{\vecnot{u}} = & -\frac{1}{\zeta}D(\dot{\vecnot{u}})-\left[\frac{1}{\tau}+J\left( \frac{v}{v_0} - \frac{v_0}{v}+\frac{D(\vecnot{u})}{\zeta v v_0}\cdot \dot{\vecnot{u}}\right) \right]\dot{\vecnot{u}} \nonumber \\
    &-\left[ \frac{1}{\tau} + J\left(\frac{v}{v_0} +\frac{D(\vecnot{u})}{\zeta v v_0}\cdot \dot{\vecnot{u}}\right)\right]D(\vecnot{u}) \nonumber \\
    &+\sqrt{\frac{2}{\tau}}\eta(t)\hatnot{z}\times \left[\dot{\vecnot{u}}+ \frac{1}{\zeta} D(\vecnot{u}) \right],
\label{eq:nonlinear_displacement_continuum}
\end{align}
where we have defined for brevity the differential operator $D(\vecnot{u})\equiv -B \nabla\left(\nabla \cdot \vecnot{u}\right) - \mu \nabla^2 \vecnot{u}$, equivalent to the dynamical matrix in the discrete treatment. The fully expanded nonlinear equation can be found in Appendix \ref{apx:nonlinear_continuum}. To help put our results into the context of active hydrodynamics \cite{marchetti2013hydrodynamics}, in the incompressible limit $\nabla \cdot \dot{\vecnot{u}}=0$, and sorted by order of derivatives, we find
 \begin{align}
    &\ddot{\vecnot{u}} =\!-\!\left[ \frac{1}{\tau }-J \left(\!\frac{v_0}{v}\!+\!\frac{v}{ v_0}\!\right)\!\right]\!\dot{\vecnot{u}}+\!\frac{\mu}{\zeta}\nabla^2 \dot{\vecnot{u}} +\!\left[\!\frac{1}{\tau}\!+\!\frac{Jv }{v_0}\!\right]\!\frac{\mu}{\zeta}\nabla^2 \vecnot{u}  \label{eq:continuum_incomp}\\
   &+  \frac{\mu J}{ \zeta v v_0}\left[\dot{\vecnot{u}}  \cdot\nabla^2 \vecnot{u}\right]\left[\mu\nabla^2 \vecnot{u} \right]   + \sqrt{\frac{2}{\tau}} \eta(t)\hatnot{z} \times \left[\dot{\vecnot{u}} - \frac{\mu}{\zeta} \nabla^2 \vecnot{u}  \right], \nonumber
\end{align}
in which we omitted the pressure term that enforces the incompressibility condition. 
Before we move on to linearising this equation, we discuss a special case of Eq.~\eqref{eq:nonlinear_displacement_continuum}: homogeneous flocking.

\subsection{Homogeneous flocking state}
Consider a state with a homogeneous displacement profile $\vecnot{u}$, so that all gradient terms vanish in Eq.~\eqref{eq:nonlinear_displacement_continuum}. We find  a single equation for the velocity field $\vecnot{v} = \dot {\vecnot{u}}$,
\begin{equation}
   \dot{\vecnot{v}} =  - \frac{1}{\tau} \left( 1-\frac{J \tau v_0}{v}+\frac{J \tau v}{v_0}  \right)\vecnot{v} + v_0\sqrt{\frac{2}{\tau}} \eta(t)\hatnot{z} \times \hatnot{n},
\end{equation}
and again with $J\rightarrow Jv$ for the system with full alignment. This corresponds to non-conservative model A dynamics \cite{chaikin2000principles} for the scaled field $\vecnot a = \vecnot{v}/v_0$ with friction coefficient $\tau^{-1}$, $\dot{\vecnot{a}} = - \frac{1}{\tau} \frac{\delta F(\tilde{J},\vecnot{a})}{\delta \vecnot{a}} + \vecnot{\tilde{\eta}}$ \cite{chaikin2000principles}. Here $F(\tilde{J},\vecnot{a})$ is the effective free energy 
\begin{align}
F_{\text{unit}}(\tilde{J}_u,\vecnot{a}) =  - \tilde{J}_u a +\frac{a^2}{2}  +  \tilde{J}_u \frac{a^3}{3} , \quad \tilde{J}_u = J\tau\\
F_{\text{full}}(\tilde{J}_f,\vecnot{a}) = \frac{1}{2}(1- \tilde{J}_f) a^2 + \tilde{J}_f \frac{a^4}{4}, \quad \tilde{J}_f = Jv_0\tau
\end{align}
for unit and full alignment respectively, and where $a = |\vecnot{a}|>0$ we have the noise term $\vecnot{\tilde{\eta}}=\sqrt{\frac{2}{\tau}} \eta(t)\hatnot{z} \times \hatnot{n}$. 

\begin{figure}
    \centering
\includegraphics[width=1\linewidth]{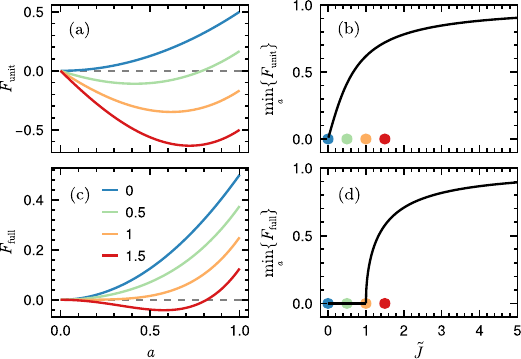}
    \caption{Free energy in the deformation free limit for different values of $\tilde{J}$, and the corresponding locations of the minimum. (a-b) Unit alignment.  (c-d) Full alignment.}
    \label{fig:free_energy}
\end{figure}

In Fig.~\eqref{fig:free_energy}, we plot both versions of the free energy. The flocking state in both models is a Goldstone mode, as there is a continuous line of minima with $|\vecnot{a}| =a_{\text{min}}$. The full alignment case is very similar to a classic Landau free energy, with minima
\begin{equation}
a_{\text{min}}=0 \text{  or  } a_{\text{min}}^2 = \frac{\tilde{J}_f-1}{\tilde{J}_f}.
\end{equation}
This system exhibits a pitchfork bifurcation at $\tilde{J}_f=1$ from a disordered $a=0$ state for $\tilde{J}_f<1$ to a flocking state with $a = \sqrt{\frac{\tilde{J}_f-1}{\tilde{J}_f}}$, with $a\rightarrow 1$, i.e. flocking motion with $v=v_0$ in the limit $\tilde{J}_f \gg 1$. This strongly resembles the drift-pitchfork bifurcation introduced in \cite{baconnier2022selective}, and such a Goldstone mode was also pointed out in crowds \cite{gu2025emergence} and theoretically for self-aligning crystals \cite{musacchio2026flocking}. 
In contrast, the unit alignment free energy only has a single minimum for $a>0$,
\begin{equation}
 a_{\text{min}} = \frac{1}{2 \tilde{J}_u}\left(\sqrt{1+ 4  \tilde{J}_u^2}-1\right),
\end{equation}
which smoothly goes between $a=0$ for $\tilde{J}_u\rightarrow 0$ and $a\rightarrow 1$ for $\tilde{J}_u\gg1$.

While the full alignment case predicts a phase transition between a disordered and a flocking state, in contrast the unit alignment state always predicts flocking, which is contradicted by the numerical results and the normal mode theory (Fig. \ref{fig:predictions_pbc}). This highlights the role of the noise term combined with the presence of elastic deformations in stabilising the disordered state. 
In this disordered state, if we make again the approximation that $\hatnot{n}$ is not spatially correlated, we can map $\vecnot{\tilde{\eta}}$ to a white noise with mean $0$ and variance $\frac{2}{\tau}$. This noise is then equivalent to a thermal noise in thermal equilibrium, with an effective temperature determined by the fluctuation-dissipation relation $D = \zeta k_bT$ where $\zeta = 1/\tau$ and $k_bT = 1$.
In contrast, in the fully aligned state, $\vecnot{a}$ and $\vecnot{n}$ are in the same direction, and so the noise is purely transverse along $\hatnot{a}^{\perp}$, which is reminiscent of the asymmetric diffusion in other flocking models such as Toner-Tu \cite{toner1995long} and spin-wave dynamics \cite{bialek2012statistical,cavagna2014bird,cavagna2013boundary}.
%\sander{ $\leftarrow$ I have added the original Toner-Tu paper, but I am not sure which spin-wave paper to add that mentions asymmetric diffusion}.

We emphasise that this free energy picture is only valid for uniform states, and that solely adding typical gradient terms $(\vecnot\nabla a)^2$ to the free energy as in \cite{musacchio2026flocking} is not consistent with the full continuum equations \ref{eq:nonlinear_displacement_continuum} and \ref{eq:continuum_incomp}. In particular, they depend on displacement gradients $\vecnot \nabla \vecnot{u}$ explicitly, not only on the velocity $\dot{\vecnot{u}}$, and so cannot be mapped to either a free energy of $\vecnot a$ only, or even to overdamped model A dynamics. Note that instead, writing two coupled, nonlinear overdamped equations for the displacement $\vecnot u$ and polarisation $\vecnot n$ dynamics is possible, and is analysed in more detail, though without the stochastic noise terms, in \cite{baconnier2022selective,baconnier2025collective}.

\subsection{Linearization}
In essence, we follow the same linearization steps as for the discrete solid. That is, we consider $\vecnot{u}$ to consist of small deviations around the noiseless climbing state (\textit{i.e.} around the elastic and active force balance), which in this case is given by
\begin{equation}
    -D(\vecnot{u}_c) =B \nabla\left(\nabla \cdot \vecnot{u}_c\right) + \mu \nabla^2 \vecnot{u}_c =  -\zeta v_0 \hatnot{n}_c,
\label{eq:noiseless_climbing_state_continuum}
\end{equation}
which results in (cf. Appendix, \ref{apx:linearization_continuum})
\begin{align}
    & \ddot{\vecnot{u}} =\frac{1}{\zeta}D(\dot{\vecnot{u}}) - \left( \frac{1}{\tau }-\frac{J v_0}{2 v}+\frac{J v}{2v_0} \right)\dot{\vecnot{u}}  \nonumber \\
   &+\left(\frac{1}{\tau}+\frac{J v}{2 v_0} \right)D(\vecnot{u})
   + v_0 \sqrt{\frac{2}{\tau}} \vecnot{w}(\vecnot{r},t),
\label{eq:continuum_decoupled_linearized}
\end{align}
where the zero mean noise field $\vecnot{w}(\vecnot{r},t)$ is $\delta$-correlated in space and time,  with $\langle \vecnot{w}(\vecnot{r}+\vecnot{r'},t+t') \cdot \vecnot{w}(\vecnot{r'},t')\rangle= b^2\delta(t-t')\delta(\vecnot{r} - \vecnot{r'})$ at the particle length scale $b$.
For a flocking state where the system is not undergoing oscillations around a elastically deformed state, we can instead simply neglect higher order nonlinear terms, and find the very similar equation
\begin{align}
    & \ddot{\vecnot{u}} =\frac{1}{\zeta}D(\dot{\vecnot{u}}) - \left( \frac{1}{\tau }-\frac{J v_0}{v}+\frac{J v}{v_0} \right)\dot{\vecnot{u}}  \nonumber \\
   &+\left(\frac{1}{\tau}+\frac{J v}{v_0} \right)D(\vecnot{u})
   + v_0 \sqrt{\frac{2}{\tau}} \vecnot{w}_F(\vecnot{r},t),
\label{eq:continuum_decoupled_linearized_flock}
\end{align}
where the noise term $\vecnot{w}_F$ is now orthogonal to the instantaneous flocking direction $\hatnot{v}$. 
In this distortion-free flocking expansion, the $v$ predicted in the undamped regime is then found by requiring $\frac{1}{\tau }-\frac{J v_0}{v}+\frac{J v}{v_0} =0$. As this prediction does not not depend on the modes, we can compare directly to simulations and we see that this prediction works well in the large $J$-limit when the system is flocking (Fig.~\ref{fig:a_Dr0.1_w_pred}c), but does not predict the transition properly, due to the neglected elastic deformations. Close to the disordered state, the linearization prescribed by Eq.~\eqref{eq:discrete_decoupled_linearized} is much better.
Similarly, the polarity spectrum is better described using the distortion-free expansion only for large $J$ in the flocking regime, see Appendix, ~\ref{axp:pbc_distortion_free_spectrum}.

Assuming a constant speed $v$, equations \eqref{eq:continuum_decoupled_linearized} and \eqref{eq:continuum_decoupled_linearized_flock} are now linearized and after diagonalizing the equation in terms of transverse and longitudinal modes in spatiotemporal Fourier space, one obtains the linearized dynamics, described next.

\subsection{Linearized dynamics}

\paragraph*{Velocity spectrum.}
We obtain the following transverse velocity spectrum for the expansion around the elastic and active force balance (cf. Appendix, \ref{apx:velocity_spectrum}),
% Made a single equation label due to space issues
\begin{equation*}
        \langle V_{\perp}(\vecnot{q},\omega) V_{\perp}(\vecnot{q'},\omega') \rangle= \frac{ (2\pi)^3 v_0^2 b^2 \frac{1}{\tau} \delta(\vecnot{q}+\vecnot{q'})\delta(\omega+\omega') \omega^2} { \left[\Omega^2_{\perp}(q)- \omega^2\right]^2 + \omega^2 \Gamma^2_{\perp}(q) } ,
\end{equation*}
where 
\begin{align}
    &\Omega^2_{\perp}(q) = \left( \frac{1}{\tau}+ \frac{Jv}{2v_0} \right)(B + \mu)q^2/\zeta, \label{eq:continuum_velocity_spectrum} \\
    &\Gamma_{\perp}(q) = (B + \mu)q^2/\zeta + \frac{1}{\tau} - \frac{Jv_0 }{2v} + \frac{Jv}{2v_0}. \nonumber
\end{align} 
The longitudinal spectrum follows directly by letting $\mu \rightarrow B + \mu$, and as before, the result for full-alignment is obtained simply by letting $J\rightarrow Jv$.

\paragraph*{Dispersion relation for acto-elastic waves.}
One can additionally write a dispersion relation for these waves by Fourier transforming the noiseless version of Eq.~\eqref{eq:continuum_decoupled_linearized} to obtain, \textit{e.g.} for the transverse waves
\begin{align}
    \omega^2 =& -i\omega\Gamma_{\perp}(q) + \Omega^2_{\perp}(q) \label{eq:dispersion_relation} \\
    =&-i\omega \left(\frac{\mu}{\zeta} q^2 + \frac{1}{\tau} - \frac{Jv_0 }{2v} + \frac{Jv}{2v_0} \right) + \left( \frac{1}{\tau}+ \frac{Jv}{2v_0}  \right)\frac{\mu}{\zeta} q^2 \nonumber.
\end{align}
 Eq.~\eqref{eq:dispersion_relation} is nothing but the classic expression for the dispersion of elastic travelling sound waves \cite{chaikin2000principles}, and of the same structure as S (transverse) and P (longitudinal) waves in seismic wave propagation \cite{sato2012seismic}.
Crucially, the possibility of longitudinal or transverse traveling waves, \textit{i.e.} when $\phi_{\parallel}(q)\equiv \Omega^2_{\parallel}(q) -\Gamma^2_{\parallel}(q)/4>0$ or $\phi_{\perp}(q)\equiv \Omega^2_{\perp}(q) -\Gamma^2_{\perp}(q)/4>0$, respectively, only happens for a sufficiently large $J$, just as in the discrete case.  The damping can also become zero, heralding the transition to the globally oscillating state. This happens first for the transverse shear waves and at the smallest $q$, that is $\Gamma_{\perp}(q_{\mathrm{min}})=0$, which for a system of size $L$ is $q_{\mathrm{min}} =q_{L} \equiv \frac{2 \pi}{L}$.  \\

\paragraph*{Critical phase transition to oscillations.}
To locate the oscillation transition, we need to find a self-consistent way to compute the mean speed $v$, or equivalently the kinetic energy of the system.
For that we first inverse-Fourier transform Eq.\eqref{eq:continuum_velocity_spectrum} back to the time domain, to obtain the Fourier space, equal-time velocity correlation functions, which will give us insight in the nature of the transition to oscillations. In the final step,  we inverse-Fourier transform back to real space and real time to obtain $v$ and construct the phase diagram.
Integrating Eq.~\ref{eq:continuum_velocity_spectrum} back to the time domain, in a calculation directly analogous to the discrete case, we find for a given $\vecnot{q}$ and while $\Gamma_{\perp}(q)>0$ the transverse equal-time velocity correlation function
\begin{equation}
    \langle |\tilde{v}_{\perp}(\vecnot{q},t)|^2 \rangle = \frac{(2\pi)^2 \frac{v_0^2}{\tau} b^2  }{2\Gamma_{\perp}(q)} 
    =\frac{2\pi^2 \frac{v_0^2}{\tau} b^2  }{\frac{1}{\tau}-  \frac{Jv_0}{2v} + \frac{Jv}{2v_0}+ \frac{\mu}{\zeta} q^2 }. 
\end{equation}
This corresponds to a classic velocity correlation function of the type $\langle |v(q)|^2 \rangle \sim \frac{1}{1+(\xi q)^2}$, with the correlation length $\xi$ given by \begin{equation}\xi^2 = \frac{\mu/\zeta}{\frac{1}{\tau}-  \frac{Jv_0}{2v} + \frac{Jv}{2v_0}}.\end{equation}
We thus have two transitions as a function of $J$: First where the lowest transverse wave turns from an overdamped wave, to a critically damped wave to finally a underdamped travelling wave with $\phi_{\perp}(q_{\mathrm{min}})>0$ while still $\Gamma_{\perp}(q_{\min})>0$. The second transition is when the damping of this travelling wave becomes zero: $\Gamma_{\perp}(q_{\min})\rightarrow0$, implying \emph{sustained} travelling, or rather standing,  waves at system scale with wave speed 
\begin{equation} c(J)=\sqrt{\left[\frac{1}{\tau} + \frac{Jv}{2v_0}\right]\frac{\mu}{\zeta}}. \end{equation}
The second transition is particularly striking, because as $\Gamma_{\perp}(q_{\min})\rightarrow0$, the velocity correlation length $\xi$ diverges and is only limited by the system size $L$, corresponding to a second order phase transition, where the correlation length diverges at the critical point.

For full alignment, we observe much the same, but now with a length scale $\xi^2 = \frac{\mu/\zeta}{\frac{1}{\tau}-  \frac{Jv_0}{2} + \frac{Jv^2}{2v_0}}$. 
Finally, for the flocking system, using the elastic and active force balance expansion, the periodic boundary conditions imply that $q_{\mathrm{min}}=0$, but the length scale relation is the same as the confined counterpart.
This contrasts the prediction by the distortion-free expansion, which predicts $\xi^2 = \frac{\mu/\zeta}{\frac{1}{\tau}-  \frac{Jv_0}{v} + \frac{v}{v_0}}$, which is of the same form but with a distinction: the factors of $\frac{1}{2}$ in the $J$ terms vanish, de facto corresponding to a transformation $J\rightarrow 2J$ in the correlation functions and in the dispersion relation. As we have stated in the previous section, for a correct prediction to underdamped oscillations in the system with periodic boundary conditions, the active and elastic force balance is the right state to expand around. In the large $J$ limit, however, it is the stress-free expansion that yields the correct prediction for $v$. In any case, both expansions predict a system size correlation length at criticality.

Our predictions still depend on the internal variable $v$, which we can also now compute through integrating the velocity correlation functions as a function of $q$
This integral diverges for $q\rightarrow \infty$ and so a large (UV) $q$-cutoff is needed. We do this at the coarse-graining scale, i.e. at $q_b=2 \pi/b$.
We also need to introduce an infrared cutoff at low $q$ when approaching the transition, assuming the absence of zero modes as is the case for confined systems. Taking a disk-shaped domain of radial size $L$, i.e. $q_L = 2 \pi/L$, we eventually obtain
\begin{align}
     \langle |v_{\perp} |^2 \rangle =&
     \frac{v_0^2 b^2 \zeta}{8 \pi \tau \mu} \left[\ln\left(1 -\tau \frac{Jv_0}{2v}+\tau\frac{Jv}{2v_0}+\tau \frac{\mu}{\zeta} q_b^2\zeta \right) \right. \nonumber \\
     & \left.-  \ln\left(1 -\tau \frac{Jv_0}{2v}+\tau\frac{Jv}{2v_0}+\tau \frac{\mu}{\zeta} q_L^2 \right)\right]
\end{align}
and similar for the longitudinal velocity (replace $\mu \rightarrow B+\mu$), as well as for systems with full alignment (replace $J\rightarrow Jv$). This leads to the transcendental self consistency relation for the kinetic energy of the system,
\begin{equation} v^2 = \langle |v_{\perp}(v) |^2 \rangle + \langle |v_{\parallel}(v) |^2 \rangle,
\label{eq:consistency_equation_continuum}
\end{equation}
which can however be solved numerically.

\subsection{Phase diagram for self-aligning solids}
\begin{figure}
    \centering
    \includegraphics[width=1\linewidth]{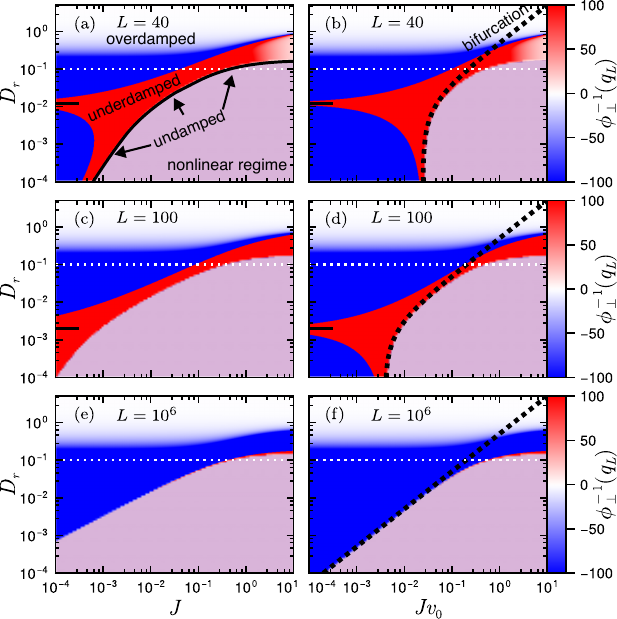}
    \caption{Phase diagram of unit and full aligning confined solids with different system sizes $L$, based on the nature $\phi_{\perp}(q_{L})$  of the lowest mode. Note the diverging colorscale for $\phi_{\perp}(q_{L})$. The black line element corresponds to the value of $D_r$ for which the lowest mode is undamped in the ABP limit ($J=0$). The dashed black line is the bifurcation line defined by $Jv_0= 2\left(D_r+\mu q_L^2/\zeta\right)$, and the white dashed line indicates the values of $D_r=0.1$ used in the simulations. $v_0=0.01$, $\mu=0.5$, $B=1$, and $\zeta=1$. Left column - unit alignment, right column - full alignment. Top row, panels (a)-(b), $L=40$, middle row, panels (c)-(d), $L=100$, bottom row, panels (e)-(f), $L=10^6$.}
    \label{fig:phase_diagrams}
\end{figure}

Solving the consistency equation Eq.~\eqref{eq:consistency_equation_continuum} numerically, we obtain the full solution to the linearized continuum theory of Eq.~\eqref{eq:continuum_decoupled_linearized}. This then yields the nature of the lowest mode, prescribed by $\phi_{\perp}(q_{L})$, for different values of system size $L$, noise $D_r=1/\tau$ and alignment $J$, which results in the phase diagrams shown in Fig.~\ref{fig:phase_diagrams}. Starting with the phase diagrams for unit alignment (left column, for increasing $L$), we see that for small $J$ and large $D_r$, $\phi_{\perp}(q_{L})>0$ and the system does not exhibit any oscillations. Then as $J$ increases, e.g. for $D_r=0.1$ as we investigated numerically (white dashed lines), the lowest mode eventually becomes critically damped, $\phi_{\perp}(q_{L})=0$ and then underdamped, $\phi_{\perp}(q_{L})>0$ admitting first damped oscillations and finally undamped, sustained, oscillations for which $\phi_{\perp}(q_{L})>0$ \emph{and} $\Gamma_{\perp}(q_{L})=0$. For even smaller $D_r$ and larger $J$, the linear theory breaks down, because the consistency equation Eq.~\ref{eq:consistency_equation_continuum} can no longer be solved. It is at this point that we expect nonlinearities to become important, just as in the earlier discussion of the simulations and discrete theory. We emphasise again that this is a noise-driven transition, and that in the simplified free energy picture, there is no transition in the unit alignment state, contradicting numerical evidence.

An interesting observation for small systems (Fig.~\ref{fig:phase_diagrams}a-b) is that the regime of damped oscillations (red) is reentrant, reminiscent of the reentrant transition found in \cite{baconnier2025reentrant}
, and sandwiched between the line where in the ABP theory ($J=0$) the lowest mode is critically damped and the freely oscillating state. Any amount of alignment $J$ then tunes that mode from critically damped towards damped oscillations, and in fact \emph{increasing} $D_r$ can make the system enter a damped oscillating state. We do note however that the reentrant region is still significantly damped, having a $Q$-factor close to $1/2$. Oscillations become more and more visible when approaching the critical line. Remarkably, at very large $L=10^6$, the damped oscillation region practically vanishes and the transition of the lowest mode is nearly directly from overdamped to undamped (Figs.~\ref{fig:phase_diagrams}e-f).

The picture in the case of full alignment is slightly more complex in regards to the transition to sustained, undamped, oscillations. From the linear theory we know that the damping of the lowest mode is given by  $\Gamma_{\perp}(q_L)=\frac{1}{\tau} - \frac{J v_0}{2} + \frac{J v^2}{2v_0} + \frac{\mu}{\zeta}q_L^2$. In the case of zero damping, $\Gamma_{\perp}(q_L)=0$, and this sets $v$ to be $v=v_0\sqrt{1 - \frac{2}{J v_0}\left(\frac{\mu}{\zeta}q_L^2+D_r\right)}$. For $v$ to be nonzero and hence for undamped oscillations to be even possible, this requires $Jv_0> 2\left(D_r+\mu q_L^2/\zeta \right)$. A continuum theory derived in Ref.~\cite{baconnier2024noise} predicts this precise condition as a Hopf bifurcation between disordered motion and the noise-induced collective actuation regime (undamped global oscillations). From Figs.~\ref{fig:phase_diagrams}b, d and especially f, we see that for small $D_r$ the transition to undamped oscillations is indeed very well described by this bifurcation line. However, at larger noise, we see that the transition to undamped oscillations happens to the right of this prediction; $J$ needs to be significantly larger to obtain undamped oscillations.

\section{Discussion}
Inspired by active self-aligning solids in biological and artificial systems, we have derived an exact nonlinear displacement equation for active, noisy, self-aligning solids. From the our linearized theory, applicable to disordered solids, we understand that their dynamics can be understood as a collection of damped, driven harmonic oscillators along normal modes that have a different nature based on the values of the alignment strength $J$. As $J$ increases, the lowest mode starts to become underdamped, exhibiting damped oscillations and for even larger $J$ this damping vanishes and the solid exhibits sustained, undamped oscillations along it lowest energy mode. The transition to the sustained undamped oscillations is paired with a diverging length scale of the spatial velocity correlations, in a second order phase transition. At even larger $J$, we learned from simulations that more low modes become excited and nonlinear mode coupling starts to become important. This expanded understanding of self-aligning solids now allows us to address where various artificial and biological systems fit into the phase diagram of self-aligning solids:

First, the small ordered networks that exhibit sustained global oscillations, i.e. collective actuation, are in the low noise and strong alignment highly nonlinear region of phase space \cite{baconnier2022selective}. This is consistent with the observed strong coupling between modes, and the very heterogeneous mode spectra that do not necessarily amplify the lowest mode. It would be instructive to numerically map the mode-frequency space as in Fig.~\ref{fig:ua_mode_spectra_strong_alignment} to gain further insight into the coupling, and to see if any features of the linear dispersion relation remain visible.  

Second, in the recent work on emergent oscillations in dense human crowds \cite{gu2025emergence}, the authors conclude that the individuals follow underdamped oscillations, in the form of local orbital trajectories, just like in Fig.~\ref{fig:schematics}c. Furthermore, the transition to undamped oscillations described here has the same set of signatures as that observed in the crowds: a peak in the kinetic energy spectrum at non-zero $\omega$, non-oscillating particle speed, equal probability for counter- and clockwise oscillations, and an oscillation frequency that decreases with system size. The crowds are therefore well described by our model near the critical transition line of the phase diagram, and we hypothesize that the crowd oscillations are of the same type as the oscillations discussed in this work.
%and a bifurcation towards oscillations depending on the strength of alignment. (SH: that's the  model, not the observations)
The authors of \cite{gu2025emergence} successfully described these metrics using a slightly different set of overdamped equations of motion for the displacement and polarity vector, with similar feedback between polarity and velocity vector, but only one system-size elastic well. We speculate that this model corresponds to a mean-field version of our model. 

In contrast to the sustained global oscillations in ordered solids \cite{baconnier2022selective} at system scale, human crowds tend to display oscillations that seem to be coherent over a smaller region size \cite{gu2025emergence}, something we also observe. Like in our disordered packings, the disordered nature of the human crowd likely shapes their normal modes to have finite correlation lengths \cite{silbert2005vibrations}.

There is a transition to oscillations based on density in crowds \cite{gu2025emergence} and similarly, collective oscillations in tissues are also only observed beyond a threshold density \cite{deforet2014emergence,petrolli2019confinement,peyretSustainedOscillationsEpithelial2019}. Both are consistent with a jamming transition towards an active solid with the collectively oscillating or flocking state developed in this paper. On a similar scale as tissues, Placozoa are flat animals formed from solid-like cell aggregates. To understand their locomotion, they are modelled as an elastic cell sheet in which self-alignment is an ingredient coordinating the activity across the animal \cite{davidescu2023growth,prakash2021motility}. These models have been studied only computationally and so the theory developed here may extend the understanding of these animals. 

The jamming and/or glass transition of self-aligning systems has been investigated both for flocking systems \cite{lam2015self,paoluzzi2024flocking,giavazzi2018flocking} and for systems in confinement \cite{henkes2011active}, using either particle-based or vertex models. The active solid state is necessarily manifest in the dense or low target shape index $p_0$, low activity, rigid region of the phase diagram. In this limit, the transition from a disordered to an oscillating or flocking state should be well described by our model. For example, the transition from disordered motion to flocking in the cell sheets of \cite{malinverno2017endocytic} occurs upon chemically induced weakening of cell-cell bonds, suggesting a change in the elastic properties of the cell sheet. In contrast, the low density region is better described by conventional liquid flocking models \cite{lam2015self,yllanes2017many}. The glass transition in the presence of self-alignment is poorly understood, with non-monotonous dependence of the $\alpha$-relaxation time on the alignment strength, and it will be very interesting how the length and time scales of oscillating modes interact with dynamical heterogeneities \cite{paoluzzi2024flocking}. Similarly, unjamming or loss of rigidity correspond to one or more floppy modes with zero eigenvalue appearing in the system, which immediately suggests an avenue for the system to move collectively along those modes.

In fact, from an engineering point of view, exploiting the condensation on the lowest mode of disordered materials only when activity is turned on, may prove fruitful in the design of  smart active materials \cite{dauchot2026active}. Alternatively, shaping the solid to condense the dynamics on a specific (\textit{i.e.} not necessarily the lowest mode) could be helpful for this and first steps have been made \cite{lazzari2024tuning}. 

A slightly different form of self-alignment has been used to modeled the bidirectional laning of epithelial cells \cite{lacroix2024emergence}. An important aspect of the model there is the use of a friction matrix, whereas in the current theory a scalar friction coefficient is used. It would therefore be interesting to add this form of anisotropic friction, which adds another coupling between polarity and velocity besides the self-alignment torque and investigate its effect on the mode selection in the framework of the current theory. Indeed, further simulation studies on the effect of anistropic friction in self-aligning systems have been performed \cite{tang2025collective,tang2026passivity}, but a complete theoretical understanding of its effects are still lacking. 

In the current work, we have only considered $J\geq0$, but research also spans systems that have negative self-alignment. For example, robots may be engineered that exhibit this form self-alignment \cite{ben2023morphological}. These systems have been explored in the low density regime \cite{casiulis2025geometric}, but our theory may apply to the high density limit where they form a solid.

\begin{acknowledgments}
We thank Y.-E. Keta for helpful early discussions. Furthermore, we also thank P. Baconnier and O. Dauchot for interesting discussions. We thank  A. Hernandez and L. Puggioni for insightful discussions about the continuum theory. This work was performed using the compute resources from the Academic Leiden Interdisciplinary Cluster Environment (ALICE) provided by Leiden University.
\end{acknowledgments}

%For now, separate the appendix onto a separate page
\appendix
\section{Derivations for the discrete theory}
\subsection{Nonlinear displacement dynamics}\label{apx:nonlinear_displacement_dynamics}
From Eqs.~\eqref{eq:discrete_displacement_eom} and \eqref{eq:discrete_polarity_eom}, and using the vector identity $\left[\hatnot{n}_i(t)\times\hatnot{v}_i (t)\right]\times \hatnot{n}_i(t) =  \left[\hatnot{n}_i(t) \cdot  \hatnot{n}_i(t)\right]\hatnot{v}_i - \left[\hatnot{n}_i(t) \cdot\hatnot{v}_i (t) \right]\hatnot{n}_i(t)= \hatnot{v}_i(t) -\left[\hatnot{n}_i(t) \cdot\hatnot{v}_i (t) \right]\hatnot{n}_i(t) $, we directly obtain Eq.~\eqref{eq:nonlinear_displacement_with_n}. Solving Eq.~\eqref{eq:discrete_displacement_eom} for $\zeta v_0\hatnot{n}$ and reserving repeated index notation for summation on index $j$ and $k$ \emph{only}, one obtains
\begin{equation}
    \zeta v_0\hatnot{n}_i = \zeta \delta \dot{\vecnot{r}}_i+  \matnot{D_{ij}} \ \delta \vecnot{r}_j.
    \label{eq:displacement_solved_for_polarity}
\end{equation}
The dot product is then 
\begin{equation}
    \zeta v_0\hatnot{n}_i \cdot \delta \dot{\vecnot{r}}_i  = \zeta v_i^2+  \left[ \matnot{D_{ij}} \ \delta \vecnot{r}_j \right] \cdot \delta \dot{\vecnot{r}}_i.
\end{equation}
Hence, we find for Eq.~\eqref{eq:nonlinear_displacement_with_n}, 
\begin{align}
    \zeta \delta\ddot{\vecnot{r}}_i = & -\matnot{D_{ij}} \ \delta \dot{\vecnot{r}}_j + \zeta \frac{J v_0}{v_i} \delta \dot{\vecnot{r}}_i- \frac{1}{\tau} \left(\zeta \delta \dot{\vecnot{r}}_i+  \matnot{D_{ij}} \ \delta \vecnot{r}_j \right) \nonumber \\
    & - \frac{J}{v_i v_0 \zeta} \left( \zeta v_i^2+  \left[ \matnot{D_{ik}} \ \delta \vecnot{r}_k \right] \cdot \delta \dot{\vecnot{r}}_i \right)\left(\zeta \delta \dot{\vecnot{r}}_i+  \matnot{D_{ij}} \ \delta \vecnot{r}_j \right) \nonumber \\
    & + \sqrt{\frac{2}{\tau}}\eta_i(t) \hatnot{z} \times \left(\zeta \delta \dot{\vecnot{r}}_i+  \matnot{D_{ij}} \ \delta \vecnot{r}_j \right).
\end{align}
Then, by grouping equal order in time derivatives, and using matrix notation, we obtain
\begin{align}
    \zeta \delta\ddot{\vecnot{r}}_i =& -\matnot{D_{ij}} \ \delta \dot{\vecnot{r}}_j + \zeta \frac{J v_0}{v_i} \delta \dot{\vecnot{r}}_i- \frac{1}{\tau} \zeta \delta \dot{\vecnot{r}}_i \nonumber \\
    & - \frac{J}{v_i v_0} \left( \zeta v_i^2+  \left[ \matnot{D_{ik}} \ \delta \vecnot{r}_k \right] \cdot \delta \dot{\vecnot{r}}_i \right)\delta \dot{\vecnot{r}}_i \nonumber \\
    &-  \left\lbrace \frac{1}{\tau} + \frac{J}{v_i v_0 \zeta} \left( \zeta v_i^2+  \left[ \matnot{D_{ik}} \ \delta \vecnot{r}_k \right] \cdot \delta \dot{\vecnot{r}}_i \right) \right\rbrace \matnot{D_{ij}} \ \delta \vecnot{r}_j \nonumber \\
    & + \sqrt{\frac{2}{\tau}}\eta_i(t) \hatnot{z} \times \left(\zeta \delta \dot{\vecnot{r}}_i+  \matnot{D_{ij}} \ \delta \vecnot{r}_j \right).
\end{align}
Finally, by writing every $\delta \vecnot{r}_i$ as a sum over the $\delta \vecnot{r}_j$, \textit{i.e.} $\delta \vecnot{r}_i= \delta_{ij} \delta \vecnot{r}_j$ to allow for matrix notation, and dividing both sides by the friction coefficient $\zeta$, we arrive at Eq.~\eqref{eq:discrete_decoupled}.

\subsection{Linearization}\label{apx:linearization}
Here we explain the linearization procedure, starting from the exact equation  Eq.~\eqref{eq:discrete_decoupled}. First we rewrite Eq.~\eqref{eq:discrete_decoupled} to show the almost-force-balance. For this we make use of the following identity,
\begin{equation}
    v_0^2 = v_i^2 + \|\zeta^{-1} \matnot{D_{ik}} \ \delta \vecnot{r}_k\|^2 + 2\zeta^{-1}  \left[\matnot{D_{ik}}\delta \vecnot{r}_k \right]\cdot \delta\dot{\vecnot{r}}_i  ,
\end{equation}
which follows directly from squaring Eq.~\eqref{eq:displacement_solved_for_polarity}.
Solving this identity for $\left[\matnot{D_{ik}}\delta \vecnot{r}_k \right]\cdot \delta\dot{\vecnot{r}}_i$, one finds:

\begin{equation}
    \frac{\left[\matnot{D_{ik}}\delta \vecnot{r}_k \right]\cdot \delta\dot{\vecnot{r}}_i}{\zeta v_iv_0} = \frac{v_0}{2v_i} - \frac{v_i}{2v_0} - \frac{1}{2 v_iv_0} \|\zeta^{-1} \matnot{D_{ik}} \ \delta \vecnot{r}_k \|^2,
 \end{equation}
then substituting it in Eq.~\eqref{eq:discrete_decoupled}, one obtains Eq.~\eqref{eq:discrete_decoupled_linearized} plus an additional nonlinear term (NLT):

\begin{align}
    \mathrm{NLT} = & \frac{J }{2v_i v_0} \left[ \delta_{ij}\|\zeta^{-1} \matnot{D_{ik}} \ \delta \vecnot{r}_k \|^2 \delta \dot{\vecnot{r}}_j \right. \nonumber\\
     &\left. - \zeta^{-1} \left(v_0^2 - \|\zeta^{-1} \matnot{D_{ik}} \ \delta \vecnot{r}_k \|^2 \right) \matnot{D_{ij}} \delta \vecnot{r}_j  \right].
\end{align}

Now, we can recognize the almost-force-balance by reintroducing $\hatnot{n}_i$ (using the original equation of motion Eq.~\eqref{eq:discrete_displacement_eom}), and, for brevity write $\vecnot{f}_i\equiv -\matnot{D_{ij}} \ \delta\vecnot{r}_j$ to find:

\begin{align}
    \mathrm{NLT} = \frac{J v_0}{2 v_i \zeta}\left[\frac{\| \vecnot{f}_i / \zeta\|^2}{v_0^2}\zeta v_0 \hatnot{n}_i+
\vecnot{f}_i \right].
\label{eq:NLT_force_balance}
\end{align}

In the noiseless climbing state, we have $\delta \dot{\vecnot{r}} = \vecnot{0}$ and the force balance $\frac{\| \vecnot{f}_i / \zeta \|^2}{v_0^2}\zeta v_0 \hatnot{n}_i+
\vecnot{f}_i= \vecnot{0}$, because the direction of the elastic force and the polarity vector are antiparallel, \textit{i.e.} $\hatnot{f}_i = -\hatnot{n}_i$, and the magnitude of the elastic force balances the active self-propulsion force: $\|\vecnot{f}_i \| = \zeta v_0$. 
And we will now show that by considering $\delta \vecnot{r}_i$ to be rotations around the noiseless climbing state, this term $\mathrm{NLT}$ vanishes, not only to zeroth order, but also to first order in these fluctuations. We will first discuss the case of a single particle in a harmonic well to explain the motivation and then apply this to the multiparticle case.
\subsubsection{Single particle}
The inspiration for the linearization procedure is the case of a single particle in a harmonic well with $\vecnot{f}_i = -k \vecnot{r}_i$. The noiseless climbing states is the set of all vectors $\{ \vecnot{r}_c, \hatnot{n}_c\}$ satisfying $\frac{\|\vecnot{f}_i / \zeta \|^2}{v_0^2} \zeta v_0 \hatnot{n}_c+ \vecnot{f}_i  = \frac{k^2 r_c^2}{v_0^2 \zeta^2} \zeta v_0 \hatnot{n}_c-k \vecnot{r}_c=0$. which represents the force balance between the elastic and active self-propulsion force. Note that this implies $\frac{ k^2 r_c^2}{v_0^2 \zeta^2} \zeta v_0=k r_{c} $.
Now consider fluctuations around any of these climbing states, in the form of rotations by $\delta \theta$, \textit{i.e.}
$\hatnot{n} = \hatnot{n}_c + \delta \theta \hatnot{n}_c^{\perp}$ (which to first order in the rotations $\delta \theta$ is still a unit vector as required), and
$\vecnot{r} = \vecnot{r}_c + r_c \delta \theta  \hatnot{n}_c^{\perp}$. Here, $\hatnot{n}_c^{\perp}$ is perpendicular to  $\vecnot{r}_c$, because $\hatnot{n}_c$ points in the same direction as $\hatnot{r}_c$.
Then, to first order in $\delta \theta$, we get
$\frac{|\vecnot{f}|^2}{v_0^2 \zeta^2} \zeta v_0 \hatnot{n}- \vecnot{f} =\frac{k^2 r_c^2}{\zeta^2 v_0^2} \zeta v_0 \hatnot{n}_c- k \vecnot{r}_c +\frac{k^2 r_c^2}{v_0^2 \zeta^2} \zeta v_0 \delta \theta \hatnot{n}_c^{\perp}-k r_c \delta \theta \hatnot{n}_c^{\perp}= 0$, because the first two terms sum to zero by definition of the climbing state and the last two terms sum to zero because $\frac{k^2 r_c^2}{v_0^2 \zeta^2 } \zeta v_0= k r_{c}$, again by definition of the climbing state.
\subsubsection{Multiparticle solid}
Going back to the multi particle case, we naturally consider then rotations $\delta \theta_i$ for each particle around its climbing state (in its local potential well), where the active and elastic forces balance eachother, $\hatnot{n}_i = \hatnot{n}_{i,c} + \delta \theta_i \hatnot{n}_{i,c}^{\perp}$ and $\delta \vecnot{r}_{i} = \delta \vecnot{r}_{i,c} + \|\delta \vecnot{r}_{i,c}\|\delta \theta_{i}\hatnot{n}_{i,c}^{\perp}$.
To first order in $\delta \theta_i$, we get
\begin{align}
    &\frac{\| \vecnot{f}_i\|^2}{v_0^2 \zeta^2} \zeta v_0 \hatnot{n}_i+\vecnot{f}_i = \\&\frac{\|\matnot{D_{ij}} \  \delta \vecnot{r}_{j,c}\|^2}{v_0^2 \zeta^2}\zeta v_0 \hatnot{n}_{i,c}- \matnot{D_{ij}} \  \delta \vecnot{r}_{j,c} \nonumber \\ 
    &+ \frac{\| \matnot{D_{ij}} \  \delta \vecnot{r}_{j,c}\|^2}{v_0^2 \zeta^2}\zeta v_0 \delta \theta_i \hatnot{n}_{i,c}^{\perp}-  \|\delta \vecnot{r}_{j,c}\| \matnot{D_{ij}}\  \delta \theta_j \hatnot{n}_{j,c}^{\perp} \nonumber \\
    &+2\frac{ \left(\matnot{D_{ij}} \  \delta \vecnot{r}_{j,c} \right) \cdot \left( \|\delta \vecnot{r}_{i,c}\| \matnot{D_{ij}} \ \delta \theta_j \hatnot{n}_{j,c}^{\perp}\right)}{v_0^2 \zeta^2} \zeta v_0\hatnot{n}_{i,c} \nonumber.
\end{align}
The second line is zero by definition of the noiseless climbing state. From the climbing state definition, we also learn that $\frac{\|\matnot{D_{ij}}  \ \delta \vecnot{r}_{j,c}\|^2}{v_0^2 \zeta^2 }\zeta v_0$ is of order $\|\matnot{D_{ij}} \ \delta \vecnot{r}_{j,c}\|$. If we consider small $\delta \vecnot{r}_{j,c}$, then the third and fourth line on the RHS are zero simply by being of higher order. One can show this more formally by writing $\delta \vecnot{r} = \epsilon \Delta R \hatnot{r}_c + \delta \theta \epsilon \Delta R \hatnot{n}^{\perp}_c$, with $\Delta R$ being the interparticle distance and with $\epsilon$ a small dimensionless number compared to 1, so that $\|\delta\vecnot{r}_{i,c}\|=\epsilon \Delta R$. We see that first order fluctuations around the climbing state rotate the polarity vector, but leave the position vector unchanged (to first order in $\delta \theta$ and $\epsilon$) and hence the third and fourth line are of order $\epsilon \delta\theta$, which is much smaller compared to 1. So, to first order in $\delta \theta$, $\mathrm{NLT}=0$.  Conceptually, the condition for this linearization to work is thus that both the distance from the reference position in force balance, \textit{i.e.} $\|\delta\vecnot{r}_{i,c}\|$ and the perpendicular component to the force balance $\delta\theta$ are both small, so that their product is of second order. Therefore, by considering the deviations $\delta \vecnot{r}_i$ to be rotations around the climbing state, that is, small deviations around the elastic and active force balance, we arrive at the simplified equation Eq.~\eqref{eq:discrete_decoupled_linearized}. Alternatively, we can employ the (approximate) rotational invariance of $\matnot{D}$, which would also make line 3 and 4 vanish. This should in fact be a much better assumption for the continuum theory, cf. the continuum linearization in (Appendix, \ref{apx:linearization_continuum}).

\subsection{Steady state dynamics}\label{apx:velocity_projections}

\subsubsection{Mode spectrum}
We solve Eq.~\eqref{eq:discrete_mode_dynamics} for steady state and go to temporal Fourier space to obtain
\begin{equation}
\begin{aligned}
    -\omega^2 \tilde{a}_{\nu}(\omega) = & i\omega \Gamma_{\nu}  \tilde{a}_{\nu}(\omega) -\Omega_{\nu}^2\tilde{a}_{\nu}(\omega) + v_0 \sqrt{\frac{2}{\tau}} \tilde{w}(\omega).
\end{aligned}
\end{equation}
Solving for the Fourier amplitude, we thus obtain

\begin{equation}
\begin{aligned}
    \tilde{a}_{\nu}(\omega) = \frac{v_0 \sqrt{\frac{2}{\tau}} \tilde{w}(\omega)}{\Omega_{\nu}^2- \omega^2 - i\omega\Gamma_{\nu}}.
\end{aligned}
\end{equation}
And so, the correlation is given by, using the variance of the noise  $\langle w_{\nu}(t) w_{\rho}(t') \rangle = \frac{1}{2} \delta_{\nu, \rho}\delta(t-t')$,
\begin{equation}
\begin{aligned}
    \langle \tilde{a}_{\nu}(\omega)  \tilde{a}_{\nu}(\omega') \rangle= \frac{v_0^2 \frac{1}{\tau} 2 \pi \delta(\omega+\omega')}{ \left(\Omega_{\nu}^2- \omega^2 \right)^2+ \omega^2 \Gamma_{\nu}^2},
\end{aligned}
\end{equation}
which means the spectrum of the time derivative of the mode amplitude is given by Eq.~\eqref{eq:discrete_mode_velocity_spectrum}.
\subsubsection{Velocity projections}
The steady state velocity projections can be found by inverse Fourier transforming equation \eqref{eq:discrete_mode_velocity_spectrum} back to
to the time-domain, i.e.
\begin{equation}
\begin{aligned}
    \langle \dot{a}_{\nu}(t) \dot{a}_{\nu}(t) \rangle= \frac{1}{(2 \pi)^2}\iint d\omega d\omega'\langle \tilde{\dot{a}}_{\nu}(\omega)  \tilde{\dot{a}}_{\nu}(\omega') \rangle e^{-i (\omega+ \omega')t}.
\end{aligned}
\end{equation}

Resolving the first integral using the Dirac delta, we are left with
\begin{equation}
    \langle \dot{a}_{\nu}(t) \dot{a}_{\nu}(t) \rangle=\frac{\frac{1}{\tau}v_0^2}{2 \pi} \int_{-\infty}^{\infty}d\omega \frac{\omega^2}{ \left(\Omega_{\nu}^2 - \omega^2 \right)^2+ \omega^2 \Gamma_{\nu}^2}.
\end{equation}
As long as $\Gamma_{\nu}>0$, this integral converges and the result is given by Eq.~\eqref{eq:discrete_mode_velocity_equal_time}, which can be verified by contour integration. We do this explicitly for the more general case of the  unequal-time correlation in section \ref{apx:velocity_autocorrelation} and the equal-time result follows by setting $t'=t$ there.

%If necessary we can include the explict calculation here, or, as we do now, just refer to the standard literature results. The poles of the integrand are given by $\omega^2 = -\frac{1}{2}\left(\Gamma_{\nu}^2 - 2 \Omega_{\nu}^2 \right) \pm \frac{1}{2}\Gamma_{\nu}\sqrt{\Gamma_{\nu}^2 - 4 \Omega_{\nu}^2}$. For $\Omega_{\nu}^2 - \Gamma_{\nu}^2/4<0$ (overdamped case) we find the poles to be purely imaginary... 

\subsubsection{Polarity spectrum}\label{apx:polarity_spectrum}
By decomposing the polarity vectors into the eigenvectors, $\hatnot{n}_1 \otimes \hatnot{n}_2 \otimes \ldots \otimes \hatnot{n}_N = \sum_{\nu}\beta_{\nu} \vecnot{\chi}_{\nu}$, it follows from the original equation of motion for the displacement (Eq.~\eqref{eq:discrete_displacement_eom}) that
\begin{equation}
    v_0 \beta_{\nu} = \dot{a}_{\nu}+ \frac{\lambda_{\nu}}{\zeta}a_{\nu},
\end{equation}
therefore, we find
\begin{equation}
    \tilde{\beta}_{\nu}(\omega) = \frac{1}{v_0} \tilde{\dot{a}}_{\nu}(\omega)\left[1+i\frac{ \lambda_{\nu}}{\zeta \omega}\right].
\end{equation}
Hence,
\begin{equation}
    \langle \tilde{\beta}_{\nu}(\omega)\tilde{\beta}_{\nu}(\omega')\rangle = \left[1+\frac{ \lambda_{\nu}^2}{\zeta^2 \omega^2}\right]\frac{\langle \tilde{\dot{a}}_{\nu}(\omega)  \tilde{\dot{a}}_{\nu}(\omega') \rangle}{v_0^2}.
\end{equation}
Since $\langle \hatnot{n}_i (t) \cdot  \hatnot{n}_i(t') \rangle_{i} = \frac{1}{N} \sum_{\nu} \langle  \beta_{\nu}(t) \beta_{\nu}(t')\rangle$, and assuming an isotropic ensemble average, in which $\langle \tilde{n}_x(\omega)  \tilde{n}_x(\omega')\rangle = \frac{1}{2} \langle \tilde{\hatnot{n}}(\omega) \cdot   \tilde{\hatnot{n}}(\omega')\rangle$, we arrive at Eq.~\eqref{eq:discrete_polarity_spectrum}.

\subsubsection{Velocity autocorrelation function}\label{apx:velocity_autocorrelation}
We will assume $\Gamma_{\nu}>0$ and start from
\begin{equation}
\begin{aligned}
    \langle \dot{a}_{\nu}(t) \dot{a}_{\nu}(t') \rangle= \frac{1}{(2 \pi)^2}\iint d\omega d\omega'\langle \tilde{\dot{a}}_{\nu}(\omega)  \tilde{\dot{a}}_{\nu}(\omega') \rangle e^{-i \omega t - i \omega' t'}.
\end{aligned}
\end{equation}
We resolve one of the $\omega$ integrals using the Dirac delta in $\langle \tilde{\dot{a}}_{\nu}(\omega)  \tilde{\dot{a}}_{\nu}(\omega') \rangle$ to find

\begin{align}
    \langle \dot{a}_{\nu}(t) \dot{a}_{\nu}(t') \rangle= \frac{\frac{1}{\tau}v_0^2}{2 \pi} \int_{-\infty}^{\infty}d\omega \frac{\omega^2  e^{-i \omega(t-t')}}{ \left(\Omega_{\nu}^2 - \omega^2 \right)^2+ \omega^2 \Gamma_{\nu}^2}.
\end{align}
The poles of the integrand are given by 
\begin{align}
        \omega^2 =& -\frac{1}{2}\left(\Gamma_{\nu}^2 - 2 \Omega_{\nu}^2 \right) \pm \frac{1}{2}\Gamma_{\nu}\sqrt{\Gamma_{\nu}^2 - 4 \Omega_{\nu}^2} \nonumber \\
    &=-\left(\frac{1}{2} \Gamma_{\nu} \ \pm \frac{1}{2}\sqrt{\Gamma_{\nu}^2 - 4\Omega_{\nu}^2} \right)^2,
\end{align}
this means the poles are, after some algebra and defining the (possibly imaginary) angular frequency $\omega_{\nu} = \sqrt{\Omega_{\nu}^2 - \Gamma_{\nu}^2 /4}$,

\begin{align}
    \omega_{\pm, \pm} = \pm \left( \pm\omega_{\nu}+ i\Gamma_{\nu}/2 \right).
\end{align}

If $\omega_{\nu}$ is real (in case of a critical damped or underdamped mode), the poles in the top half of the complex plane are $\omega_{+, \pm}$, \textit{i.e.} $\omega_{+, +}= +(\omega_{\nu} + i\Gamma_{\nu}/2 )$ and $\omega_{+, -}= +(-\omega_{\nu} + i\Gamma_{\nu}/2 )$. In case $\omega_{\nu}$ is imaginary, we can write $\omega_{\nu}= i|\omega_{\nu}|=i\sqrt{\Gamma_{\nu}^2/4 - \Omega_{\nu}^2}$ and $|\omega_{\nu}|<\Gamma_{\nu}/2$, which means the poles in the top half of the complex plane are still $\omega_{+, +}$ and $\omega_{+, -}$.
Considering $t'>t$ (the other cases follow analogously), we take a contour in the upper half of the complex plane, and write the integrand as $\frac{\omega^2 e^{-i\omega(t-t')}}{(\omega - \omega_{+,+})(\omega - \omega_{+,-})(\omega - \omega_{-,+})(\omega - \omega_{-,-})}$ with the residues
\begin{align}
    \mathrm{Res}_{ \omega_{+,+}} &= \frac{\omega_{+,+}^2 e^{-\Gamma_{\nu}(t'-t)/2} e^{-i\omega_{\nu}(t'-t)}}{2\omega_{\nu}\left(2\omega_{\nu} +i\Gamma_{\nu} \right)i\Gamma_{\nu}}  \\
    &= \frac{\left(\omega_{\nu}/2 + i\Gamma_{\nu}/4 \right) e^{-\Gamma_{\nu}(t'-t)/2} e^{-i\omega_{\nu}(t'-t)}}{2\omega_{\nu}i\Gamma_{\nu}} \nonumber,
\end{align}
and 
\begin{align}
    \mathrm{Res}_{ \omega_{+,-}} &= \frac{\omega_{+,-}^2 e^{-\Gamma_{\nu}(t'-t)/2} e^{i\omega_{\nu}(t'-t)}}{-2\omega_{\nu}i\Gamma_{\nu}\left(-2\omega_{\nu} +i\Gamma_{\nu} \right)} \\
    &=\frac{\omega_{+,-}^2 e^{-\Gamma_{\nu}(t'-t)/2} e^{i\omega_{\nu}(t'-t)}}{2\omega_{\nu}i\Gamma_{\nu}\left(2\omega_{\nu} -i\Gamma_{\nu} \right)} \nonumber \\
    &=\frac{\left(\omega_{\nu}/2 - i\Gamma_{\nu}/4 \right)e^{-\Gamma_{\nu}(t'-t)/2} e^{i\omega_{\nu}(t'-t)}}{2\omega_{\nu}i\Gamma_{\nu}} \nonumber,
\end{align}
so that one obtains, with the help of the residue theorem \cite{byron2012mathematics},
\begin{align}
&\langle \dot{a}_{\nu}(t) \dot{a}_{\nu}(t') \rangle=\frac{v_0^2}{2\tau \Gamma_{\nu}}e^{-\Gamma_{\nu}|t-t'|/2} \cdot  \\ &\left[\frac{e^{i\omega_{\nu}|t-t'|}+e^{-i\omega_{\nu}|t-t'|}}{2}  -  \frac{\Gamma_{\nu}}{2\omega_{\nu}}\frac{e^{i\omega_{\nu}|t-t'|}- e^{-i\omega_{\nu}|t-t'|}}{2i} \right]. \nonumber
\end{align}
From this we see that for an overdamped mode with $\omega_{\nu}=i|\omega_{\nu}|$, we have 
\begin{align}
&\langle \dot{a}_{\nu}(t) \dot{a}_{\nu}(t') \rangle= \frac{v_0^2}{2\tau \Gamma_{\nu}}e^{-\Gamma_{\nu}|t-t'|/2}\cdot \nonumber\\
&\left[ \cosh(|\omega_{\nu}||t-t'|)- \frac{\Gamma_{\nu}}{2|\omega_{\nu}|}\sinh(|\omega_{\nu}||t-t'|) \right],
\label{eq:overdamped unequal time correlation}
\end{align}
or for a critically damped or underdamped mode, with $\omega_{\nu}$ real,
\begin{align}
&\langle \dot{a}_{\nu}(t) \dot{a}_{\nu}(t') \rangle=\frac{v_0^2}{2\tau \Gamma_{\nu}}e^{-\Gamma_{\nu}|t-t'|/2} \cdot \\
&\left[ \cos(\omega_{\nu}|t-t'|)- \frac{\Gamma_{\nu}}{2\omega_{\nu}}\sin(\omega_{\nu}|t-t'|) \right] \nonumber.
\label{eq:underdamped unequal time correlation}
\end{align}
For zero modes, which have $\Omega_{\nu}=0$, and are thus by necessity either overdamped or undamped, Eq.\eqref{eq:overdamped unequal time correlation} still holds. This is expected from continuity, but can also be shown explicitly: in this case, the integrand simplifies significantly and we get
\begin{align}
    \langle \dot{a}_{\nu}(t) \dot{a}_{\nu}(t') \rangle= \frac{\frac{1}{\tau}v_0^2}{2 \pi} \int_{-\infty}^{\infty}d\omega \frac{  e^{-i \omega(t-t')}}{ \omega^2+  \Gamma_{\nu}^2}.
\end{align}
The poles are $\omega_{+} = i \Gamma_{\nu}$ and $\omega_{-} = -i \Gamma_{\nu}$. Considering $t'>t$, we choose a contour enclosing the pole in the positive complex plane. Employing the residue theorem, we calculate the residue of the integrand
\begin{align}
    \mathrm{Res} = \frac{e^{-\Gamma_{\nu} (t'-t)}}{2 i \Gamma_{\nu}},
\end{align}
and so we get in the case of a zero mode
\begin{align}
    \langle \dot{a}_{\nu}(t) \dot{a}_{\nu}(t') \rangle= \frac{v_0^2}{2\tau \Gamma_{\nu}} e^{-\Gamma_{\nu} |t-t'|}.
\end{align}

In all of the above cases, the velocity autocorrelation is given by 
\begin{align}
    C_{vv}(\Delta t) = \frac{\frac{1}{N} \sum_{\nu=1}^{2N}  \langle \dot{a}_{\nu}(t) \dot{a}_{\nu}(t+\Delta t) \rangle}{\frac{1}{N} \sum_{\nu=1}^{2N}\langle \dot{a}_{\nu}(t) \dot{a}_{\nu}(t) \rangle}.
\end{align}

\subsection{Simulations}\label{apx:simulations}
\subsubsection{Extracting the dynamical matrix}
The potential of the repulsive springs is defined as 

\begin{equation}
    V = \sum_{\langle i,j \rangle}\frac{k_{ij}}{2}\left( \|\vecnot{r}_i - \vecnot{r}_j\| - R_i - R_j \right)^2,
\label{eq:potential}
\end{equation}
here, the sum runs over all unique particle pairs that are in contact with each other in the equilibrium reference state: $\|\vecnot{r}_{i,0} - \vecnot{r}_{j,0} \| \leq R_i +R_j$. $k_{ij}=1$ in the simulations between interior particles and $k_{ij}=2$ for interactions between a boundary particle and an interior particle, and $R_i$ and $R_j$ are the radii of the particles.
The dynamical matrix subblock $\matnot{D_{ij}}$ can be found by Taylor expanding the force on particle $i$ derived from this potential around the reference equilibrium configuration $\{\vecnot{r}_{i,0} \}$, \textit{i.e.} 
\begin{align}
    \vecnot{f}_i =& - \nabla_{\vecnot{r}_i}V \\
    =&-\nabla_{\vecnot{r}_i}V\rvert_{\{\vecnot{r}_{i,0} \}} - \sum_{j=1}^{N}\left[\nabla_{\vecnot{r}_j}\nabla_{\vecnot{r}_i}V\rvert_{\{\vecnot{r}_{i,0} \} }\right] \cdot \delta \vecnot{r}_j \nonumber\\&+\mathrm{h.o.t.} \nonumber,
\label{eq:gradient_expansion}
\end{align} where h.o.t. stands for higher order terms.
By definition of the reference state $-\nabla_{\vecnot{r}_i}V\rvert_{\{\vecnot{r}_{i,0} \}}=\vecnot{0}$, and so to lowest (linear) order in the displacements around the reference state, the elastic response is encoded in the submatrices $\matnot{D_{ij}} = \nabla_{\vecnot{r}_j}\nabla_{\vecnot{r}_i}V\rvert_{\{\vecnot{r}_{i,0} \} }$, where it is good to note that $j$ can be equal to $i$.
For the potential $V$ of Eq.~\eqref{eq:potential}, we have
\begin{align}
    \nabla_{\vecnot{r}_i}V =& \nabla_{\vecnot{r}_i}\sum_{j \in n_i} \frac{k_{ij}}{2} \left( \|\vecnot{r}_i - \vecnot{r}_{j}\| - R_i - R_j \right)^2 \\ \nonumber
    =&  -\sum_{j \in n_i}k_{ij} \left(\|\vecnot{r}_i - \vecnot{r}_{j}\| - R_i - R_j  \right)\hatnot{r}_{ij},
    \label{eq:gradient_V}
\end{align}
where $n_i$ is the set of neighbours of particle $i$, and $\hatnot{r}_{ij}$ is the unit vector pointing from particle $i$ to particle $j$.
Then, when $j\neq i$ in the sum of Eq.~\eqref{eq:gradient_expansion}, we get from Eq.~\eqref{eq:gradient_V},

\begin{align}
    \nabla_{\vecnot{r}_j} \nabla_{\vecnot{r}_i}V  \stackrel{j\neq i}{=} &-k_{ij} \hatnot{r}_{ij} \otimes\hatnot{r}_{ij} \\
    &- k_{ij} \left(1 - \frac{R_i+R_j}{\|\vecnot{r}_ i - \vecnot{r}_j \|}\right) \left(\matnot{I} -\hatnot{r}_{ij} \otimes\hatnot{r}_{ij} \right) \nonumber,
        \label{eq:D_{ij}}
\end{align}
in which $\otimes$ denotes the outer product and $\matnot{I}$ is the identity matrix in two dimensions. In the jamming literature, it is common to denote $\hatnot{r}_{ij} = \hatnot{n}_{ij}$ (which is normal to the contact surface) and the perpendicular direction as $\hatnot{t}_{ij}$ (which in the discussion here can be either of the two tangential vectors to the contact surface). In this convention, $\matnot{I} = \hatnot{n}_{ij} \otimes \hatnot{n}_{ij} + \hatnot{t}_{ij} \otimes \hatnot{t}_{ij}$, so that for $i\neq j$, we get

\begin{align}
    \matnot{D_{ij}} \stackrel{i\neq j}{=} &-k_{ij} \hatnot{n}_{ij,0} \otimes\hatnot{n}_{ij,0} \\
    &- k_{ij}\left(1 - \frac{R_i+R_j}{\|\vecnot{r}_{i,0} - \vecnot{r}_{j,0} \|}\right) \hatnot{t}_{ij,0} \otimes \hatnot{t}_{ij,0} \nonumber.
    \label{eq:D_{ij}}
\end{align}
For $j=i$ in the sum Eq.~\eqref{eq:gradient_expansion}, we find
\begin{equation}
    \nabla_{\vecnot{r}_i} \nabla_{\vecnot{r}_i}V = - \nabla_{\vecnot{r}_i}\sum_{j \in n_i}k_{ij} \left(\|\vecnot{r}_i - \vecnot{r}_{j}\| - R_i - R_j  \right)\hatnot{r}_{ij},
\end{equation}
which leads to
\begin{equation}
    \matnot{D_{ii}} = -\sum_{j \neq i}\matnot{D_{ij}}.
    \label{eq:D_{ii}}
\end{equation}
By using Eqs.~\eqref{eq:D_{ij}} and \eqref{eq:D_{ii}}, we calculate the dynamical matrix in simulations. The dynamical matrix is then diagonalized to find its eigenvalues $\{ \lambda_{\nu}\}$ and orthonormal eigenvectors $\{ \vecnot{\chi}_{\nu}\}$.

\section{Derivations for the continuum theory}
\subsection{Full nonlinear expression}\label{apx:nonlinear_continuum}

\begin{widetext}
    \begin{eqnarray}
    \ddot{\vecnot{u}} =&& \frac{B}{\zeta}\nabla\left( \nabla \cdot \dot{\vecnot{u}}\right) + \frac{\mu}{\zeta}\nabla^2 \dot{\vecnot{u}} - \left( \frac{1}{\tau }-\frac{J v_0}{v}+\frac{J}{v v_0} \left[ v^2 -  \frac{B}{\zeta} \dot{\vecnot{u}}  \cdot \nabla\left( \nabla \cdot \vecnot{u}\right) -\frac{\mu}{\zeta} \dot{\vecnot{u}}  \cdot\nabla^2 \vecnot{u} \right]\right)\dot{\vecnot{u}}  \nonumber \\
   &&+\left(\frac{1}{\tau}+\frac{J}{v v_0} \left[ v^2 -  \frac{B}{\zeta} \dot{\vecnot{u}}  \cdot \nabla\left( \nabla \cdot \vecnot{u}\right) -\frac{\mu}{\zeta} \dot{\vecnot{u}}  \cdot\nabla^2 \vecnot{u} \right]\right)\left(\frac{B}{\zeta} \nabla\left( \nabla \cdot \vecnot{u}\right) + \frac{\mu}{\zeta}\nabla^2 \vecnot{u} \right) \nonumber \\ 
   &&+ \sqrt{\frac{2}{\tau}} \eta(t)\hatnot{z} \times \left(\dot{\vecnot{u}} -\frac{B}{\zeta}\nabla\left( \nabla \cdot \vecnot{u}\right) - \frac{\mu}{\zeta} \nabla^2 \vecnot{u}  \right).
\end{eqnarray}
\end{widetext}

\subsection{Linearization}\label{apx:linearization_continuum}
Just as in Section \ref{apx:linearization},
we use the identity derived from squaring the displacement equation of motion (Eq.~\eqref{eq:co_displacement_eom}),
\begin{eqnarray}
    v_0^2 = v^2 + \|\zeta^{-1}D(\vecnot{u}) \|^2 + 2\zeta^{-1}D(\vecnot{u})\cdot \dot{\vecnot{u}},
\end{eqnarray}
and solve this for $D(\vecnot{u})\cdot \dot{\vecnot{u}}$, to get,
\begin{align}
    \frac{D(\vecnot{u})\cdot \dot{\vecnot{u}}}{\zeta v v_0}=\frac{v_0}{2v} - \frac{v}{2v_0} - \frac{1}{2vv_0}\|\zeta^{-1}D(\vecnot{u}) \|^2
\end{align}
which we then substitute in Eq.~\eqref{eq:nonlinear_displacement_continuum}. This leads to, following the equivalent steps that lead to Eq.~\eqref{eq:NLT_force_balance},  Eq.~\eqref{eq:continuum_decoupled_linearized}
with the addition of the nonlinearity (NLT)

\begin{align}
    \mathrm{NLT}=\frac{Jv_0}{2v \zeta}\left[\frac{\|D(\vecnot{u})/\zeta\|^2}{v_0^2}\zeta v_0 \hatnot{n}- D(\vecnot{u})\right].
\end{align}
Now, considering $\hatnot{n} = \hatnot{n}_c + \delta \theta \hatnot{n}_{\perp}$ and $\vecnot{u} = \vecnot{u}_c + u_c \delta \theta \hatnot{n}_{\perp}$  to be deviations around the noiseless climbing states $\{\vecnot{u}_c, \hatnot{n}_c \}$ defined by Eq.~\eqref{eq:noiseless_climbing_state_continuum}, we get to first order 
\begin{align}
    &\frac{\|D(\vecnot{u})/\zeta \|^2}{v_0^2}\zeta v_0 \hatnot{n}- D(\vecnot{u}) = \\&\frac{\|D(\vecnot{u}_c)\|^2}{v_0^2 \zeta^2}\zeta v_0 \hatnot{n}_{c}- D(\vecnot{u}_c) \nonumber \\ 
    &+ \frac{\|D(\vecnot{u_{c}})\|^2}{v_0^2 \zeta^2}\zeta v_0 \delta \theta \hatnot{n}_{c}^{\perp}-  D(u_c \delta \theta \hatnot{n}_{c}^{\perp})\   \nonumber \\
    &+2\frac{ D(\vecnot{u}_c) \cdot D(u_c \delta \theta \hatnot{n}_{c}^{\perp})}{v_0^2 \zeta^2} \zeta v_0\hatnot{n}_{c} \nonumber,
\end{align}
the first term on the right hand side is zero by definition of the noiseless climbing state, and the last two terms are zero by considering small $u_c$ and small $\delta\theta$ (cf. the discussion of the discrete treatment) or exactly zero by rotational invariance of the operator $D(\vecnot{u})\equiv -B \nabla\left(\nabla \cdot \vecnot{u}\right) - \mu \nabla^2 \vecnot{u}$.

\subsection{Velocity spectrum}\label{apx:velocity_spectrum}
Taking the spatiotemporal Fourier transform $\vecnot{u}(\vecnot{r},t)\rightarrow \vecnot{U}(\vecnot{q},\omega)$of the linearized equation, Eq.~\ref{eq:continuum_decoupled_linearized}, and defining the longitudinal component as $U_{\parallel}(\vecnot{q},\omega) \equiv \vecnot{U}(\vecnot{q},\omega) \cdot \hatnot{q}$, and the transversal component $U_{\perp}(\vecnot{q},\omega) \equiv \vecnot{U}(\vecnot{q},\omega)-U_{\parallel}(\vecnot{q},\omega)\hatnot{q}$, allows to diagonalize the equation of motion in Fourier space in terms of these components.
Then using  $\vecnot{v} = \partial_t \vecnot{u} \leftrightarrow \vecnot{V}(\vecnot{q},\omega) = -i\omega\vecnot{U}(\vecnot{q},\omega)$, one arrives at Eq.~\eqref{eq:continuum_velocity_spectrum}.
\subsection{Polarity spectrum distortion-free expansion}\label{axp:pbc_distortion_free_spectrum}
\begin{figure}[H]
    \centering
\includegraphics{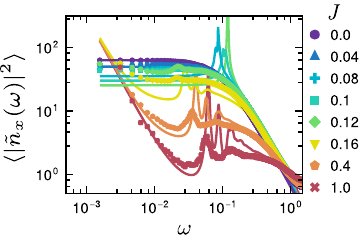}
    \caption{Polarity spectrum for the system with periodic boundary conditions for different values of $J$ (log-log scale). Scatters represent the simulations, the lines correspond to the linearized theory, but using the distortion-free expansion. Notably, this expansion only works well, and better than the elastic and active force balance expansion, in the flocking regime $J>0.4$.}
    \label{fig:pbc_distortion_free_spectrum}
\end{figure}

\bibliography{references} % Produces the bibliography via BibTeX.

\providecommand{\noopsort}[1]{}\providecommand{\singleletter}[1]{#1}%
\begin{thebibliography}{80}%
\makeatletter
\providecommand \@ifxundefined [1]{%
 \@ifx{#1\undefined}
}%
\providecommand \@ifnum [1]{%
 \ifnum #1\expandafter \@firstoftwo
 \else \expandafter \@secondoftwo
 \fi
}%
\providecommand \@ifx [1]{%
 \ifx #1\expandafter \@firstoftwo
 \else \expandafter \@secondoftwo
 \fi
}%
\providecommand \natexlab [1]{#1}%
\providecommand \enquote  [1]{``#1''}%
\providecommand \bibnamefont  [1]{#1}%
\providecommand \bibfnamefont [1]{#1}%
\providecommand \citenamefont [1]{#1}%
\providecommand \href@noop [0]{\@secondoftwo}%
\providecommand \href [0]{\begingroup \@sanitize@url \@href}%
\providecommand \@href[1]{\@@startlink{#1}\@@href}%
\providecommand \@@href[1]{\endgroup#1\@@endlink}%
\providecommand \@sanitize@url [0]{\catcode `\\12\catcode `\$12\catcode
  `\&12\catcode `\#12\catcode `\^12\catcode `\_12\catcode `\%12\relax}%
\providecommand \@@startlink[1]{}%
\providecommand \@@endlink[0]{}%
\providecommand \url  [0]{\begingroup\@sanitize@url \@url }%
\providecommand \@url [1]{\endgroup\@href {#1}{\urlprefix }}%
\providecommand \urlprefix  [0]{URL }%
\providecommand \Eprint [0]{\href }%
\providecommand \doibase [0]{https://doi.org/}%
\providecommand \selectlanguage [0]{\@gobble}%
\providecommand \bibinfo  [0]{\@secondoftwo}%
\providecommand \bibfield  [0]{\@secondoftwo}%
\providecommand \translation [1]{[#1]}%
\providecommand \BibitemOpen [0]{}%
\providecommand \bibitemStop [0]{}%
\providecommand \bibitemNoStop [0]{.\EOS\space}%
\providecommand \EOS [0]{\spacefactor3000\relax}%
\providecommand \BibitemShut  [1]{\csname bibitem#1\endcsname}%
\let\auto@bib@innerbib\@empty
%</preamble>
\bibitem [{\citenamefont {Marchetti}\ \emph {et~al.}(2013)\citenamefont
  {Marchetti}, \citenamefont {Joanny}, \citenamefont {Ramaswamy}, \citenamefont
  {Liverpool}, \citenamefont {Prost}, \citenamefont {Rao},\ and\ \citenamefont
  {Simha}}]{marchetti2013hydrodynamics}%
  \BibitemOpen
  \bibfield  {author} {\bibinfo {author} {\bibfnamefont {M.~C.}\ \bibnamefont
  {Marchetti}}, \bibinfo {author} {\bibfnamefont {J.~F.}\ \bibnamefont
  {Joanny}}, \bibinfo {author} {\bibfnamefont {S.}~\bibnamefont {Ramaswamy}},
  \bibinfo {author} {\bibfnamefont {T.~B.}\ \bibnamefont {Liverpool}}, \bibinfo
  {author} {\bibfnamefont {J.}~\bibnamefont {Prost}}, \bibinfo {author}
  {\bibfnamefont {M.}~\bibnamefont {Rao}},\ and\ \bibinfo {author}
  {\bibfnamefont {R.~A.}\ \bibnamefont {Simha}},\ }\href
  {https://doi.org/10.1103/RevModPhys.85.1143} {\bibfield  {journal} {\bibinfo
  {journal} {Rev. Mod. Phys.}\ }\textbf {\bibinfo {volume} {85}},\ \bibinfo
  {pages} {1143} (\bibinfo {year} {2013})}\BibitemShut {NoStop}%
\bibitem [{\citenamefont {Gompper}\ \emph {et~al.}(2025)\citenamefont
  {Gompper}, \citenamefont {Stone}, \citenamefont {Kurzthaler}, \citenamefont
  {Saintillan}, \citenamefont {Peruani}, \citenamefont {Fedosov}, \citenamefont
  {Auth}, \citenamefont {Cottin-Bizonne}, \citenamefont {Ybert}, \citenamefont
  {Cl{\'e}ment} \emph {et~al.}}]{gompper20252025}%
  \BibitemOpen
  \bibfield  {author} {\bibinfo {author} {\bibfnamefont {G.}~\bibnamefont
  {Gompper}}, \bibinfo {author} {\bibfnamefont {H.~A.}\ \bibnamefont {Stone}},
  \bibinfo {author} {\bibfnamefont {C.}~\bibnamefont {Kurzthaler}}, \bibinfo
  {author} {\bibfnamefont {D.}~\bibnamefont {Saintillan}}, \bibinfo {author}
  {\bibfnamefont {F.}~\bibnamefont {Peruani}}, \bibinfo {author} {\bibfnamefont
  {D.~A.}\ \bibnamefont {Fedosov}}, \bibinfo {author} {\bibfnamefont
  {T.}~\bibnamefont {Auth}}, \bibinfo {author} {\bibfnamefont {C.}~\bibnamefont
  {Cottin-Bizonne}}, \bibinfo {author} {\bibfnamefont {C.}~\bibnamefont
  {Ybert}}, \bibinfo {author} {\bibfnamefont {E.}~\bibnamefont {Cl{\'e}ment}},
  \emph {et~al.},\ }\href@noop {} {\bibfield  {journal} {\bibinfo  {journal}
  {Journal of Physics: Condensed Matter}\ }\textbf {\bibinfo {volume} {37}},\
  \bibinfo {pages} {143501} (\bibinfo {year} {2025})}\BibitemShut {NoStop}%
\bibitem [{\citenamefont {Cavagna}\ and\ \citenamefont
  {Giardina}(2014)}]{cavagna2014bird}%
  \BibitemOpen
  \bibfield  {author} {\bibinfo {author} {\bibfnamefont {A.}~\bibnamefont
  {Cavagna}}\ and\ \bibinfo {author} {\bibfnamefont {I.}~\bibnamefont
  {Giardina}},\ }\href@noop {} {\bibfield  {journal} {\bibinfo  {journal}
  {Annu. Rev. Condens. Matter Phys.}\ }\textbf {\bibinfo {volume} {5}},\
  \bibinfo {pages} {183} (\bibinfo {year} {2014})}\BibitemShut {NoStop}%
\bibitem [{\citenamefont {Dauchot}(2026)}]{dauchot2026active}%
  \BibitemOpen
  \bibfield  {author} {\bibinfo {author} {\bibfnamefont {O.}~\bibnamefont
  {Dauchot}},\ }\href@noop {} {\bibfield  {journal} {\bibinfo  {journal}
  {Physical Review E}\ }\textbf {\bibinfo {volume} {114}},\ \bibinfo {pages}
  {011001} (\bibinfo {year} {2026})}\BibitemShut {NoStop}%
\bibitem [{\citenamefont {Petridou}\ and\ \citenamefont
  {Heisenberg}(2019)}]{petridou2019tissue}%
  \BibitemOpen
  \bibfield  {author} {\bibinfo {author} {\bibfnamefont {N.~I.}\ \bibnamefont
  {Petridou}}\ and\ \bibinfo {author} {\bibfnamefont {C.-P.}\ \bibnamefont
  {Heisenberg}},\ }\href@noop {} {\bibfield  {journal} {\bibinfo  {journal}
  {The EMBO journal}\ }\textbf {\bibinfo {volume} {38}},\ \bibinfo {pages}
  {EMBJ2019102497} (\bibinfo {year} {2019})}\BibitemShut {NoStop}%
\bibitem [{\citenamefont {Wei}\ and\ \citenamefont
  {Kraft}(2026)}]{wei2026life}%
  \BibitemOpen
  \bibfield  {author} {\bibinfo {author} {\bibfnamefont {M.}~\bibnamefont
  {Wei}}\ and\ \bibinfo {author} {\bibfnamefont {D.~J.}\ \bibnamefont
  {Kraft}},\ }\href@noop {} {\bibfield  {journal} {\bibinfo  {journal}
  {Proceedings of the National Academy of Sciences}\ }\textbf {\bibinfo
  {volume} {123}},\ \bibinfo {pages} {e2531743123} (\bibinfo {year}
  {2026})}\BibitemShut {NoStop}%
\bibitem [{\citenamefont {Xi}\ \emph {et~al.}(2024)\citenamefont {Xi},
  \citenamefont {Marzin}, \citenamefont {Huang}, \citenamefont {Jones},\ and\
  \citenamefont {Brun}}]{xi2024emergent}%
  \BibitemOpen
  \bibfield  {author} {\bibinfo {author} {\bibfnamefont {Y.}~\bibnamefont
  {Xi}}, \bibinfo {author} {\bibfnamefont {T.}~\bibnamefont {Marzin}}, \bibinfo
  {author} {\bibfnamefont {R.~B.}\ \bibnamefont {Huang}}, \bibinfo {author}
  {\bibfnamefont {T.~J.}\ \bibnamefont {Jones}},\ and\ \bibinfo {author}
  {\bibfnamefont {P.-T.}\ \bibnamefont {Brun}},\ }\href@noop {} {\bibfield
  {journal} {\bibinfo  {journal} {Proceedings of the National Academy of
  Sciences}\ }\textbf {\bibinfo {volume} {121}},\ \bibinfo {pages}
  {e2410654121} (\bibinfo {year} {2024})}\BibitemShut {NoStop}%
\bibitem [{\citenamefont {Janssen}(2019)}]{janssen2019active}%
  \BibitemOpen
  \bibfield  {author} {\bibinfo {author} {\bibfnamefont {L.~M.}\ \bibnamefont
  {Janssen}},\ }\href@noop {} {\bibfield  {journal} {\bibinfo  {journal}
  {Journal of Physics: Condensed Matter}\ }\textbf {\bibinfo {volume} {31}},\
  \bibinfo {pages} {503002} (\bibinfo {year} {2019})}\BibitemShut {NoStop}%
\bibitem [{\citenamefont {Henkes}\ \emph {et~al.}(2020)\citenamefont {Henkes},
  \citenamefont {Kostanjevec}, \citenamefont {Collinson}, \citenamefont
  {Sknepnek},\ and\ \citenamefont {Bertin}}]{henkesDenseActiveMatter2020}%
  \BibitemOpen
  \bibfield  {author} {\bibinfo {author} {\bibfnamefont {S.}~\bibnamefont
  {Henkes}}, \bibinfo {author} {\bibfnamefont {K.}~\bibnamefont {Kostanjevec}},
  \bibinfo {author} {\bibfnamefont {J.~M.}\ \bibnamefont {Collinson}}, \bibinfo
  {author} {\bibfnamefont {R.}~\bibnamefont {Sknepnek}},\ and\ \bibinfo
  {author} {\bibfnamefont {E.}~\bibnamefont {Bertin}},\ }\href
  {https://doi.org/10.1038/s41467-020-15164-5} {\bibfield  {journal} {\bibinfo
  {journal} {Nature Communications}\ }\textbf {\bibinfo {volume} {11}},\
  \bibinfo {pages} {1405} (\bibinfo {year} {2020})},\ \bibinfo {note} {number:
  1}\BibitemShut {NoStop}%
\bibitem [{\citenamefont {Caprini}\ \emph {et~al.}(2020)\citenamefont
  {Caprini}, \citenamefont {Marini Bettolo~Marconi},\ and\ \citenamefont
  {Puglisi}}]{caprini2020spontaneous}%
  \BibitemOpen
  \bibfield  {author} {\bibinfo {author} {\bibfnamefont {L.}~\bibnamefont
  {Caprini}}, \bibinfo {author} {\bibfnamefont {U.}~\bibnamefont {Marini
  Bettolo~Marconi}},\ and\ \bibinfo {author} {\bibfnamefont {A.}~\bibnamefont
  {Puglisi}},\ }\href@noop {} {\bibfield  {journal} {\bibinfo  {journal}
  {Physical review letters}\ }\textbf {\bibinfo {volume} {124}},\ \bibinfo
  {pages} {078001} (\bibinfo {year} {2020})}\BibitemShut {NoStop}%
\bibitem [{\citenamefont {Keta}\ \emph {et~al.}(2024)\citenamefont {Keta},
  \citenamefont {Klamser}, \citenamefont {Jack},\ and\ \citenamefont
  {Berthier}}]{keta2024emerging}%
  \BibitemOpen
  \bibfield  {author} {\bibinfo {author} {\bibfnamefont {Y.-E.}\ \bibnamefont
  {Keta}}, \bibinfo {author} {\bibfnamefont {J.~U.}\ \bibnamefont {Klamser}},
  \bibinfo {author} {\bibfnamefont {R.~L.}\ \bibnamefont {Jack}},\ and\
  \bibinfo {author} {\bibfnamefont {L.}~\bibnamefont {Berthier}},\ }\href@noop
  {} {\bibfield  {journal} {\bibinfo  {journal} {Physical Review Letters}\
  }\textbf {\bibinfo {volume} {132}},\ \bibinfo {pages} {218301} (\bibinfo
  {year} {2024})}\BibitemShut {NoStop}%
\bibitem [{\citenamefont {Berthier}(2014)}]{berthier2014nonequilibrium}%
  \BibitemOpen
  \bibfield  {author} {\bibinfo {author} {\bibfnamefont {L.}~\bibnamefont
  {Berthier}},\ }\href@noop {} {\bibfield  {journal} {\bibinfo  {journal}
  {Physical review letters}\ }\textbf {\bibinfo {volume} {112}},\ \bibinfo
  {pages} {220602} (\bibinfo {year} {2014})}\BibitemShut {NoStop}%
\bibitem [{\citenamefont {Nandi}\ \emph {et~al.}(2018)\citenamefont {Nandi},
  \citenamefont {Mandal}, \citenamefont {Bhuyan}, \citenamefont {Dasgupta},
  \citenamefont {Rao},\ and\ \citenamefont {Gov}}]{nandi2018random}%
  \BibitemOpen
  \bibfield  {author} {\bibinfo {author} {\bibfnamefont {S.~K.}\ \bibnamefont
  {Nandi}}, \bibinfo {author} {\bibfnamefont {R.}~\bibnamefont {Mandal}},
  \bibinfo {author} {\bibfnamefont {P.~J.}\ \bibnamefont {Bhuyan}}, \bibinfo
  {author} {\bibfnamefont {C.}~\bibnamefont {Dasgupta}}, \bibinfo {author}
  {\bibfnamefont {M.}~\bibnamefont {Rao}},\ and\ \bibinfo {author}
  {\bibfnamefont {N.~S.}\ \bibnamefont {Gov}},\ }\href@noop {} {\bibfield
  {journal} {\bibinfo  {journal} {Proceedings of the National Academy of
  Sciences}\ }\textbf {\bibinfo {volume} {115}},\ \bibinfo {pages} {7688}
  (\bibinfo {year} {2018})}\BibitemShut {NoStop}%
\bibitem [{\citenamefont {Mandal}\ \emph {et~al.}(2020)\citenamefont {Mandal},
  \citenamefont {Bhuyan}, \citenamefont {Chaudhuri}, \citenamefont {Dasgupta},\
  and\ \citenamefont {Rao}}]{mandal2020extreme}%
  \BibitemOpen
  \bibfield  {author} {\bibinfo {author} {\bibfnamefont {R.}~\bibnamefont
  {Mandal}}, \bibinfo {author} {\bibfnamefont {P.~J.}\ \bibnamefont {Bhuyan}},
  \bibinfo {author} {\bibfnamefont {P.}~\bibnamefont {Chaudhuri}}, \bibinfo
  {author} {\bibfnamefont {C.}~\bibnamefont {Dasgupta}},\ and\ \bibinfo
  {author} {\bibfnamefont {M.}~\bibnamefont {Rao}},\ }\href@noop {} {\bibfield
  {journal} {\bibinfo  {journal} {Nature communications}\ }\textbf {\bibinfo
  {volume} {11}},\ \bibinfo {pages} {2581} (\bibinfo {year}
  {2020})}\BibitemShut {NoStop}%
\bibitem [{\citenamefont {Morse}\ \emph {et~al.}(2021)\citenamefont {Morse},
  \citenamefont {Roy}, \citenamefont {Agoritsas}, \citenamefont {Stanifer},
  \citenamefont {Corwin},\ and\ \citenamefont {Manning}}]{morse2021direct}%
  \BibitemOpen
  \bibfield  {author} {\bibinfo {author} {\bibfnamefont {P.~K.}\ \bibnamefont
  {Morse}}, \bibinfo {author} {\bibfnamefont {S.}~\bibnamefont {Roy}}, \bibinfo
  {author} {\bibfnamefont {E.}~\bibnamefont {Agoritsas}}, \bibinfo {author}
  {\bibfnamefont {E.}~\bibnamefont {Stanifer}}, \bibinfo {author}
  {\bibfnamefont {E.~I.}\ \bibnamefont {Corwin}},\ and\ \bibinfo {author}
  {\bibfnamefont {M.~L.}\ \bibnamefont {Manning}},\ }\href@noop {} {\bibfield
  {journal} {\bibinfo  {journal} {Proceedings of the National Academy of
  Sciences}\ }\textbf {\bibinfo {volume} {118}},\ \bibinfo {pages}
  {e2019909118} (\bibinfo {year} {2021})}\BibitemShut {NoStop}%
\bibitem [{\citenamefont {Keta}\ \emph {et~al.}(2023)\citenamefont {Keta},
  \citenamefont {Mandal}, \citenamefont {Sollich}, \citenamefont {Jack},\ and\
  \citenamefont {Berthier}}]{keta2023intermittent}%
  \BibitemOpen
  \bibfield  {author} {\bibinfo {author} {\bibfnamefont {Y.-E.}\ \bibnamefont
  {Keta}}, \bibinfo {author} {\bibfnamefont {R.}~\bibnamefont {Mandal}},
  \bibinfo {author} {\bibfnamefont {P.}~\bibnamefont {Sollich}}, \bibinfo
  {author} {\bibfnamefont {R.~L.}\ \bibnamefont {Jack}},\ and\ \bibinfo
  {author} {\bibfnamefont {L.}~\bibnamefont {Berthier}},\ }\href@noop {}
  {\bibfield  {journal} {\bibinfo  {journal} {Soft Matter}\ }\textbf {\bibinfo
  {volume} {19}},\ \bibinfo {pages} {3871} (\bibinfo {year}
  {2023})}\BibitemShut {NoStop}%
\bibitem [{\citenamefont {Doostmohammadi}\ \emph {et~al.}(2018)\citenamefont
  {Doostmohammadi}, \citenamefont {Ign{\'e}s-Mullol}, \citenamefont {Yeomans},\
  and\ \citenamefont {Sagu{\'e}s}}]{doostmohammadi2018active}%
  \BibitemOpen
  \bibfield  {author} {\bibinfo {author} {\bibfnamefont {A.}~\bibnamefont
  {Doostmohammadi}}, \bibinfo {author} {\bibfnamefont {J.}~\bibnamefont
  {Ign{\'e}s-Mullol}}, \bibinfo {author} {\bibfnamefont {J.~M.}\ \bibnamefont
  {Yeomans}},\ and\ \bibinfo {author} {\bibfnamefont {F.}~\bibnamefont
  {Sagu{\'e}s}},\ }\href@noop {} {\bibfield  {journal} {\bibinfo  {journal}
  {Nature communications}\ }\textbf {\bibinfo {volume} {9}},\ \bibinfo {pages}
  {3246} (\bibinfo {year} {2018})}\BibitemShut {NoStop}%
\bibitem [{\citenamefont {Br{\"u}ckner}\ \emph {et~al.}(2022)\citenamefont
  {Br{\"u}ckner}, \citenamefont {Schmitt}, \citenamefont {Fink}, \citenamefont
  {Ladurner}, \citenamefont {Flommersfeld}, \citenamefont {Arlt}, \citenamefont
  {Hannezo}, \citenamefont {R{\"a}dler},\ and\ \citenamefont
  {Broedersz}}]{bruckner2022geometry}%
  \BibitemOpen
  \bibfield  {author} {\bibinfo {author} {\bibfnamefont {D.~B.}\ \bibnamefont
  {Br{\"u}ckner}}, \bibinfo {author} {\bibfnamefont {M.}~\bibnamefont
  {Schmitt}}, \bibinfo {author} {\bibfnamefont {A.}~\bibnamefont {Fink}},
  \bibinfo {author} {\bibfnamefont {G.}~\bibnamefont {Ladurner}}, \bibinfo
  {author} {\bibfnamefont {J.}~\bibnamefont {Flommersfeld}}, \bibinfo {author}
  {\bibfnamefont {N.}~\bibnamefont {Arlt}}, \bibinfo {author} {\bibfnamefont
  {E.}~\bibnamefont {Hannezo}}, \bibinfo {author} {\bibfnamefont {J.~O.}\
  \bibnamefont {R{\"a}dler}},\ and\ \bibinfo {author} {\bibfnamefont {C.~P.}\
  \bibnamefont {Broedersz}},\ }\href@noop {} {\bibfield  {journal} {\bibinfo
  {journal} {Physical Review X}\ }\textbf {\bibinfo {volume} {12}},\ \bibinfo
  {pages} {031041} (\bibinfo {year} {2022})}\BibitemShut {NoStop}%
\bibitem [{\citenamefont {Vagne}\ and\ \citenamefont
  {Salbreux}(2025)}]{vagne2025generic}%
  \BibitemOpen
  \bibfield  {author} {\bibinfo {author} {\bibfnamefont {Q.}~\bibnamefont
  {Vagne}}\ and\ \bibinfo {author} {\bibfnamefont {G.}~\bibnamefont
  {Salbreux}},\ }\href@noop {} {\bibfield  {journal} {\bibinfo  {journal}
  {Physical Review E}\ }\textbf {\bibinfo {volume} {111}},\ \bibinfo {pages}
  {014423} (\bibinfo {year} {2025})}\BibitemShut {NoStop}%
\bibitem [{\citenamefont {Fruchart}\ \emph {et~al.}(2023)\citenamefont
  {Fruchart}, \citenamefont {Scheibner},\ and\ \citenamefont
  {Vitelli}}]{fruchart2023odd}%
  \BibitemOpen
  \bibfield  {author} {\bibinfo {author} {\bibfnamefont {M.}~\bibnamefont
  {Fruchart}}, \bibinfo {author} {\bibfnamefont {C.}~\bibnamefont
  {Scheibner}},\ and\ \bibinfo {author} {\bibfnamefont {V.}~\bibnamefont
  {Vitelli}},\ }\href@noop {} {\bibfield  {journal} {\bibinfo  {journal}
  {Annual Review of Condensed Matter Physics}\ }\textbf {\bibinfo {volume}
  {14}},\ \bibinfo {pages} {471} (\bibinfo {year} {2023})}\BibitemShut
  {NoStop}%
\bibitem [{\citenamefont {Saarloos}\ \emph {et~al.}(2024)\citenamefont
  {Saarloos}, \citenamefont {Vitelli},\ and\ \citenamefont
  {Zeravcic}}]{saarloos2024soft}%
  \BibitemOpen
  \bibfield  {author} {\bibinfo {author} {\bibfnamefont {W.~v.}\ \bibnamefont
  {Saarloos}}, \bibinfo {author} {\bibfnamefont {V.}~\bibnamefont {Vitelli}},\
  and\ \bibinfo {author} {\bibfnamefont {Z.}~\bibnamefont {Zeravcic}},\
  }\href@noop {} {\emph {\bibinfo {title} {Soft {Matter}: {Concepts},
  {Phenomena}, and {Applications}}}}\ (\bibinfo  {publisher} {Princeton
  University Press},\ \bibinfo {year} {2024})\ \bibinfo {note}
  {google-Books-ID: K0XLEAAAQBAJ}\BibitemShut {NoStop}%
\bibitem [{\citenamefont {Scheibner}\ \emph {et~al.}(2020)\citenamefont
  {Scheibner}, \citenamefont {Souslov}, \citenamefont {Banerjee}, \citenamefont
  {Sur{\'o}wka}, \citenamefont {Irvine},\ and\ \citenamefont
  {Vitelli}}]{scheibner2020odd}%
  \BibitemOpen
  \bibfield  {author} {\bibinfo {author} {\bibfnamefont {C.}~\bibnamefont
  {Scheibner}}, \bibinfo {author} {\bibfnamefont {A.}~\bibnamefont {Souslov}},
  \bibinfo {author} {\bibfnamefont {D.}~\bibnamefont {Banerjee}}, \bibinfo
  {author} {\bibfnamefont {P.}~\bibnamefont {Sur{\'o}wka}}, \bibinfo {author}
  {\bibfnamefont {W.~T.}\ \bibnamefont {Irvine}},\ and\ \bibinfo {author}
  {\bibfnamefont {V.}~\bibnamefont {Vitelli}},\ }\href@noop {} {\bibfield
  {journal} {\bibinfo  {journal} {Nature Physics}\ }\textbf {\bibinfo {volume}
  {16}},\ \bibinfo {pages} {475} (\bibinfo {year} {2020})}\BibitemShut
  {NoStop}%
\bibitem [{\citenamefont {Huang}\ \emph {et~al.}(2023)\citenamefont {Huang},
  \citenamefont {Mandal}, \citenamefont {Scheibner},\ and\ \citenamefont
  {Vitelli}}]{huang2023odd}%
  \BibitemOpen
  \bibfield  {author} {\bibinfo {author} {\bibfnamefont {R.}~\bibnamefont
  {Huang}}, \bibinfo {author} {\bibfnamefont {R.}~\bibnamefont {Mandal}},
  \bibinfo {author} {\bibfnamefont {C.}~\bibnamefont {Scheibner}},\ and\
  \bibinfo {author} {\bibfnamefont {V.}~\bibnamefont {Vitelli}},\ }\href@noop
  {} {\bibfield  {journal} {\bibinfo  {journal} {arXiv preprint
  arXiv:2311.18720}\ } (\bibinfo {year} {2023})}\BibitemShut {NoStop}%
\bibitem [{\citenamefont {Tiwari}\ \emph {et~al.}(2026)\citenamefont {Tiwari},
  \citenamefont {Arora}, \citenamefont {Sood}, \citenamefont {Ramaswamy},
  \citenamefont {Mandal},\ and\ \citenamefont
  {Ganapathy}}]{tiwari2026reentrant}%
  \BibitemOpen
  \bibfield  {author} {\bibinfo {author} {\bibfnamefont {U.}~\bibnamefont
  {Tiwari}}, \bibinfo {author} {\bibfnamefont {P.}~\bibnamefont {Arora}},
  \bibinfo {author} {\bibfnamefont {A.}~\bibnamefont {Sood}}, \bibinfo {author}
  {\bibfnamefont {S.}~\bibnamefont {Ramaswamy}}, \bibinfo {author}
  {\bibfnamefont {R.}~\bibnamefont {Mandal}},\ and\ \bibinfo {author}
  {\bibfnamefont {R.}~\bibnamefont {Ganapathy}},\ }\href@noop {} {\bibfield
  {journal} {\bibinfo  {journal} {Nature Communications}\ }\textbf {\bibinfo
  {volume} {17}},\ \bibinfo {pages} {1802} (\bibinfo {year}
  {2026})}\BibitemShut {NoStop}%
\bibitem [{\citenamefont {Du}\ \emph {et~al.}(2026)\citenamefont {Du},
  \citenamefont {van Mastrigt}, \citenamefont {Veenstra},\ and\ \citenamefont
  {Coulais}}]{du2026metamaterials}%
  \BibitemOpen
  \bibfield  {author} {\bibinfo {author} {\bibfnamefont {Y.}~\bibnamefont
  {Du}}, \bibinfo {author} {\bibfnamefont {R.}~\bibnamefont {van Mastrigt}},
  \bibinfo {author} {\bibfnamefont {J.}~\bibnamefont {Veenstra}},\ and\
  \bibinfo {author} {\bibfnamefont {C.}~\bibnamefont {Coulais}},\ }\href@noop
  {} {\bibfield  {journal} {\bibinfo  {journal} {Nature Physics}\ ,\ \bibinfo
  {pages} {1}} (\bibinfo {year} {2026})}\BibitemShut {NoStop}%
\bibitem [{\citenamefont {Fruchart}\ \emph {et~al.}(2021)\citenamefont
  {Fruchart}, \citenamefont {Hanai}, \citenamefont {Littlewood},\ and\
  \citenamefont {Vitelli}}]{fruchart2021non}%
  \BibitemOpen
  \bibfield  {author} {\bibinfo {author} {\bibfnamefont {M.}~\bibnamefont
  {Fruchart}}, \bibinfo {author} {\bibfnamefont {R.}~\bibnamefont {Hanai}},
  \bibinfo {author} {\bibfnamefont {P.~B.}\ \bibnamefont {Littlewood}},\ and\
  \bibinfo {author} {\bibfnamefont {V.}~\bibnamefont {Vitelli}},\ }\href@noop
  {} {\bibfield  {journal} {\bibinfo  {journal} {Nature}\ }\textbf {\bibinfo
  {volume} {592}},\ \bibinfo {pages} {363} (\bibinfo {year}
  {2021})}\BibitemShut {NoStop}%
\bibitem [{\citenamefont {Tan}\ \emph {et~al.}(2022)\citenamefont {Tan},
  \citenamefont {Mietke}, \citenamefont {Li}, \citenamefont {Chen},
  \citenamefont {Higinbotham}, \citenamefont {Foster}, \citenamefont {Gokhale},
  \citenamefont {Dunkel},\ and\ \citenamefont {Fakhri}}]{tan2022odd}%
  \BibitemOpen
  \bibfield  {author} {\bibinfo {author} {\bibfnamefont {T.~H.}\ \bibnamefont
  {Tan}}, \bibinfo {author} {\bibfnamefont {A.}~\bibnamefont {Mietke}},
  \bibinfo {author} {\bibfnamefont {J.}~\bibnamefont {Li}}, \bibinfo {author}
  {\bibfnamefont {Y.}~\bibnamefont {Chen}}, \bibinfo {author} {\bibfnamefont
  {H.}~\bibnamefont {Higinbotham}}, \bibinfo {author} {\bibfnamefont {P.~J.}\
  \bibnamefont {Foster}}, \bibinfo {author} {\bibfnamefont {S.}~\bibnamefont
  {Gokhale}}, \bibinfo {author} {\bibfnamefont {J.}~\bibnamefont {Dunkel}},\
  and\ \bibinfo {author} {\bibfnamefont {N.}~\bibnamefont {Fakhri}},\
  }\href@noop {} {\bibfield  {journal} {\bibinfo  {journal} {Nature}\ }\textbf
  {\bibinfo {volume} {607}},\ \bibinfo {pages} {287} (\bibinfo {year}
  {2022})}\BibitemShut {NoStop}%
\bibitem [{\citenamefont {Loos}\ \emph {et~al.}(2023)\citenamefont {Loos},
  \citenamefont {Klapp},\ and\ \citenamefont {Martynec}}]{loos2023long}%
  \BibitemOpen
  \bibfield  {author} {\bibinfo {author} {\bibfnamefont {S.~A.}\ \bibnamefont
  {Loos}}, \bibinfo {author} {\bibfnamefont {S.~H.}\ \bibnamefont {Klapp}},\
  and\ \bibinfo {author} {\bibfnamefont {T.}~\bibnamefont {Martynec}},\
  }\href@noop {} {\bibfield  {journal} {\bibinfo  {journal} {Physical review
  letters}\ }\textbf {\bibinfo {volume} {130}},\ \bibinfo {pages} {198301}
  (\bibinfo {year} {2023})}\BibitemShut {NoStop}%
\bibitem [{\citenamefont {Veenstra}\ \emph {et~al.}(2025)\citenamefont
  {Veenstra}, \citenamefont {Binysh}, \citenamefont {Seinen}, \citenamefont
  {Naber}, \citenamefont {Robledo-Poisson}, \citenamefont {Hunt}, \citenamefont
  {van Saarloos}, \citenamefont {Souslov},\ and\ \citenamefont
  {Coulais}}]{veenstra2025wave}%
  \BibitemOpen
  \bibfield  {author} {\bibinfo {author} {\bibfnamefont {J.}~\bibnamefont
  {Veenstra}}, \bibinfo {author} {\bibfnamefont {J.}~\bibnamefont {Binysh}},
  \bibinfo {author} {\bibfnamefont {V.}~\bibnamefont {Seinen}}, \bibinfo
  {author} {\bibfnamefont {R.}~\bibnamefont {Naber}}, \bibinfo {author}
  {\bibfnamefont {D.}~\bibnamefont {Robledo-Poisson}}, \bibinfo {author}
  {\bibfnamefont {A.}~\bibnamefont {Hunt}}, \bibinfo {author} {\bibfnamefont
  {W.}~\bibnamefont {van Saarloos}}, \bibinfo {author} {\bibfnamefont
  {A.}~\bibnamefont {Souslov}},\ and\ \bibinfo {author} {\bibfnamefont
  {C.}~\bibnamefont {Coulais}},\ }\href@noop {} {\bibfield  {journal} {\bibinfo
   {journal} {arXiv preprint arXiv:2508.20052}\ } (\bibinfo {year}
  {2025})}\BibitemShut {NoStop}%
\bibitem [{\citenamefont {Baconnier}\ \emph
  {et~al.}(2025{\natexlab{a}})\citenamefont {Baconnier}, \citenamefont
  {Dauchot}, \citenamefont {D{\'e}mery}, \citenamefont {D{\"u}ring},
  \citenamefont {Henkes}, \citenamefont {Huepe},\ and\ \citenamefont
  {Shee}}]{baconnier2025self}%
  \BibitemOpen
  \bibfield  {author} {\bibinfo {author} {\bibfnamefont {P.}~\bibnamefont
  {Baconnier}}, \bibinfo {author} {\bibfnamefont {O.}~\bibnamefont {Dauchot}},
  \bibinfo {author} {\bibfnamefont {V.}~\bibnamefont {D{\'e}mery}}, \bibinfo
  {author} {\bibfnamefont {G.}~\bibnamefont {D{\"u}ring}}, \bibinfo {author}
  {\bibfnamefont {S.}~\bibnamefont {Henkes}}, \bibinfo {author} {\bibfnamefont
  {C.}~\bibnamefont {Huepe}},\ and\ \bibinfo {author} {\bibfnamefont
  {A.}~\bibnamefont {Shee}},\ }\href@noop {} {\bibfield  {journal} {\bibinfo
  {journal} {Reviews of Modern Physics}\ }\textbf {\bibinfo {volume} {97}},\
  \bibinfo {pages} {015007} (\bibinfo {year} {2025}{\natexlab{a}})}\BibitemShut
  {NoStop}%
\bibitem [{\citenamefont {Helbing}\ \emph {et~al.}(2007)\citenamefont
  {Helbing}, \citenamefont {Johansson},\ and\ \citenamefont
  {Al-Abideen}}]{helbing2007dynamics}%
  \BibitemOpen
  \bibfield  {author} {\bibinfo {author} {\bibfnamefont {D.}~\bibnamefont
  {Helbing}}, \bibinfo {author} {\bibfnamefont {A.}~\bibnamefont {Johansson}},\
  and\ \bibinfo {author} {\bibfnamefont {H.~Z.}\ \bibnamefont {Al-Abideen}},\
  }\href@noop {} {\bibfield  {journal} {\bibinfo  {journal} {Physical Review
  E—Statistical, Nonlinear, and Soft Matter Physics}\ }\textbf {\bibinfo
  {volume} {75}},\ \bibinfo {pages} {046109} (\bibinfo {year}
  {2007})}\BibitemShut {NoStop}%
\bibitem [{\citenamefont {Gu}\ \emph {et~al.}(2025)\citenamefont {Gu},
  \citenamefont {Guiselin}, \citenamefont {Bain}, \citenamefont {Zuriguel},\
  and\ \citenamefont {Bartolo}}]{gu2025emergence}%
  \BibitemOpen
  \bibfield  {author} {\bibinfo {author} {\bibfnamefont {F.}~\bibnamefont
  {Gu}}, \bibinfo {author} {\bibfnamefont {B.}~\bibnamefont {Guiselin}},
  \bibinfo {author} {\bibfnamefont {N.}~\bibnamefont {Bain}}, \bibinfo {author}
  {\bibfnamefont {I.}~\bibnamefont {Zuriguel}},\ and\ \bibinfo {author}
  {\bibfnamefont {D.}~\bibnamefont {Bartolo}},\ }\href
  {https://doi.org/10.1038/s41586-024-08514-6} {\bibfield  {journal} {\bibinfo
  {journal} {Nature}\ }\textbf {\bibinfo {volume} {638}},\ \bibinfo {pages}
  {112} (\bibinfo {year} {2025})}\BibitemShut {NoStop}%
\bibitem [{\citenamefont {Peyret}\ \emph {et~al.}(2019)\citenamefont {Peyret},
  \citenamefont {Mueller}, \citenamefont {{d'Alessandro}}, \citenamefont
  {Begnaud}, \citenamefont {Marcq}, \citenamefont {M{\`e}ge}, \citenamefont
  {Yeomans}, \citenamefont {Doostmohammadi},\ and\ \citenamefont
  {Ladoux}}]{peyretSustainedOscillationsEpithelial2019}%
  \BibitemOpen
  \bibfield  {author} {\bibinfo {author} {\bibfnamefont {G.}~\bibnamefont
  {Peyret}}, \bibinfo {author} {\bibfnamefont {R.}~\bibnamefont {Mueller}},
  \bibinfo {author} {\bibfnamefont {J.}~\bibnamefont {{d'Alessandro}}},
  \bibinfo {author} {\bibfnamefont {S.}~\bibnamefont {Begnaud}}, \bibinfo
  {author} {\bibfnamefont {P.}~\bibnamefont {Marcq}}, \bibinfo {author}
  {\bibfnamefont {R.-M.}\ \bibnamefont {M{\`e}ge}}, \bibinfo {author}
  {\bibfnamefont {J.~M.}\ \bibnamefont {Yeomans}}, \bibinfo {author}
  {\bibfnamefont {A.}~\bibnamefont {Doostmohammadi}},\ and\ \bibinfo {author}
  {\bibfnamefont {B.}~\bibnamefont {Ladoux}},\ }\href
  {https://doi.org/10.1016/j.bpj.2019.06.013} {\bibfield  {journal} {\bibinfo
  {journal} {Biophysical Journal}\ }\textbf {\bibinfo {volume} {117}},\
  \bibinfo {pages} {464} (\bibinfo {year} {2019})}\BibitemShut {NoStop}%
\bibitem [{\citenamefont {Petrolli}\ \emph {et~al.}(2019)\citenamefont
  {Petrolli}, \citenamefont {Le~Goff}, \citenamefont {Tadrous}, \citenamefont
  {Martens}, \citenamefont {Allier}, \citenamefont {Mandula}, \citenamefont
  {Herv{\'e}}, \citenamefont {Henkes}, \citenamefont {Sknepnek}, \citenamefont
  {Boudou} \emph {et~al.}}]{petrolli2019confinement}%
  \BibitemOpen
  \bibfield  {author} {\bibinfo {author} {\bibfnamefont {V.}~\bibnamefont
  {Petrolli}}, \bibinfo {author} {\bibfnamefont {M.}~\bibnamefont {Le~Goff}},
  \bibinfo {author} {\bibfnamefont {M.}~\bibnamefont {Tadrous}}, \bibinfo
  {author} {\bibfnamefont {K.}~\bibnamefont {Martens}}, \bibinfo {author}
  {\bibfnamefont {C.}~\bibnamefont {Allier}}, \bibinfo {author} {\bibfnamefont
  {O.}~\bibnamefont {Mandula}}, \bibinfo {author} {\bibfnamefont
  {L.}~\bibnamefont {Herv{\'e}}}, \bibinfo {author} {\bibfnamefont
  {S.}~\bibnamefont {Henkes}}, \bibinfo {author} {\bibfnamefont
  {R.}~\bibnamefont {Sknepnek}}, \bibinfo {author} {\bibfnamefont
  {T.}~\bibnamefont {Boudou}}, \emph {et~al.},\ }\href@noop {} {\bibfield
  {journal} {\bibinfo  {journal} {Physical review letters}\ }\textbf {\bibinfo
  {volume} {122}},\ \bibinfo {pages} {168101} (\bibinfo {year}
  {2019})}\BibitemShut {NoStop}%
\bibitem [{\citenamefont {Xu}\ \emph {et~al.}(2023)\citenamefont {Xu},
  \citenamefont {Huang}, \citenamefont {Zhang},\ and\ \citenamefont
  {Wu}}]{xu2023autonomous}%
  \BibitemOpen
  \bibfield  {author} {\bibinfo {author} {\bibfnamefont {H.}~\bibnamefont
  {Xu}}, \bibinfo {author} {\bibfnamefont {Y.}~\bibnamefont {Huang}}, \bibinfo
  {author} {\bibfnamefont {R.}~\bibnamefont {Zhang}},\ and\ \bibinfo {author}
  {\bibfnamefont {Y.}~\bibnamefont {Wu}},\ }\href@noop {} {\bibfield  {journal}
  {\bibinfo  {journal} {Nature Physics}\ }\textbf {\bibinfo {volume} {19}},\
  \bibinfo {pages} {46} (\bibinfo {year} {2023})}\BibitemShut {NoStop}%
\bibitem [{\citenamefont {Baconnier}\ \emph {et~al.}(2022)\citenamefont
  {Baconnier}, \citenamefont {Shohat}, \citenamefont {López}, \citenamefont
  {Coulais}, \citenamefont {Démery}, \citenamefont {Düring},\ and\
  \citenamefont {Dauchot}}]{baconnier2022selective}%
  \BibitemOpen
  \bibfield  {author} {\bibinfo {author} {\bibfnamefont {P.}~\bibnamefont
  {Baconnier}}, \bibinfo {author} {\bibfnamefont {D.}~\bibnamefont {Shohat}},
  \bibinfo {author} {\bibfnamefont {C.~H.}\ \bibnamefont {López}}, \bibinfo
  {author} {\bibfnamefont {C.}~\bibnamefont {Coulais}}, \bibinfo {author}
  {\bibfnamefont {V.}~\bibnamefont {Démery}}, \bibinfo {author} {\bibfnamefont
  {G.}~\bibnamefont {Düring}},\ and\ \bibinfo {author} {\bibfnamefont
  {O.}~\bibnamefont {Dauchot}},\ }\href
  {https://doi.org/10.1038/s41567-022-01704-x} {\bibfield  {journal} {\bibinfo
  {journal} {Nature Physics}\ }\textbf {\bibinfo {volume} {18}},\ \bibinfo
  {pages} {1234} (\bibinfo {year} {2022})}\BibitemShut {NoStop}%
\bibitem [{\citenamefont {Baconnier}\ \emph {et~al.}(2023)\citenamefont
  {Baconnier}, \citenamefont {Shohat},\ and\ \citenamefont
  {Dauchot}}]{baconnier2023discontinuous}%
  \BibitemOpen
  \bibfield  {author} {\bibinfo {author} {\bibfnamefont {P.}~\bibnamefont
  {Baconnier}}, \bibinfo {author} {\bibfnamefont {D.}~\bibnamefont {Shohat}},\
  and\ \bibinfo {author} {\bibfnamefont {O.}~\bibnamefont {Dauchot}},\
  }\href@noop {} {\bibfield  {journal} {\bibinfo  {journal} {Physical Review
  Letters}\ }\textbf {\bibinfo {volume} {130}},\ \bibinfo {pages} {028201}
  (\bibinfo {year} {2023})}\BibitemShut {NoStop}%
\bibitem [{\citenamefont {Baconnier}\ \emph {et~al.}(2024)\citenamefont
  {Baconnier}, \citenamefont {D{\'e}mery},\ and\ \citenamefont
  {Dauchot}}]{baconnier2024noise}%
  \BibitemOpen
  \bibfield  {author} {\bibinfo {author} {\bibfnamefont {P.}~\bibnamefont
  {Baconnier}}, \bibinfo {author} {\bibfnamefont {V.}~\bibnamefont
  {D{\'e}mery}},\ and\ \bibinfo {author} {\bibfnamefont {O.}~\bibnamefont
  {Dauchot}},\ }\href@noop {} {\bibfield  {journal} {\bibinfo  {journal}
  {Physical Review E}\ }\textbf {\bibinfo {volume} {109}},\ \bibinfo {pages}
  {024606} (\bibinfo {year} {2024})}\BibitemShut {NoStop}%
\bibitem [{\citenamefont {Baconnier}\ \emph
  {et~al.}(2025{\natexlab{b}})\citenamefont {Baconnier}, \citenamefont
  {D{\'e}mery},\ and\ \citenamefont {Dauchot}}]{baconnier2025collective}%
  \BibitemOpen
  \bibfield  {author} {\bibinfo {author} {\bibfnamefont {P.}~\bibnamefont
  {Baconnier}}, \bibinfo {author} {\bibfnamefont {V.}~\bibnamefont
  {D{\'e}mery}},\ and\ \bibinfo {author} {\bibfnamefont {O.}~\bibnamefont
  {Dauchot}},\ }\href@noop {} {\bibfield  {journal} {\bibinfo  {journal}
  {Physical Review E}\ }\textbf {\bibinfo {volume} {112}},\ \bibinfo {pages}
  {045505} (\bibinfo {year} {2025}{\natexlab{b}})}\BibitemShut {NoStop}%
\bibitem [{\citenamefont {Baconnier}\ \emph
  {et~al.}(2025{\natexlab{c}})\citenamefont {Baconnier}, \citenamefont {Aksil},
  \citenamefont {D{\'e}mery},\ and\ \citenamefont
  {Dauchot}}]{baconnier2025reentrant}%
  \BibitemOpen
  \bibfield  {author} {\bibinfo {author} {\bibfnamefont {P.}~\bibnamefont
  {Baconnier}}, \bibinfo {author} {\bibfnamefont {M.}~\bibnamefont {Aksil}},
  \bibinfo {author} {\bibfnamefont {V.}~\bibnamefont {D{\'e}mery}},\ and\
  \bibinfo {author} {\bibfnamefont {O.}~\bibnamefont {Dauchot}},\ }\href@noop
  {} {\bibfield  {journal} {\bibinfo  {journal} {Physical Review Letters}\
  }\textbf {\bibinfo {volume} {135}},\ \bibinfo {pages} {188302} (\bibinfo
  {year} {2025}{\natexlab{c}})}\BibitemShut {NoStop}%
\bibitem [{\citenamefont {Szab\'o}\ \emph {et~al.}(2006)\citenamefont
  {Szab\'o}, \citenamefont {Sz\"oll\"osi}, \citenamefont {G\"onci},
  \citenamefont {Jur\'anyi}, \citenamefont {Selmeczi},\ and\ \citenamefont
  {Vicsek}}]{szabo2006collective}%
  \BibitemOpen
  \bibfield  {author} {\bibinfo {author} {\bibfnamefont {B.}~\bibnamefont
  {Szab\'o}}, \bibinfo {author} {\bibfnamefont {G.~J.}\ \bibnamefont
  {Sz\"oll\"osi}}, \bibinfo {author} {\bibfnamefont {B.}~\bibnamefont
  {G\"onci}}, \bibinfo {author} {\bibfnamefont {Z.}~\bibnamefont {Jur\'anyi}},
  \bibinfo {author} {\bibfnamefont {D.}~\bibnamefont {Selmeczi}},\ and\
  \bibinfo {author} {\bibfnamefont {T.}~\bibnamefont {Vicsek}},\ }\href
  {https://doi.org/10.1103/PhysRevE.74.061908} {\bibfield  {journal} {\bibinfo
  {journal} {Phys. Rev. E}\ }\textbf {\bibinfo {volume} {74}},\ \bibinfo
  {pages} {061908} (\bibinfo {year} {2006})}\BibitemShut {NoStop}%
\bibitem [{\citenamefont {Henkes}\ \emph {et~al.}(2011)\citenamefont {Henkes},
  \citenamefont {Fily},\ and\ \citenamefont {Marchetti}}]{henkes2011active}%
  \BibitemOpen
  \bibfield  {author} {\bibinfo {author} {\bibfnamefont {S.}~\bibnamefont
  {Henkes}}, \bibinfo {author} {\bibfnamefont {Y.}~\bibnamefont {Fily}},\ and\
  \bibinfo {author} {\bibfnamefont {M.~C.}\ \bibnamefont {Marchetti}},\ }\href
  {https://doi.org/10.1103/PhysRevE.84.040301} {\bibfield  {journal} {\bibinfo
  {journal} {Physical Review E}\ }\textbf {\bibinfo {volume} {84}},\ \bibinfo
  {pages} {040301} (\bibinfo {year} {2011})}\BibitemShut {NoStop}%
\bibitem [{\citenamefont {Malinverno}\ \emph {et~al.}(2017)\citenamefont
  {Malinverno}, \citenamefont {Corallino}, \citenamefont {Giavazzi},
  \citenamefont {Bergert}, \citenamefont {Li}, \citenamefont {Leoni},
  \citenamefont {Disanza}, \citenamefont {Frittoli}, \citenamefont {Oldani},
  \citenamefont {Martini} \emph {et~al.}}]{malinverno2017endocytic}%
  \BibitemOpen
  \bibfield  {author} {\bibinfo {author} {\bibfnamefont {C.}~\bibnamefont
  {Malinverno}}, \bibinfo {author} {\bibfnamefont {S.}~\bibnamefont
  {Corallino}}, \bibinfo {author} {\bibfnamefont {F.}~\bibnamefont {Giavazzi}},
  \bibinfo {author} {\bibfnamefont {M.}~\bibnamefont {Bergert}}, \bibinfo
  {author} {\bibfnamefont {Q.}~\bibnamefont {Li}}, \bibinfo {author}
  {\bibfnamefont {M.}~\bibnamefont {Leoni}}, \bibinfo {author} {\bibfnamefont
  {A.}~\bibnamefont {Disanza}}, \bibinfo {author} {\bibfnamefont
  {E.}~\bibnamefont {Frittoli}}, \bibinfo {author} {\bibfnamefont
  {A.}~\bibnamefont {Oldani}}, \bibinfo {author} {\bibfnamefont
  {E.}~\bibnamefont {Martini}}, \emph {et~al.},\ }\href@noop {} {\bibfield
  {journal} {\bibinfo  {journal} {Nature materials}\ }\textbf {\bibinfo
  {volume} {16}},\ \bibinfo {pages} {587} (\bibinfo {year} {2017})}\BibitemShut
  {NoStop}%
\bibitem [{\citenamefont {Giavazzi}\ \emph {et~al.}(2018)\citenamefont
  {Giavazzi}, \citenamefont {Paoluzzi}, \citenamefont {Macchi}, \citenamefont
  {Bi}, \citenamefont {Scita}, \citenamefont {Manning}, \citenamefont
  {Cerbino},\ and\ \citenamefont {Marchetti}}]{giavazzi2018flocking}%
  \BibitemOpen
  \bibfield  {author} {\bibinfo {author} {\bibfnamefont {F.}~\bibnamefont
  {Giavazzi}}, \bibinfo {author} {\bibfnamefont {M.}~\bibnamefont {Paoluzzi}},
  \bibinfo {author} {\bibfnamefont {M.}~\bibnamefont {Macchi}}, \bibinfo
  {author} {\bibfnamefont {D.}~\bibnamefont {Bi}}, \bibinfo {author}
  {\bibfnamefont {G.}~\bibnamefont {Scita}}, \bibinfo {author} {\bibfnamefont
  {M.~L.}\ \bibnamefont {Manning}}, \bibinfo {author} {\bibfnamefont
  {R.}~\bibnamefont {Cerbino}},\ and\ \bibinfo {author} {\bibfnamefont {M.~C.}\
  \bibnamefont {Marchetti}},\ }\href@noop {} {\bibfield  {journal} {\bibinfo
  {journal} {Soft matter}\ }\textbf {\bibinfo {volume} {14}},\ \bibinfo {pages}
  {3471} (\bibinfo {year} {2018})}\BibitemShut {NoStop}%
\bibitem [{\citenamefont {Barton}\ \emph {et~al.}(2017)\citenamefont {Barton},
  \citenamefont {Henkes}, \citenamefont {Weijer},\ and\ \citenamefont
  {Sknepnek}}]{barton2017active}%
  \BibitemOpen
  \bibfield  {author} {\bibinfo {author} {\bibfnamefont {D.~L.}\ \bibnamefont
  {Barton}}, \bibinfo {author} {\bibfnamefont {S.}~\bibnamefont {Henkes}},
  \bibinfo {author} {\bibfnamefont {C.~J.}\ \bibnamefont {Weijer}},\ and\
  \bibinfo {author} {\bibfnamefont {R.}~\bibnamefont {Sknepnek}},\ }\href@noop
  {} {\bibfield  {journal} {\bibinfo  {journal} {PLoS computational biology}\
  }\textbf {\bibinfo {volume} {13}},\ \bibinfo {pages} {e1005569} (\bibinfo
  {year} {2017})}\BibitemShut {NoStop}%
\bibitem [{\citenamefont {Paoluzzi}\ \emph {et~al.}(2024)\citenamefont
  {Paoluzzi}, \citenamefont {Levis},\ and\ \citenamefont
  {Pagonabarraga}}]{paoluzzi2024flocking}%
  \BibitemOpen
  \bibfield  {author} {\bibinfo {author} {\bibfnamefont {M.}~\bibnamefont
  {Paoluzzi}}, \bibinfo {author} {\bibfnamefont {D.}~\bibnamefont {Levis}},\
  and\ \bibinfo {author} {\bibfnamefont {I.}~\bibnamefont {Pagonabarraga}},\
  }\href@noop {} {\bibfield  {journal} {\bibinfo  {journal} {Communications
  Physics}\ }\textbf {\bibinfo {volume} {7}},\ \bibinfo {pages} {57} (\bibinfo
  {year} {2024})}\BibitemShut {NoStop}%
\bibitem [{\citenamefont {Musacchio}\ \emph {et~al.}(2026)\citenamefont
  {Musacchio}, \citenamefont {Antonov}, \citenamefont {L{\"o}wen},\ and\
  \citenamefont {Caprini}}]{musacchio2026flocking}%
  \BibitemOpen
  \bibfield  {author} {\bibinfo {author} {\bibfnamefont {M.}~\bibnamefont
  {Musacchio}}, \bibinfo {author} {\bibfnamefont {A.~P.}\ \bibnamefont
  {Antonov}}, \bibinfo {author} {\bibfnamefont {H.}~\bibnamefont {L{\"o}wen}},\
  and\ \bibinfo {author} {\bibfnamefont {L.}~\bibnamefont {Caprini}},\
  }\href@noop {} {\bibfield  {journal} {\bibinfo  {journal} {The Journal of
  Chemical Physics}\ }\textbf {\bibinfo {volume} {164}} (\bibinfo {year}
  {2026})}\BibitemShut {NoStop}%
\bibitem [{SI()}]{SI}%
  \BibitemOpen
  \href@noop {} {}\bibinfo {howpublished} {See Supplemental Material at [URL
  will be inserted by publisher]}\BibitemShut {NoStop}%
\bibitem [{\citenamefont {Marchetti}\ \emph {et~al.}(2016)\citenamefont
  {Marchetti}, \citenamefont {Fily}, \citenamefont {Henkes}, \citenamefont
  {Patch},\ and\ \citenamefont {Yllanes}}]{marchetti2016minimal}%
  \BibitemOpen
  \bibfield  {author} {\bibinfo {author} {\bibfnamefont {M.~C.}\ \bibnamefont
  {Marchetti}}, \bibinfo {author} {\bibfnamefont {Y.}~\bibnamefont {Fily}},
  \bibinfo {author} {\bibfnamefont {S.}~\bibnamefont {Henkes}}, \bibinfo
  {author} {\bibfnamefont {A.}~\bibnamefont {Patch}},\ and\ \bibinfo {author}
  {\bibfnamefont {D.}~\bibnamefont {Yllanes}},\ }\href@noop {} {\bibfield
  {journal} {\bibinfo  {journal} {Current Opinion in Colloid \& Interface
  Science}\ }\textbf {\bibinfo {volume} {21}},\ \bibinfo {pages} {34} (\bibinfo
  {year} {2016})}\BibitemShut {NoStop}%
\bibitem [{\citenamefont {Jung}\ and\ \citenamefont
  {H{\"a}nggi}(1987)}]{jung1987dynamical}%
  \BibitemOpen
  \bibfield  {author} {\bibinfo {author} {\bibfnamefont {P.}~\bibnamefont
  {Jung}}\ and\ \bibinfo {author} {\bibfnamefont {P.}~\bibnamefont
  {H{\"a}nggi}},\ }\href@noop {} {\bibfield  {journal} {\bibinfo  {journal}
  {Physical review A}\ }\textbf {\bibinfo {volume} {35}},\ \bibinfo {pages}
  {4464} (\bibinfo {year} {1987})}\BibitemShut {NoStop}%
\bibitem [{\citenamefont {Caprini}\ \emph {et~al.}(2023)\citenamefont
  {Caprini}, \citenamefont {Löwen},\ and\ \citenamefont {Marini
  Bettolo~Marconi}}]{capriniChiralActiveMatter2023}%
  \BibitemOpen
  \bibfield  {author} {\bibinfo {author} {\bibfnamefont {L.}~\bibnamefont
  {Caprini}}, \bibinfo {author} {\bibfnamefont {H.}~\bibnamefont {Löwen}},\
  and\ \bibinfo {author} {\bibfnamefont {U.}~\bibnamefont {Marini
  Bettolo~Marconi}},\ }\href {https://doi.org/10.1039/D3SM00793F} {\bibfield
  {journal} {\bibinfo  {journal} {Soft Matter}\ }\textbf {\bibinfo {volume}
  {19}},\ \bibinfo {pages} {6234} (\bibinfo {year} {2023})}\BibitemShut
  {NoStop}%
\bibitem [{\citenamefont {Caprini}\ and\ \citenamefont {Marini
  Bettolo~Marconi}(2025)}]{caprini2025odd}%
  \BibitemOpen
  \bibfield  {author} {\bibinfo {author} {\bibfnamefont {L.}~\bibnamefont
  {Caprini}}\ and\ \bibinfo {author} {\bibfnamefont {U.}~\bibnamefont {Marini
  Bettolo~Marconi}},\ }\href {https://doi.org/10.1088/1367-2630/add366}
  {\bibfield  {journal} {\bibinfo  {journal} {New Journal of Physics}\ }\textbf
  {\bibinfo {volume} {27}},\ \bibinfo {pages} {054401} (\bibinfo {year}
  {2025})}\BibitemShut {NoStop}%
\bibitem [{\citenamefont {Lacroix}\ \emph {et~al.}(2024)\citenamefont
  {Lacroix}, \citenamefont {Smeets}, \citenamefont {Blanch-Mercader},
  \citenamefont {Bell}, \citenamefont {Giuglaris}, \citenamefont {Chen},
  \citenamefont {Prost},\ and\ \citenamefont
  {Silberzan}}]{lacroix2024emergence}%
  \BibitemOpen
  \bibfield  {author} {\bibinfo {author} {\bibfnamefont {M.}~\bibnamefont
  {Lacroix}}, \bibinfo {author} {\bibfnamefont {B.}~\bibnamefont {Smeets}},
  \bibinfo {author} {\bibfnamefont {C.}~\bibnamefont {Blanch-Mercader}},
  \bibinfo {author} {\bibfnamefont {S.}~\bibnamefont {Bell}}, \bibinfo {author}
  {\bibfnamefont {C.}~\bibnamefont {Giuglaris}}, \bibinfo {author}
  {\bibfnamefont {H.-Y.}\ \bibnamefont {Chen}}, \bibinfo {author}
  {\bibfnamefont {J.}~\bibnamefont {Prost}},\ and\ \bibinfo {author}
  {\bibfnamefont {P.}~\bibnamefont {Silberzan}},\ }\href@noop {} {\bibfield
  {journal} {\bibinfo  {journal} {Nature Physics}\ }\textbf {\bibinfo {volume}
  {20}},\ \bibinfo {pages} {1324} (\bibinfo {year} {2024})}\BibitemShut
  {NoStop}%
\bibitem [{\citenamefont {Dauchot}\ and\ \citenamefont
  {D{\'e}mery}(2019)}]{dauchot2019dynamics}%
  \BibitemOpen
  \bibfield  {author} {\bibinfo {author} {\bibfnamefont {O.}~\bibnamefont
  {Dauchot}}\ and\ \bibinfo {author} {\bibfnamefont {V.}~\bibnamefont
  {D{\'e}mery}},\ }\href@noop {} {\bibfield  {journal} {\bibinfo  {journal}
  {Physical review letters}\ }\textbf {\bibinfo {volume} {122}},\ \bibinfo
  {pages} {068002} (\bibinfo {year} {2019})}\BibitemShut {NoStop}%
\bibitem [{\citenamefont {Chaikin}\ and\ \citenamefont
  {Lubensky}(2000)}]{chaikin2000principles}%
  \BibitemOpen
  \bibfield  {author} {\bibinfo {author} {\bibfnamefont {P.}~\bibnamefont
  {Chaikin}}\ and\ \bibinfo {author} {\bibfnamefont {T.}~\bibnamefont
  {Lubensky}},\ }\href {https://books.google.nl/books?id=P9YjNjzr9OIC} {\emph
  {\bibinfo {title} {Principles of Condensed Matter Physics}}}\ (\bibinfo
  {publisher} {Cambridge University Press},\ \bibinfo {year}
  {2000})\BibitemShut {NoStop}%
\bibitem [{\citenamefont {N{\o}rrelykke}\ and\ \citenamefont
  {Flyvbjerg}(2011)}]{norrelykke2011harmonic}%
  \BibitemOpen
  \bibfield  {author} {\bibinfo {author} {\bibfnamefont {S.~F.}\ \bibnamefont
  {N{\o}rrelykke}}\ and\ \bibinfo {author} {\bibfnamefont {H.}~\bibnamefont
  {Flyvbjerg}},\ }\href@noop {} {\bibfield  {journal} {\bibinfo  {journal}
  {Physical Review E—Statistical, Nonlinear, and Soft Matter Physics}\
  }\textbf {\bibinfo {volume} {83}},\ \bibinfo {pages} {041103} (\bibinfo
  {year} {2011})}\BibitemShut {NoStop}%
\bibitem [{\citenamefont {Silbert}\ \emph {et~al.}(2005)\citenamefont
  {Silbert}, \citenamefont {Liu},\ and\ \citenamefont
  {Nagel}}]{silbert2005vibrations}%
  \BibitemOpen
  \bibfield  {author} {\bibinfo {author} {\bibfnamefont {L.~E.}\ \bibnamefont
  {Silbert}}, \bibinfo {author} {\bibfnamefont {A.~J.}\ \bibnamefont {Liu}},\
  and\ \bibinfo {author} {\bibfnamefont {S.~R.}\ \bibnamefont {Nagel}},\
  }\href@noop {} {\bibfield  {journal} {\bibinfo  {journal} {Physical review
  letters}\ }\textbf {\bibinfo {volume} {95}},\ \bibinfo {pages} {098301}
  (\bibinfo {year} {2005})}\BibitemShut {NoStop}%
\bibitem [{\citenamefont {Henkes}\ \emph {et~al.}(2012)\citenamefont {Henkes},
  \citenamefont {Brito},\ and\ \citenamefont {Dauchot}}]{henkes2012extracting}%
  \BibitemOpen
  \bibfield  {author} {\bibinfo {author} {\bibfnamefont {S.}~\bibnamefont
  {Henkes}}, \bibinfo {author} {\bibfnamefont {C.}~\bibnamefont {Brito}},\ and\
  \bibinfo {author} {\bibfnamefont {O.}~\bibnamefont {Dauchot}},\ }\href@noop
  {} {\bibfield  {journal} {\bibinfo  {journal} {Soft Matter}\ }\textbf
  {\bibinfo {volume} {8}},\ \bibinfo {pages} {6092} (\bibinfo {year}
  {2012})}\BibitemShut {NoStop}%
\bibitem [{\citenamefont {Kammeraat}()}]{Kammeraat_JAMs}%
  \BibitemOpen
  \bibfield  {author} {\bibinfo {author} {\bibfnamefont {S.~C.}\ \bibnamefont
  {Kammeraat}},\ }\href@noop {} {\bibinfo {title} {{JAMs: Julia Active Matter
  simulations}}},\ \bibinfo {howpublished}
  {\url{https://github.com/sanderkammeraat/JAMs/}},\ \bibinfo {note}
  {software}\BibitemShut {NoStop}%
\bibitem [{\citenamefont {O’hern}\ \emph {et~al.}(2003)\citenamefont
  {O’hern}, \citenamefont {Silbert}, \citenamefont {Liu},\ and\ \citenamefont
  {Nagel}}]{o2003jamming}%
  \BibitemOpen
  \bibfield  {author} {\bibinfo {author} {\bibfnamefont {C.~S.}\ \bibnamefont
  {O’hern}}, \bibinfo {author} {\bibfnamefont {L.~E.}\ \bibnamefont
  {Silbert}}, \bibinfo {author} {\bibfnamefont {A.~J.}\ \bibnamefont {Liu}},\
  and\ \bibinfo {author} {\bibfnamefont {S.~R.}\ \bibnamefont {Nagel}},\
  }\href@noop {} {\bibfield  {journal} {\bibinfo  {journal} {Physical Review
  E}\ }\textbf {\bibinfo {volume} {68}},\ \bibinfo {pages} {011306} (\bibinfo
  {year} {2003})}\BibitemShut {NoStop}%
\bibitem [{\citenamefont {Wyart}\ \emph {et~al.}(2005)\citenamefont {Wyart},
  \citenamefont {Silbert}, \citenamefont {Nagel},\ and\ \citenamefont
  {Witten}}]{wyart2005effects}%
  \BibitemOpen
  \bibfield  {author} {\bibinfo {author} {\bibfnamefont {M.}~\bibnamefont
  {Wyart}}, \bibinfo {author} {\bibfnamefont {L.~E.}\ \bibnamefont {Silbert}},
  \bibinfo {author} {\bibfnamefont {S.~R.}\ \bibnamefont {Nagel}},\ and\
  \bibinfo {author} {\bibfnamefont {T.~A.}\ \bibnamefont {Witten}},\
  }\href@noop {} {\bibfield  {journal} {\bibinfo  {journal} {Physical Review
  E—Statistical, Nonlinear, and Soft Matter Physics}\ }\textbf {\bibinfo
  {volume} {72}},\ \bibinfo {pages} {051306} (\bibinfo {year}
  {2005})}\BibitemShut {NoStop}%
\bibitem [{\citenamefont {van Hecke}(2010)}]{van2010jamming}%
  \BibitemOpen
  \bibfield  {author} {\bibinfo {author} {\bibfnamefont {M.}~\bibnamefont {van
  Hecke}},\ }\href@noop {} {\bibfield  {journal} {\bibinfo  {journal} {Journal
  of Physics: Condensed Matter}\ }\textbf {\bibinfo {volume} {22}},\ \bibinfo
  {pages} {033101} (\bibinfo {year} {2010})}\BibitemShut {NoStop}%
\bibitem [{\citenamefont {Fily}\ \emph {et~al.}(2014)\citenamefont {Fily},
  \citenamefont {Henkes},\ and\ \citenamefont {Marchetti}}]{fily2014freezing}%
  \BibitemOpen
  \bibfield  {author} {\bibinfo {author} {\bibfnamefont {Y.}~\bibnamefont
  {Fily}}, \bibinfo {author} {\bibfnamefont {S.}~\bibnamefont {Henkes}},\ and\
  \bibinfo {author} {\bibfnamefont {M.~C.}\ \bibnamefont {Marchetti}},\
  }\href@noop {} {\bibfield  {journal} {\bibinfo  {journal} {Soft matter}\
  }\textbf {\bibinfo {volume} {10}},\ \bibinfo {pages} {2132} (\bibinfo {year}
  {2014})}\BibitemShut {NoStop}%
\bibitem [{\citenamefont {Hern{\'a}ndez-L{\'o}pez}\ \emph
  {et~al.}(2024)\citenamefont {Hern{\'a}ndez-L{\'o}pez}, \citenamefont
  {Baconnier}, \citenamefont {Coulais}, \citenamefont {Dauchot},\ and\
  \citenamefont {D{\"u}ring}}]{hernandez2024model}%
  \BibitemOpen
  \bibfield  {author} {\bibinfo {author} {\bibfnamefont {C.}~\bibnamefont
  {Hern{\'a}ndez-L{\'o}pez}}, \bibinfo {author} {\bibfnamefont
  {P.}~\bibnamefont {Baconnier}}, \bibinfo {author} {\bibfnamefont
  {C.}~\bibnamefont {Coulais}}, \bibinfo {author} {\bibfnamefont
  {O.}~\bibnamefont {Dauchot}},\ and\ \bibinfo {author} {\bibfnamefont
  {G.}~\bibnamefont {D{\"u}ring}},\ }\href@noop {} {\bibfield  {journal}
  {\bibinfo  {journal} {Physical Review Letters}\ }\textbf {\bibinfo {volume}
  {132}},\ \bibinfo {pages} {238303} (\bibinfo {year} {2024})}\BibitemShut
  {NoStop}%
\bibitem [{\citenamefont {Melio}\ \emph {et~al.}(2024)\citenamefont {Melio},
  \citenamefont {Henkes},\ and\ \citenamefont {Kraft}}]{melio2024soft}%
  \BibitemOpen
  \bibfield  {author} {\bibinfo {author} {\bibfnamefont {J.}~\bibnamefont
  {Melio}}, \bibinfo {author} {\bibfnamefont {S.~E.}\ \bibnamefont {Henkes}},\
  and\ \bibinfo {author} {\bibfnamefont {D.~J.}\ \bibnamefont {Kraft}},\
  }\href@noop {} {\bibfield  {journal} {\bibinfo  {journal} {Physical Review
  Letters}\ }\textbf {\bibinfo {volume} {132}},\ \bibinfo {pages} {078202}
  (\bibinfo {year} {2024})}\BibitemShut {NoStop}%
\bibitem [{\citenamefont {Toner}\ and\ \citenamefont
  {Tu}(1995)}]{toner1995long}%
  \BibitemOpen
  \bibfield  {author} {\bibinfo {author} {\bibfnamefont {J.}~\bibnamefont
  {Toner}}\ and\ \bibinfo {author} {\bibfnamefont {Y.}~\bibnamefont {Tu}},\
  }\href@noop {} {\bibfield  {journal} {\bibinfo  {journal} {Physical review
  letters}\ }\textbf {\bibinfo {volume} {75}},\ \bibinfo {pages} {4326}
  (\bibinfo {year} {1995})}\BibitemShut {NoStop}%
\bibitem [{\citenamefont {Bialek}\ \emph {et~al.}(2012)\citenamefont {Bialek},
  \citenamefont {Cavagna}, \citenamefont {Giardina}, \citenamefont {Mora},
  \citenamefont {Silvestri}, \citenamefont {Viale},\ and\ \citenamefont
  {Walczak}}]{bialek2012statistical}%
  \BibitemOpen
  \bibfield  {author} {\bibinfo {author} {\bibfnamefont {W.}~\bibnamefont
  {Bialek}}, \bibinfo {author} {\bibfnamefont {A.}~\bibnamefont {Cavagna}},
  \bibinfo {author} {\bibfnamefont {I.}~\bibnamefont {Giardina}}, \bibinfo
  {author} {\bibfnamefont {T.}~\bibnamefont {Mora}}, \bibinfo {author}
  {\bibfnamefont {E.}~\bibnamefont {Silvestri}}, \bibinfo {author}
  {\bibfnamefont {M.}~\bibnamefont {Viale}},\ and\ \bibinfo {author}
  {\bibfnamefont {A.~M.}\ \bibnamefont {Walczak}},\ }\href@noop {} {\bibfield
  {journal} {\bibinfo  {journal} {Proceedings of the National Academy of
  Sciences}\ }\textbf {\bibinfo {volume} {109}},\ \bibinfo {pages} {4786}
  (\bibinfo {year} {2012})}\BibitemShut {NoStop}%
\bibitem [{\citenamefont {Cavagna}\ \emph {et~al.}(2013)\citenamefont
  {Cavagna}, \citenamefont {Giardina},\ and\ \citenamefont
  {Ginelli}}]{cavagna2013boundary}%
  \BibitemOpen
  \bibfield  {author} {\bibinfo {author} {\bibfnamefont {A.}~\bibnamefont
  {Cavagna}}, \bibinfo {author} {\bibfnamefont {I.}~\bibnamefont {Giardina}},\
  and\ \bibinfo {author} {\bibfnamefont {F.}~\bibnamefont {Ginelli}},\
  }\href@noop {} {\bibfield  {journal} {\bibinfo  {journal} {Physical review
  letters}\ }\textbf {\bibinfo {volume} {110}},\ \bibinfo {pages} {168107}
  (\bibinfo {year} {2013})}\BibitemShut {NoStop}%
\bibitem [{\citenamefont {Sato}\ \emph {et~al.}(2012)\citenamefont {Sato},
  \citenamefont {Fehler},\ and\ \citenamefont {Maeda}}]{sato2012seismic}%
  \BibitemOpen
  \bibfield  {author} {\bibinfo {author} {\bibfnamefont {H.}~\bibnamefont
  {Sato}}, \bibinfo {author} {\bibfnamefont {M.~C.}\ \bibnamefont {Fehler}},\
  and\ \bibinfo {author} {\bibfnamefont {T.}~\bibnamefont {Maeda}},\
  }\href@noop {} {\emph {\bibinfo {title} {Seismic wave propagation and
  scattering in the heterogeneous earth}}},\ Vol.\ \bibinfo {volume} {496}\
  (\bibinfo  {publisher} {Springer},\ \bibinfo {year} {2012})\BibitemShut
  {NoStop}%
\bibitem [{\citenamefont {Deforet}\ \emph {et~al.}(2014)\citenamefont
  {Deforet}, \citenamefont {Hakim}, \citenamefont {Yevick}, \citenamefont
  {Duclos},\ and\ \citenamefont {Silberzan}}]{deforet2014emergence}%
  \BibitemOpen
  \bibfield  {author} {\bibinfo {author} {\bibfnamefont {M.}~\bibnamefont
  {Deforet}}, \bibinfo {author} {\bibfnamefont {V.}~\bibnamefont {Hakim}},
  \bibinfo {author} {\bibfnamefont {H.~G.}\ \bibnamefont {Yevick}}, \bibinfo
  {author} {\bibfnamefont {G.}~\bibnamefont {Duclos}},\ and\ \bibinfo {author}
  {\bibfnamefont {P.}~\bibnamefont {Silberzan}},\ }\href@noop {} {\bibfield
  {journal} {\bibinfo  {journal} {Nature communications}\ }\textbf {\bibinfo
  {volume} {5}},\ \bibinfo {pages} {3747} (\bibinfo {year} {2014})}\BibitemShut
  {NoStop}%
\bibitem [{\citenamefont {Davidescu}\ \emph {et~al.}(2023)\citenamefont
  {Davidescu}, \citenamefont {Romanczuk}, \citenamefont {Gregor},\ and\
  \citenamefont {Couzin}}]{davidescu2023growth}%
  \BibitemOpen
  \bibfield  {author} {\bibinfo {author} {\bibfnamefont {M.~R.}\ \bibnamefont
  {Davidescu}}, \bibinfo {author} {\bibfnamefont {P.}~\bibnamefont
  {Romanczuk}}, \bibinfo {author} {\bibfnamefont {T.}~\bibnamefont {Gregor}},\
  and\ \bibinfo {author} {\bibfnamefont {I.~D.}\ \bibnamefont {Couzin}},\
  }\href@noop {} {\bibfield  {journal} {\bibinfo  {journal} {Proceedings of the
  National Academy of Sciences}\ }\textbf {\bibinfo {volume} {120}},\ \bibinfo
  {pages} {e2206163120} (\bibinfo {year} {2023})}\BibitemShut {NoStop}%
\bibitem [{\citenamefont {Prakash}\ \emph {et~al.}(2021)\citenamefont
  {Prakash}, \citenamefont {Bull},\ and\ \citenamefont
  {Prakash}}]{prakash2021motility}%
  \BibitemOpen
  \bibfield  {author} {\bibinfo {author} {\bibfnamefont {V.~N.}\ \bibnamefont
  {Prakash}}, \bibinfo {author} {\bibfnamefont {M.~S.}\ \bibnamefont {Bull}},\
  and\ \bibinfo {author} {\bibfnamefont {M.}~\bibnamefont {Prakash}},\
  }\href@noop {} {\bibfield  {journal} {\bibinfo  {journal} {Nature Physics}\
  }\textbf {\bibinfo {volume} {17}},\ \bibinfo {pages} {504} (\bibinfo {year}
  {2021})}\BibitemShut {NoStop}%
\bibitem [{\citenamefont {Lam}\ \emph {et~al.}(2015)\citenamefont {Lam},
  \citenamefont {Schindler},\ and\ \citenamefont {Dauchot}}]{lam2015self}%
  \BibitemOpen
  \bibfield  {author} {\bibinfo {author} {\bibfnamefont {K.-D. N.~T.}\
  \bibnamefont {Lam}}, \bibinfo {author} {\bibfnamefont {M.}~\bibnamefont
  {Schindler}},\ and\ \bibinfo {author} {\bibfnamefont {O.}~\bibnamefont
  {Dauchot}},\ }\href@noop {} {\bibfield  {journal} {\bibinfo  {journal} {New
  Journal of Physics}\ }\textbf {\bibinfo {volume} {17}},\ \bibinfo {pages}
  {113056} (\bibinfo {year} {2015})}\BibitemShut {NoStop}%
\bibitem [{\citenamefont {Yllanes}\ \emph {et~al.}(2017)\citenamefont
  {Yllanes}, \citenamefont {Leoni},\ and\ \citenamefont
  {Marchetti}}]{yllanes2017many}%
  \BibitemOpen
  \bibfield  {author} {\bibinfo {author} {\bibfnamefont {D.}~\bibnamefont
  {Yllanes}}, \bibinfo {author} {\bibfnamefont {M.}~\bibnamefont {Leoni}},\
  and\ \bibinfo {author} {\bibfnamefont {M.}~\bibnamefont {Marchetti}},\
  }\href@noop {} {\bibfield  {journal} {\bibinfo  {journal} {New Journal of
  Physics}\ }\textbf {\bibinfo {volume} {19}},\ \bibinfo {pages} {103026}
  (\bibinfo {year} {2017})}\BibitemShut {NoStop}%
\bibitem [{\citenamefont {Lazzari}\ \emph {et~al.}(2024)\citenamefont
  {Lazzari}, \citenamefont {Dauchot},\ and\ \citenamefont
  {Brito}}]{lazzari2024tuning}%
  \BibitemOpen
  \bibfield  {author} {\bibinfo {author} {\bibfnamefont {D.}~\bibnamefont
  {Lazzari}}, \bibinfo {author} {\bibfnamefont {O.}~\bibnamefont {Dauchot}},\
  and\ \bibinfo {author} {\bibfnamefont {C.}~\bibnamefont {Brito}},\
  }\href@noop {} {\bibfield  {journal} {\bibinfo  {journal} {Soft Matter}\
  }\textbf {\bibinfo {volume} {20}},\ \bibinfo {pages} {8570} (\bibinfo {year}
  {2024})}\BibitemShut {NoStop}%
\bibitem [{\citenamefont {Tang}\ \emph {et~al.}(2025)\citenamefont {Tang},
  \citenamefont {Zheng}, \citenamefont {Shee}, \citenamefont {Lin},
  \citenamefont {Han}, \citenamefont {Romanczuk},\ and\ \citenamefont
  {Huepe}}]{tang2025collective}%
  \BibitemOpen
  \bibfield  {author} {\bibinfo {author} {\bibfnamefont {W.}~\bibnamefont
  {Tang}}, \bibinfo {author} {\bibfnamefont {Y.}~\bibnamefont {Zheng}},
  \bibinfo {author} {\bibfnamefont {A.}~\bibnamefont {Shee}}, \bibinfo {author}
  {\bibfnamefont {G.}~\bibnamefont {Lin}}, \bibinfo {author} {\bibfnamefont
  {Z.}~\bibnamefont {Han}}, \bibinfo {author} {\bibfnamefont {P.}~\bibnamefont
  {Romanczuk}},\ and\ \bibinfo {author} {\bibfnamefont {C.}~\bibnamefont
  {Huepe}},\ }\href@noop {} {\bibfield  {journal} {\bibinfo  {journal} {SciPost
  Physics}\ }\textbf {\bibinfo {volume} {19}},\ \bibinfo {pages} {012}
  (\bibinfo {year} {2025})}\BibitemShut {NoStop}%
\bibitem [{\citenamefont {Tang}\ \emph {et~al.}(2026)\citenamefont {Tang},
  \citenamefont {Shee}, \citenamefont {Han}, \citenamefont {Romanczuk},
  \citenamefont {Zheng},\ and\ \citenamefont {Huepe}}]{tang2026passivity}%
  \BibitemOpen
  \bibfield  {author} {\bibinfo {author} {\bibfnamefont {W.}~\bibnamefont
  {Tang}}, \bibinfo {author} {\bibfnamefont {A.}~\bibnamefont {Shee}}, \bibinfo
  {author} {\bibfnamefont {Z.}~\bibnamefont {Han}}, \bibinfo {author}
  {\bibfnamefont {P.}~\bibnamefont {Romanczuk}}, \bibinfo {author}
  {\bibfnamefont {Y.}~\bibnamefont {Zheng}},\ and\ \bibinfo {author}
  {\bibfnamefont {C.}~\bibnamefont {Huepe}},\ }\href@noop {} {\bibfield
  {journal} {\bibinfo  {journal} {arXiv preprint arXiv:2604.15105}\ } (\bibinfo
  {year} {2026})}\BibitemShut {NoStop}%
\bibitem [{\citenamefont {Ben~Zion}\ \emph {et~al.}(2023)\citenamefont
  {Ben~Zion}, \citenamefont {Fersula}, \citenamefont {Bredeche},\ and\
  \citenamefont {Dauchot}}]{ben2023morphological}%
  \BibitemOpen
  \bibfield  {author} {\bibinfo {author} {\bibfnamefont {M.~Y.}\ \bibnamefont
  {Ben~Zion}}, \bibinfo {author} {\bibfnamefont {J.}~\bibnamefont {Fersula}},
  \bibinfo {author} {\bibfnamefont {N.}~\bibnamefont {Bredeche}},\ and\
  \bibinfo {author} {\bibfnamefont {O.}~\bibnamefont {Dauchot}},\ }\href@noop
  {} {\bibfield  {journal} {\bibinfo  {journal} {Science Robotics}\ }\textbf
  {\bibinfo {volume} {8}},\ \bibinfo {pages} {eabo6140} (\bibinfo {year}
  {2023})}\BibitemShut {NoStop}%
\bibitem [{\citenamefont {Casiulis}\ \emph {et~al.}(2025)\citenamefont
  {Casiulis}, \citenamefont {Arbel}, \citenamefont {van Waes}, \citenamefont
  {Lahini}, \citenamefont {Martiniani}, \citenamefont {Oppenheimer},\ and\
  \citenamefont {Zion}}]{casiulis2025geometric}%
  \BibitemOpen
  \bibfield  {author} {\bibinfo {author} {\bibfnamefont {M.}~\bibnamefont
  {Casiulis}}, \bibinfo {author} {\bibfnamefont {E.}~\bibnamefont {Arbel}},
  \bibinfo {author} {\bibfnamefont {C.}~\bibnamefont {van Waes}}, \bibinfo
  {author} {\bibfnamefont {Y.}~\bibnamefont {Lahini}}, \bibinfo {author}
  {\bibfnamefont {S.}~\bibnamefont {Martiniani}}, \bibinfo {author}
  {\bibfnamefont {N.}~\bibnamefont {Oppenheimer}},\ and\ \bibinfo {author}
  {\bibfnamefont {M.~Y.~B.}\ \bibnamefont {Zion}},\ }\href@noop {} {\bibfield
  {journal} {\bibinfo  {journal} {Proceedings of the National Academy of
  Sciences}\ }\textbf {\bibinfo {volume} {122}},\ \bibinfo {pages}
  {e2502211122} (\bibinfo {year} {2025})}\BibitemShut {NoStop}%
\bibitem [{\citenamefont {Byron}\ and\ \citenamefont
  {Fuller}(2012)}]{byron2012mathematics}%
  \BibitemOpen
  \bibfield  {author} {\bibinfo {author} {\bibfnamefont {F.~W.}\ \bibnamefont
  {Byron}}\ and\ \bibinfo {author} {\bibfnamefont {R.~W.}\ \bibnamefont
  {Fuller}},\ }\href@noop {} {\emph {\bibinfo {title} {Mathematics of
  {Classical} and {Quantum} {Physics}}}}\ (\bibinfo  {publisher} {Courier
  Corporation},\ \bibinfo {year} {2012})\ \bibinfo {note} {google-Books-ID:
  kzeiyDgckvIC}\BibitemShut {NoStop}%
\end{thebibliography}%
 
\end{document}